\documentclass[hyper]{JHEP3}

\usepackage{axodraw}
\usepackage[dvips]{graphicx}
\usepackage{epsfig}
\usepackage{multicol}
\usepackage{bbm}
\usepackage{amsmath}
\usepackage{cite}
\usepackage{lscape}
\usepackage{multirow}

\newcommand{\beq}{\begin{equation}}
\newcommand{\eeq}{\end{equation}}
\newcommand{\bea}{\begin{eqnarray}}
\newcommand{\eea}{\end{eqnarray}}

\newcommand{\lsim}{
\mathrel{\hbox{\rlap{\hbox{\lower4pt\hbox{$\sim$}}}\hbox{$<$}}}}
\newcommand{\gsim}{
\mathrel{\hbox{\rlap{\hbox{\lower4pt\hbox{$\sim$}}}\hbox{$>$}}}}

\title{Using kinematic boundary lines for particle mass measurements
and disambiguation in SUSY-like events with missing energy}

\author{
Michael Burns, Konstantin T.~Matchev, Myeonghun Park \\
Physics Department, University of Florida,
Gainesville, FL 32611, USA 
}

\abstract{We revisit the method of kinematical endpoints
for particle mass determination, applied to the popular 
SUSY decay chain $\tilde q\to \tilde\chi^0_2\to \tilde\ell\to\tilde\chi^0_1$.
We analyze the uniqueness of the solutions for the mass 
spectrum in terms of the measured endpoints in the observable
invariant mass distributions. We provide simple analytical 
inversion formulas for the masses in terms of the measured endpoints.
We show that in a sizable portion of the SUSY mass parameter space 
the solutions always suffer from a two-fold ambiguity,
due to the fact that the original relations between the masses 
and the endpoints are piecewise-defined functions. The ambiguity
persists even in the ideal case of a perfect detector 
and infinite statistics. We delineate the corresponding 
dangerous regions of parameter space and identify the sets
of ``twin'' mass spectra. In order to resolve
the ambiguity, we propose a generalization of the endpoint method, 
from single-variable distributions to two-variable distributions.
In particular, we study analytically the boundaries of the 
$\{m_{j\ell (lo)},m_{j\ell (hi)}\}$ and $\{m_{\ell\ell},m_{j\ell\ell}\}$
distributions and prove that their shapes are in principle
sufficient to resolve the ambiguity in the mass determination.
We identify several additional independent measurements
which can be obtained from the boundary lines of these bivariate
distributions. 
The purely kinematical nature of our method makes it generally 
applicable to any model that exhibits a SUSY-like cascade decay.
}


\preprint{UFIFT-HEP-09-03 \\
          March 25, 2009
          } % OR: \preprint{Aaaa/Mm/Yy\\Aaa-aa/Nnnnnn}
\begin{document}
%%%%%%%%%%%%%%%%%%%%%%%%%%%%%%%%%%%%%%%%%%%%%%%%%%%%%%%

\section{Introduction}
\label{introduction_section}

The dark matter problem of astroparticle physics \cite{Bertone:2004pz}
greatly motivates the search for neutral, stable
and weakly interacting massive particles (WIMPs)
at colliders \cite{Hubisz:2008gg}. WIMPs are also rather 
ubiquitous in Beyond Standard Model (BSM) physics at the TeV scale.
Unfortunately, at the hadron colliders
of the current energy frontier (the Fermilab Tevatron and 
the Large Hadron Collider (LHC) at CERN), the process
of {\em direct} WIMP production, tagged with a jet or 
a photon from initial state radiation, suffers from 
insurmountable backgrounds \cite{Birkedal:2004xn,Feng:2005gj,Konar:2009ae}.
In contrast, the chances of a discovery are typically  
greatly enhanced in case of {\em indirect} production, where
the WIMPs are produced in the decays of heavier,
more strongly interacting particles.

Since the WIMPs are usually stable due to some new conserved quantum number,
they cannot be singly produced in collisions of light SM particles.
The prototypical WIMP example is the lightest superpartner (LSP)
(typically the lightest neutralino $\tilde \chi^0_1$)
in low-energy supersymmetry (SUSY) with $R$-parity conservation 
\cite{Chung:2003fi}\footnote{More recently, it was realized that 
many of the features of $R$-parity conserving SUSY are also shared 
by other model frameworks, such as Universal Extra Dimensions (UED) 
\cite{Appelquist:2000nn,Cheng:2002iz,Cheng:2002ab}, Warped Extra Dimensions 
\cite{Agashe:2004ci,Agashe:2007jb,Panico:2008bx}, 
Little Higgs theory with $T$-parity \cite{Cheng:2003ju,Cheng:2004yc}, etc.}. 
The superpartners are produced in pairs and each one decays through a
cascade decay chain down to the $\tilde\chi^0_1$ WIMP, which 
does not interact in the detector and can only manifest itself 
as missing energy (MET). Since each event contains two 
unobserved WIMPs (with unknown mass), measuring
the masses, spins, etc. of the new particles is a very challenging task. 
In recognition of this fact, in recent years 
there has been an increased interest in developing new techniques for mass
\cite{Hinchliffe:1996iu,Lester:1999tx,
Bachacou:1999zb,Hinchliffe:1999zc,atlas,Nojiri:2000wq,Allanach:2000kt,Barr:2003rg,Nojiri:2003tu,lester,
Kawagoe:2004rz,Gjelsten:2004ki,Gjelsten:2005aw,Birkedal:2005cm,Lester:2005je,Miller:2005zp,
Meade:2006dw,Lester:2006yw,Lester:2006cf,Gjelsten:2006tg,Matsumoto:2006ws,Cheng:2007xv,Lester:2007fq,
Cho:2007qv,Gripaios:2007is,Barr:2007hy,Cho:2007dh,Ross:2007rm,Nojiri:2007pq,Huang:2008ae,
Nojiri:2008hy,Tovey:2008ui,Nojiri:2008ir,Cheng:2008mg,Cho:2008cu,Serna:2008zk,Bisset:2008hm,
Barr:2008ba,Kersting:2008qn,Nojiri:2008vq,Cheng:2008hk,Burns:2008va,Barr:2008hv,Konar:2008ei}
and spin \cite{Barr:2004ze,Battaglia:2005zf,Smillie:2005ar,Battaglia:2005ma,Datta:2005zs,
Datta:2005vx,Barr:2005dz,Alves:2006df,Athanasiou:2006ef,Wang:2006hk,Athanasiou:2006hv,SA:2006jm,
Smillie:2006cd,Kong:2006pi,Kilic:2007zk,Alves:2007xt,Csaki:2007xm,Datta:2007xy,Buckley:2007th,
Buckley:2008pp,Kane:2008kw,Burns:2008cp,Cho:2008tj}
measurements in such SUSY-like missing energy events.

\FIGURE[t]{
\begin{picture}(200,80)(0,-30)
\Line(0,0)(50,0)			% D
\Text(25,-5)[t]{$D$}			% D
\Vertex(50,0){2}
\Line(50,0)(80,30)			% j
\Text(82,32)[bl]{$j$}			% j
\Line(50,0)(100,0)			% C
\Text(75,-5)[t]{$C$}			% C
\Vertex(100,0){2}
\Line(100,0)(130,30)		% l_n
\Text(132,32)[bl]{$l_n^{\pm}$}	% l_n
\Line(100,0)(150,0)		% B
\Text(125,-5)[t]{$B$}			% B
\Vertex(150,0){2}
\Line(150,0)(180,30)		% l_f
\Text(182,30)[bl]{$l_f^{\mp}$}	% l_f
\Line(150,0)(200,0)		% A
\Text(175,-5)[t]{$A$}			% A
\Text(100,-30)[c]{a) two-body (on-shell) scenario}			% A
\end{picture}
\begin{picture}(200,80)(0,-30)
\Line(25,0)(75,0)			% D
\Text(50,-5)[t]{$D$}			% D
\Vertex(75,0){2}
\Line(75,0)(105,30)		% j
\Text(107,32)[bl]{$j$}			% j
\Line(75,0)(125,0)			% C
\Text(100,-5)[t]{$C$}			% C
\Vertex(125,0){2}
\Line(125,0)(155,30)		% l_n
\Text(157,32)[bl]{$l_n^{\pm}$}	% l_n
\Line(125,0)(160,20)		% l_f
\Text(164,22)[l]{$l_f^{\mp}$}		% l_f
\Line(125,0)(170,-10)		% A
\Text(147,-9)[t]{$A$}			% A
\Text(100,-30)[c]{b) three-body (off-shell) scenario}			% A
\end{picture}
\caption{The generic decay chain considered in this paper:
$D\rightarrow{}jC\rightarrow{}jl_n^{\pm}B\rightarrow{}jl_n^{\pm}l_f^{\mp}A$.
Particles $A$, $B$, $C$ and $D$ are new BSM particles, while the
corresponding SM decay products consist of a jet $j$, 
a ``near'' lepton $l_n^\pm$ and a ``far'' lepton $l_f^\mp$.
a) In the two-body (on-shell) scenario, $C$ is kinematically allowed to 
decay to $B$, which then decays to $A$.
b) In the three-body (off-shell) scenario, $C$ is kinematically prohibited 
to decay to $B$, and decays directly to $A$.
\label{fig:chain}
}
}

%%%%%%%%%%%%%%%%%%%%%%%%%%%%%%%%
% explanation of decay schemes %
%%%%%%%%%%%%%%%%%%%%%%%%%%%%%%%%

There are three basic types of mass determination methods, 
which are reviewed and contrasted in Ref.~\cite{Burns:2008va}.
In this paper we concentrate on the classic method of
kinematical endpoints \cite{Hinchliffe:1996iu}. 
Following the previous SUSY studies,
for illustration of our results we shall use the generic decay chain 
$D\rightarrow{}jC\rightarrow{}jl_n^{\pm}B\rightarrow{}jl_n^{\pm}l_f^{\mp}A$
shown in Fig.~\ref{fig:chain}.
Here $A$, $B$, $C$ and $D$ are heavy BSM particles, while the
corresponding SM decay products are: a QCD jet $j$, 
a ``near'' lepton $l_n^\pm$ and a ``far'' lepton $l_f^\mp$.
This chain is quite common in SUSY, with the identification
$D=\tilde q$, $C=\tilde\chi^0_2$, $B=\tilde l$ and $A=\tilde\chi^0_1$,
where $\tilde q$ is a squark, $\tilde l$ is a slepton, and
$\tilde\chi^0_1$ ($\tilde\chi^0_2$) is the first (second) 
lightest neutralino. However, our analysis is not limited to SUSY,
since the decay chain in Fig.~\ref{fig:chain} is rather typical for
other BSM models as well, e.g. UED \cite{Cheng:2002ab}.
For completeness, we shall consider the
two different cases shown in Figs.~\ref{fig:chain}(a) and
\ref{fig:chain}(b), correspondingly. In Fig.~\ref{fig:chain}(a)
$m_B<m_C$, so that the $C\rightarrow l_n^\pm B$ decay is two-body.
In what follows, we shall refer to this case as the  
``two-body'' or ``on-shell'' scenario. On the other hand,
in Fig.~\ref{fig:chain}(b) $m_B>m_C$ and the decay 
$C\rightarrow l_n^\pm l_f^\mp A$ is three-body, leading 
to a ``three-body'' or ``off-shell'' scenario \cite{Lester:2006cf}.
In the two-body scenario, the goal is to determine all four 
unknown masses, $m_D$, $m_C$, $m_B$, and $m_A$.
In the three-body scenario, the goal is to determine the 
three\footnote{In the three-body scenario of Fig.~\ref{fig:chain}(b),
it may still be possible to extract the fourth mass $m_B$ from the data, e.g.
by analyzing the shapes of the invariant mass distributions
\cite{Birkedal:2005cm}. However, those approaches are 
quite challenging, since the shapes depend on a number of 
additional factors: the experimental resolution, the spins of the
new particles, the shape of the underlying backgrounds, etc.
In contrast, here we are concentrating on methods which 
use only kinematic endpoint information and are thus
immune to those detrimental factors.} 
unknown masses, $m_D$, $m_C$, and $m_A$.
Of course, the scenario is not known until the data are examined,
thus an additional goal of our analysis will be to identify the
particular scenario at hand.

%%%%%%%%%%%%%%%%%%%%%%%%%%%%%%%%%%%%%%%%
% explanation of particle combinations %
%%%%%%%%%%%%%%%%%%%%%%%%%%%%%%%%%%%%%%%%

The idea of the kinematic endpoint method is very simple.
Given the SM decay products exhibited in Fig.~\ref{fig:chain},
form the invariant mass\footnote{We shall see below that the
formulas simplify if we consider invariant masses {\em squared} 
instead. This distinction is not central to our analysis.}
of every possible combination, 
$m_{ll}$, $m_{jl_n}$, $m_{jl_f}$, and $m_{jll}$.
Since $l_n^{\pm}$ and $l_f^{\mp}$ cannot be distinguished 
on an event-by-event basis, 
one has to introduce an alternative definition of the $jl$
distributions. For example, one can identify the two leptons 
by their charge and consider the samples $\{j l^+\}$ and
$\{jl^- \}$, which are experimentally well defined.
This approach turned out to be very useful for spin studies
\cite{Barr:2004ze,Smillie:2005ar,Athanasiou:2006ef,Burns:2008cp},
since spin effects are encoded in the {\em difference}
between those two distributions. However, for mass determination,
it is more suitable to use an $m_{jl}$ ordering by invariant mass:
\begin{eqnarray}
m_{jl(lo)}
&\equiv& \min \left\{m_{jl_n}, m_{jl_f} \right\}  , \label{mjllodef} \\ [2mm]
m_{jl(hi)}
&\equiv& \max \left\{m_{jl_n}, m_{jl_f} \right\}  . \label{mjlhidef}
\end{eqnarray}
Both of the newly defined quantities $m_{jl(lo)}$ and $m_{jl(hi)}$ 
also exhibit upper kinematic endpoints ($m_{jl(lo)}^{max}$ and $m_{jl(hi)}^{max}$, correspondingly), 
which are experimentally measurable. Together with the measured 
kinematic endpoints $m_{ll}^{max}$ and $m_{jll}^{max}$
of the $m_{ll}$ and $m_{jll}$ distributions, this gives
4 measurements 
\begin{equation}
m_{ll}^{max}, m_{jll}^{max}, m_{jl(lo)}^{max}, m_{jl(hi)}^{max},
\label{4meas}
\end{equation}
which are known functions\footnote{See Section \ref{sec:forward} below.} 
of only 4 unknown parameters ($m_A$, $m_B$, $m_C$ and $m_D$).
Therefore, by inverting those relations, i.e. solving the so 
called ``inverse problem'' at the LHC \cite{ArkaniHamed:2005px}, 
one would expect to be able 
to determine the complete spectrum, at least as a matter of principle.

%%%%%%%%%%%%%%%
% the problem %
%%%%%%%%%%%%%%%
However, this determination can be ambiguous, 
and several alternative solutions for the masses may emerge,
as already recognized in, 
for example \cite{Gjelsten:2004ki,ArkaniHamed:2005px,Gjelsten:2005sv,Gjelsten:2006as,MB1,MP1,KM1,Luc,Costanzo:2009mq}.
This may happen for one of the following reasons.
\begin{enumerate}
\item {\em Insufficient number of measurements.}
The four measurements (\ref{4meas}) may not all be independent from each other.
Indeed, there are certain regions of parameter space (reviewed explicitly 
below in Sec.~\ref{sec:forward}) where one finds the following correlation
\cite{Gjelsten:2004ki}
\begin{equation}
\left( m_{jll}^{max}\right)^2 = \left( m_{jl(hi)}^{max}\right)^2 + \left( m_{ll}^{max}\right)^2.
\label{mjllcorrelation}
\end{equation}
In this case, the four measurements (\ref{4meas}) are clearly insufficient and one has
to supplement them with an additional measurement. To this end, it has been suggested 
to consider the constrained distribution $m_{jll(\theta>\frac{\pi}{2})}$, which exhibits
a useful {\em lower} kinematic endpoint $m_{jll(\theta>\frac{\pi}{2})}^{min}$ \cite{Allanach:2000kt}.
The distribution $m_{jll(\theta>\frac{\pi}{2})}$ is nothing but the
usual $m_{jll}$ distribution over a subset of the original events,
subject to the additional dilepton mass constraint 
\begin{equation}
\frac{m_{ll}^{max}}{\sqrt{2}} < m_{ll} < m_{ll}^{max}\, .
\end{equation}
In the rest frame of particle $B$, this cut implies the following 
restriction on the opening angle $\theta$ between the two leptons \cite{Nojiri:2000wq}
\begin{equation}
\theta > \frac{\pi}{2}\, ,
\end{equation}
thus justifying the notation for $m_{jll(\theta>\frac{\pi}{2})}$.
In what follows, we shall therefore always supplement the 
original set of 4 measurements (\ref{4meas}) with the additional
measurement of $m_{jll(\theta>\frac{\pi}{2})}^{min}$:
\begin{equation}
m_{ll}^{max}, m_{jll}^{max}, m_{jl(lo)}^{max}, m_{jl(hi)}^{max}, m_{jll(\theta>\frac{\pi}{2})}^{min},
\label{5meas}
\end{equation}
so that in principle there is sufficient information to determine 
the four unknown masses. Then, we shall concentrate on the question 
whether this determination is unique or not, i.e. we shall be concerned 
only with discrete ambiguities. As discussed in more detail below in
Sections~\ref{sec:analytical} and \ref{sec:duplication}, these discrete ambiguities
arise due to the very nature of the mathematical problem:
the relations giving the endpoints in terms of the masses are
{\em piecewise-defined} functions, i.e. their definitions depend on the
values of the independent variables (the masses $m_A$, $m_B$, $m_C$ and $m_D$).
Since the masses are unknown, it is not clear which definition 
is the relevant one, and one must consider all possibilities, obtain each solution,
and test for consistency at the very end.
\item {\em Experimental resolution.} 
Ideally, the procedure just described would yield a single 
consistent solution. Indeed, this is what happens throughout a large
portion of the parameter space. One should keep in mind 
that the measurements (\ref{5meas}) inevitably come with some
experimental errors, so that within those experimental uncertainties, 
two or more solutions are possible \cite{Gjelsten:2004ki,Gjelsten:2005aw,Gjelsten:2005sv}. 
One specific example of this type is shown in Table~\ref{table:dup} for
the SPS1a($\alpha$) mass spectrum, which was extensively studied in
\cite{Gjelsten:2004ki}. Even with $300\ {\rm fb}^{-1}$ of data at the LHC,
the residual experimental uncertainties (due to the finite 
detector resolution, statistical and systematic errors, etc.)
will still allow two solutions: a ``true'' and a ``false'' one,
as shown in the table.
%
% BEGINNING OF NEW TABLE
\TABULAR[ht]{|c||r|r|r||r|r||r|r|}{
\hline
         &  \multicolumn{3}{c||}{SPS1a($\alpha$) \cite{Gjelsten:2004ki,Gjelsten:2005sv}}
         &  \multicolumn{2}{c|}{SU1 \cite{Gjelsten:2006as}}
         &  \multicolumn{2}{c|}{SU3 \cite{Gjelsten:2006as}}    \\ [1mm]
\cline{2-8}
Variable &  Nominal  &   True   &  False   &  True  &   False  &  True  &   False   \\  [1mm] \hline \hline
$m_{\tilde\chi^0_1}$   
         &   96.1    &   96.3   &  85.3    &  137.0 &  122.1   &  118.0 &   346.8   \\ [1mm] \hline
$m_{\tilde l_R }$     
         &  143.0    &  143.2   & 130.4    &  254.0 &  127.5   &  155.0 &   411.1   \\ [1mm] \hline
$m_{\tilde\chi^0_2}$     
         &  176.8    &  177.0   & 165.5    &  264.0 &  245.9   &  219.0 &   451.6   \\ [1mm] \hline
$m_{\tilde q_L}$     
         &  537.2    &  537.5   & 523.5    &  760.0 &  743.6   &  631.0 &   899.9   \\ [1mm] \hline \hline
Region   & (1,1)     &  (1,1)   &  (1,2)   & (1,1)  &  (1,3)   &  (1,3) &   (1,1) \\ [1mm] \hline
\hline
$m_{ll}^{max}$    
         & 77.0      &   77.0   &   77.1   & \multicolumn{2}{c||}{61} 
                                                               & \multicolumn{2}{c|}{100} \\ [1mm] \hline
$m_{jl(lo)}^{max}$ 
         & 298.3     &  298.3   &  299.6   & \multicolumn{2}{c||}{194} 
                                                               & \multicolumn{2}{c|}{322} \\ [1mm] \hline
$m_{jl(hi)}^{max}$ 
         & 375.6     &  375.6   &  375.7   & \multicolumn{2}{c||}{600} 
                                                               & \multicolumn{2}{c|}{418} \\ [1mm] \hline
$m_{jll}^{max}$    
         & 425.8     &  425.8   &  425.6   & \multicolumn{2}{c||}{609} 
                                                               & \multicolumn{2}{c|}{499} \\ [1mm] \hline
$m_{jll(\theta>\frac{\pi}{2})}^{min}$ 
        &  200.6     &  200.6   &  205.1   & 143    & 148      & 247     & 214 \\ [1mm] \hline
% END OF NEW TABLE
}{\label{table:dup} Examples of ``mass ambiguities'' previously reported in 
\cite{Gjelsten:2004ki,Gjelsten:2005sv,Gjelsten:2006as}. The nominal values for the mass
spectrum are given in the leftmost column in each case. The analysis 
of \cite{Gjelsten:2004ki,Gjelsten:2005sv} for the SPS1a($\alpha$)
study point used all five available measurements (\ref{5meas}), and
included detector resolution effects and statistical and 
systematic errors. As a result, in the case of SPS1a($\alpha$) there are
two solutions: the ``true'' one is in the correct region (1,1) and 
is close to the nominal values, while the ``false'' one is in the wrong 
region (1,2), but nevertheless matches all of the observed invariant mass 
endpoints (\ref{5meas}) within the experimental uncertainties. 
The study points SU1 and SU3 are taken from \cite{Gjelsten:2006as},
where one requires a perfect match to only the four kinematic endpoints 
(\ref{4meas}), ignoring any experimental errors.
In this case the true and false spectra predict different values of 
$m_{jll(\theta>\frac{\pi}{2})}^{min}$.
}
However, this is not a true ambiguity in the sense that
it arises simply due to limitations in the experimental precision.
With time, the latter would be expected to improve 
and the ambiguity may eventually get resolved.  
For example, the statistical errors would be reduced with even 
more data\footnote{It is also worth noting that ref.~\cite{Gjelsten:2004ki}
conservatively assigned a rather large systematic 
error for the $m_{jll(\theta>\frac{\pi}{2})}^{min}$ measurement, 
since the analytical shape of its edge was unknown at the time.
Since then, the shape was derived in \cite{Lester:2006yw}, 
so that by now the threshold measurement $m_{jll(\theta>\frac{\pi}{2})}^{min}$
should be considered on equal footing with the other
measurements in (\ref{5meas}).}.
\item {\em Non-uniqueness of the inversion.} Even in the ideal case 
of a perfect experiment, which would yield results for
all the five measurements (\ref{5meas}) with zero error bars, 
there may still be multiple solutions to the inversion problem.
One of the main goals of this paper is to identify the specific 
circumstances when this takes place. In Section~\ref{sec:duplication}
we shall analyze the physical mass parameter space of SUSY  
and find a sizable portion in it where an {\em exact} duplication occurs,
i.e. if Nature chooses a SUSY spectrum from that region, the 
measurements (\ref{5meas}) will be consistent with two and 
only two SUSY mass spectra: the nominal one, plus a ``fake''.
We emphasize the fact that the duplicate solutions we find
yield mathematically {\em identical} values for {\em all five} experimental
observables in (\ref{5meas}). Therefore, neither improvements 
in the experimental resolution, nor increased statistics will 
be able to resolve this duplication. Several studies in the 
literature \cite{Gjelsten:2004ki,Gjelsten:2006tg,Gjelsten:2005sv,Gjelsten:2006as}
have already raised the issue of a potential ambiguity in the SUSY 
mass determination. Some representative examples from those works 
are shown in Table~\ref{table:dup}. As already mentioned, the
duplication found in \cite{Gjelsten:2004ki,Gjelsten:2005sv} 
in the case of SPS1a($\alpha$) was simply due to the experimental
uncertainties, and would be resolved in a perfect experiment.
On the other hand, the duplication in the case of SU1 and SU3
found in Ref.~\cite{Gjelsten:2006as} is exact, but relies on only
four (namely, the set (\ref{4meas})) out of the five available 
measurements (\ref{5meas}).
As seen from Table~\ref{table:dup}, the inclusion 
of the threshold $m_{jll(\theta>\frac{\pi}{2})}^{min}$ will
in principle resolve the ambiguity. In contrast,
we use the full set of measurements (\ref{5meas}), 
including $m_{jll(\theta>\frac{\pi}{2})}^{min}$,
and we still find exact duplication. In this sense, 
our findings, first reported in \cite{MP1,KM1}, are new,
and extend the results of \cite{Gjelsten:2004ki,Gjelsten:2006tg,Gjelsten:2005sv,Gjelsten:2006as}.
For example, we find that exact duplication occurs in the
(2,3), (3,1) and (3,2) parameter space regions (according to the
classification in Section~\ref{sec:analytical}), while
the examples in Table \ref{table:dup} belong to regions
(1,1), (1,2) and (1,3). It is also worth pointing out that 
the duplicated regions of parameter space that we find 
are {\em not} consistent with a typical MSUGRA-type scenario, 
which may explain why this problem has not been more broadly 
appreciated earlier. Numerical examples of a duplication similar
to ours have previously been presented in \cite{ArkaniHamed:2005px}, 
and our analytical results below in Section~\ref{sec:duplication}
now help understand their origin.
\end{enumerate}

Having identified the problem of duplication in the measured mass spectrum, 
in the second part of the paper we present a new method for its solution.
As already emphasized, the two-fold ambiguity in the spectrum is exact, 
so it cannot be resolved by simply improving the experimental precision on
the kinematical endpoint measurements (\ref{5meas}). Instead, additional
experimental input is needed. One option is to consider a longer
decay chain, which would yield several additional endpoint measurements.
For example, the decay chains in Fig.~\ref{fig:chain} may 
begin with an even heavier particle (say, $E$), at the
expense of a single new parameter (the mass of particle $E$)
\cite{Gjelsten:2005aw}. However, the presence of such a decay 
chain in the data is a model-dependent assumption and 
is by no means guaranteed. Alternatively, one may supplement
(\ref{5meas}) with data from a future lepton collider \cite{Gjelsten:2005sv},
but its existence is also an assumption and is by no means 
guaranteed. Therefore we do not consider these possibilities here.

Instead, we concentrate on the question: What additional information,
which is already present in the hadron collider data, can be used to
resolve the ambiguity? It is important to realize that in very
general terms, the kinematics of the decay in Fig.~\ref{fig:chain} 
is governed by some three-dimensional differential distribution 
\begin{equation}
\frac{d^3 \Gamma}{d\alpha\, d\beta\, d\gamma},
\label{d3Gdalpha}
\end{equation}
where $\alpha$, $\beta$ and $\gamma$ are some suitably chosen angles
specifying the particular decay configuration (see, e.g. \cite{Smillie:2005ar}). 
Through a change of variables, these angles can be traded for three 
invariant mass combinations of the visible decay products in Fig.~\ref{fig:chain}, 
e.g. $m_{ll}, m_{jl^+}, m_{jl^-}$ \cite{Athanasiou:2006hv}, but other
sets are equally possible, let us denote a generic such set by
$\{m_1,m_2,m_3\}$. In place of (\ref{d3Gdalpha}) one then has
\begin{equation}
\frac{d^3 \Gamma}{dm_1 dm_2 dm_3}\ .
\label{d3Gdm}
\end{equation}
The distribution (\ref{d3Gdm}) is experimentally observable and
is nothing but a three-dimensional histogram. It contains the full 
information about the decay in Fig.~\ref{fig:chain}, including 
the particle mass and spin information. The only disadvantage 
of (\ref{d3Gdm}) is that it cannot be easily visualized.

In order to obtain a kinematic endpoint for some mass parameter, 
say $m_1$, one then simply integrates over the other two degrees of freedom, 
and builds the one-dimensional distribution
\begin{equation}
\frac{d \Gamma}{dm_1} \equiv \int dm_2 dm_3\, \frac{d^3 \Gamma}{dm_1 dm_2 dm_3}\ .
\label{dGdm}
\end{equation}
This, being a one-dimensional distribution, exhibits an 
upper {\em endpoint} $m_1^{max}$. However, in the process
of integration in (\ref{dGdm}), one is losing a certain 
amount of the original information contained in (\ref{d3Gdm}).
Some of this information can be recovered if we consider
a two-dimensional (bivariate) distribution, e.g. in $(m_1,m_2)$:
\begin{equation}
\frac{d^2 \Gamma}{dm_1 dm_2} \equiv \int dm_3\, \frac{d^3 \Gamma}{dm_1 dm_2 dm_3}\ .
\label{d2Gdm}
\end{equation}
This, being a two-dimensional distribution, will exhibit not an endpoint, but
a {\em boundary line}, which can be parameterized by a single parameter $t$ as
$(m_1(t), m_2(t))$. Finally, if we stick to the original
three-dimensional distribution (\ref{d3Gdm}), we will obtain 
a {\em boundary surface}, parameterized by two parameters, $t_1$ and $t_2$,
as $(m_1(t_1,t_2), m_2(t_1,t_2), m_3(t_1,t_2))$.
Given that bivariate and trivariate distributions 
are more informative than the simple one-dimensional histograms, it
is rather surprising that they have not been used more often in 
the previous analyses of SUSY mass determination.

%%%%%%%%%%%%%%%%%%%%%%%%%%%
% summarize rest of paper %
%%%%%%%%%%%%%%%%%%%%%%%%%%%

The second part of the paper is thus devoted to the analysis of 
bivariate distributions of the type (\ref{d2Gdm}) \footnote{Preliminary 
results of our work were reported in \cite{MB1,KM1}. Similar ideas were
discussed more recently in \cite{Luc,Costanzo:2009mq}.}.
In particular, in Section~\ref{sec:hivslo} (Section~\ref{sec:jllvsll})
we analyze the boundaries of the bivariate distributions 
in terms of $m_{jl(lo)}^2$ and $m_{jl(hi)}^2$
($m_{jll}^2$ and $m_{ll}^2$).
We show that the shapes of those distributions are very distinct 
and can be used to identify qualitatively the type of spectrum at hand,
thus resolving the duplication discussed above.
We also provide analytical formulas for the 
boundaries of the kinematically allowed regions, which can be used
to further quantitatively improve on the mass determination 
(see also \cite{Costanzo:2009mq}). Clearly, fitting to 
a line would yield a better precision of determining the mass 
parameters than simply fitting to a point. What is more, 
we shall show that the bivariate distributions offer the possibility
of several additional measurements, in addition to those in (\ref{5meas}).
These are the locations of some special points on the boundary lines,
for which we provide analytic expressions in terms of the 
masses $m_A$, $m_B$, $m_C$ and $m_D$. 
These special points are typically hidden as subtle features of the 
one-dimensional distributions but are transparent on the bivariate distributions
which we are advertising here.

In conclusion of this section, we summarize the main goals 
and results of this paper and point to the sections where 
those results can be found.
\begin{itemize}
\item {\em Analytical solution of the inverse problem.}
In Section~\ref{sec:inverse} we provide analytical formulas which allow
one to calculate directly the BSM mass spectrum $m_A$, $m_B$, $m_C$ and $m_D$
in terms of the experimental inputs (\ref{5meas}). Our formulas 
are completely general, for example, they are valid for both
the on-shell scenario of Fig.~\ref{fig:chain}(a) as well as 
the off-shell scenario of Fig.~\ref{fig:chain}(b).
In addition, they can be applied to {\em all} regions in parameter space.
The availability of exact analytical expressions for the mass
spectrum in terms of the observed kinematical endpoints makes 
numerical fitting (e.g.~with a program like Fittino \cite{Bechtle:2004pc})
unnecessary. An important simplification in our approach is that we only 
need to consider four different cases, as opposed to the 11 cases
usually discussed in the literature. 
\item In Section~\ref{sec:duplication} we identify the {\em complete}
SUSY mass parameter space where {\em exact} duplication occurs, i.e.
two very different mass spectra predict identical values for all 
five endpoint measurements (\ref{5meas}).
\item In Section~\ref{sec:hivslo} we analyze the shape of the
bivariate distribution in terms of $m_{jl(lo)}^2$ and $m_{jl(hi)}^2$.
We identify the characteristic shape of the boundary lines of 
the distribution for each of 
our four parameter space regions. The shape not only allows to
resolve the ambiguity discovered in Section~\ref{sec:duplication},
but also contains a lot of additional useful information.
For example, the shape analysis yields an additional measurement 
of an ``edge'' point, $m_{jl_f}^{(p)}$, and also allows to determine
the endpoints $m_{jl_n}^{max}$ and $m_{jl_f}^{max}$ 
of the underlying $m_{jl_n}$ and $m_{jl_f}$ distributions.
The analytic solution to the inverse problem (presented in Appendix~\ref{app:nearfar})
takes a particularly simple form if we make use of these new measurements 
and consider the alternative set
$\{m_{ll}^{max},m_{jl_f}^{(p)},m_{jl_n}^{max},m_{jl_f}^{max}\}$.
\item In Section~\ref{sec:jllvsll} we perform a similar shape analysis
of the bivariate distribution in terms of $m_{jll}^2$ versus $m_{ll}^2$.
\end{itemize}

\section{Analytical results}
\label{sec:analytical}

In this section we present the analytical formulas which 
allow one to go from the mass spectrum to the experimentally 
observable endpoints (Sec.~\ref{sec:forward}) and vice versa
(Sec.~\ref{sec:inverse}).

%%%%%%%%%%%%%%%%%%%%%%%%%%%%%%%%%
% definition of mass parameters %
%%%%%%%%%%%%%%%%%%%%%%%%%%%%%%%%%

Before we begin, we introduce some notation.
Following existing studies in the literature
\cite{Miller:2005zp,Smillie:2005ar,Athanasiou:2006ef,Burns:2008cp},
we shall redefine the original mass parameter space 
\begin{equation}
\left\{m_A,m_B,m_C,m_D\right\}
\label{mspace}
\end{equation} 
in terms of an overall squared mass scale, $m_D^2$, 
and squared mass ratios\footnote{The practice of redefining the 
parameter space in terms of squared mass ratios is quite common in the
literature. For example, our variables $\{R_{CD}, R_{BC}, R_{AB}\}$
exactly correspond to the variables $\{x,y,z\}$ used in
\cite{Smillie:2005ar,Athanasiou:2006ef,Burns:2008cp} 
and the parameters $\{R_C,R_B,R_A\}$ used in \cite{Miller:2005zp}.}
\begin{equation}
R_{ij} \equiv \frac{m_i^2}{m_j^2}
\ ,
\label{RABdef}
\end{equation}
where $i,j\in\left\{A,B,C,D\right\}$.
Note that there are only three independent squared mass ratios in (\ref{RABdef}), 
which we shall take as the set $\{R_{AB}, R_{BC}, R_{CD}\}$. However, in what follows we shall 
also make use of the other ratios, e.g. $R_{AC}$, $R_{AD}$ and $R_{BD}$,
whenever this will lead to a simplification of our formulas. Of course, 
the latter are related to our preferred set  $\{R_{AB}, R_{BC}, R_{CD}\}$
due to the transitivity property
\beq
R_{ij}R_{jk}=R_{ik}\ .
\eeq
Notice also the useful identity
\beq
R_{ij}R_{kl}=R_{il}R_{kj}.
\eeq
We also require all of the mass parameters (\ref{RABdef}) to be positive semidefinite.
Our analysis assumes three additional absolute conditions on these parameters.
\begin{equation}
R_{AB} < 1\ , \qquad
R_{AC} < 1\ , \qquad
R_{CD} < 1\ .
\label{Rcond}
\end{equation}
This imposes a general mass hierarchy, 
\beq 
0<m_A<m_C<m_D,
\label{masshierarchy}
\eeq
while for the mass of $B$ the only constraint is $m_A<m_B$. 
Depending on the mass of $B$, we can obtain either 
the on-shell scenario of Fig.~\ref{fig:chain}(a), in which
$m_A<m_B<m_C$, or the off-shell scenario of Fig.~\ref{fig:chain}(b), 
in which $m_C<m_B$ and possibly even $m_D<m_B$. 
In summary, we shall use 
\begin{equation}
\left\{m_D, R_{AB}, R_{BC}, R_{CD}\right\}
\label{Rspace}
\end{equation}
as our default parametrization of the 4 dimensional 
mass parameter space (\ref{mspace}). 

\subsection{Forward formulas}
\label{sec:forward}

Here we list the well known formulas for the endpoints (\ref{5meas})
in terms of the parameters (\ref{Rspace}) introduced above.

\subsubsection{On-shell scenario}

In the on-shell scenario the kinematical endpoints are given by the following formulas:
\bea
a &\equiv& \left(m_{ll}^{max}\right)^2 = m_D^2\, R_{CD}\, (1-R_{BC})\, (1-R_{AB}); 
\label{lldef}
\\ [4mm]
b &\equiv& \left(m_{jll}^{max}\right)^2 = 
\left\{ 
\begin{array}{lll}
m_D^2 (1-R_{CD})(1-R_{AC}),       & ~{\rm for}\ R_{CD}<R_{AC},       & {\rm case}\ (1,-),~~ \\[4mm] 
m_D^2 (1-R_{BC})(1-R_{AB}R_{CD}), & ~{\rm for}\ R_{BC}<R_{AB}R_{CD}, & {\rm case}\ (2,-),~~ \\[4mm] 
m_D^2 (1-R_{AB})(1-R_{BD}),       & ~{\rm for}\ R_{AB}<R_{BD},       & {\rm case}\ (3,-),~~ \\[4mm] 
m_D^2\left(1-\sqrt{R_{AD}}\,\right)^2,   & ~{\rm otherwise}, & {\rm case}\ (4,-);~~
\end{array}
\right.
\label{jlldef}
\\ [4mm]
c &\equiv& \left(m_{jl(lo)}^{max}\right)^2=\left\{ 
\begin{array}{lll}
\left(m_{jl_n}^{max}\right)^2,   & ~{\rm for}\ (2-R_{AB})^{-1} < R_{BC} < 1,   & {\rm case}\ (-,1), \\[4mm] 
\left(m_{jl(eq)}^{max}\right)^2, & ~{\rm for}\ R_{AB}< R_{BC}<(2-R_{AB})^{-1}, & {\rm case}\ (-,2),\\[4mm] 
\left(m_{jl(eq)}^{max}\right)^2, & ~{\rm for}\ 0< R_{BC}<R_{AB},               & {\rm case}\ (-,3);
\end{array}%
\right .
\label{jllodef}
\\ [4mm]
d &\equiv& \left( m_{jl(hi)}^{max}\right)^2 =\left\{ 
\begin{array}{lll}
\left(m_{jl_f}^{max}\right)^2, & ~{\rm for}\ (2-R_{AB})^{-1} < R_{BC} < 1,   & {\rm case}\ (-,1), \\[4mm] 
\left(m_{jl_f}^{max}\right)^2, & ~{\rm for}\ R_{AB}< R_{BC}<(2-R_{AB})^{-1}, & {\rm case}\ (-,2), \\[4mm] 
\left(m_{jl_n}^{max}\right)^2, & ~{\rm for}\ 0< R_{BC}<R_{AB},               & {\rm case}\ (-,3);
\end{array}%
\right .
\label{jlhidef}
\eea
where
\bea
\left(m_{jl_n}^{max}\right)^2   &=& m_D^2\, (1-R_{CD})\, (1-R_{BC})\, , \label{mjlnmax}\\
\left(m_{jl_f}^{max}\right)^2   &=& m_D^2\, (1-R_{CD})\, (1-R_{AB})\, , \label{mjlfmax}\\
\left(m_{jl(eq)}^{max}\right)^2 &=& m_D^2\, (1-R_{CD})\, (1-R_{AB})\, (2-R_{AB})^{-1} \, . \label{mjleqmax}
\eea
The physical meaning of the latter three quantities will become
clear in the course of the discussion in Section~\ref{sec:hivslo}.
Finally, the endpoint $m_{jll(\theta>\frac{\pi}{2})}^{min}$ 
introduced earlier in the Introduction, is given by 
\begin{eqnarray}
e &\equiv& \left( m_{jll(\theta>\frac{\pi}{2})}^{min}\right)^2 = 
\frac{1}{4}m_D^2 \Biggl\{ (1-R_{AB})(1-R_{BC})(1+R_{CD})  \label{jllthetadef} 
\\ \nonumber
&+& 2\, (1-R_{AC})(1-R_{CD})
-(1-R_{CD})\sqrt{(1+R_{AB})^2 (1+R_{BC})^2-16 R_{AC}}\Biggr\} .
\end{eqnarray}
The physical meaning of the latter quantity will be revealed in 
Section~\ref{sec:jllvsll}.
In (\ref{lldef}-\ref{jllthetadef}) we have introduced some convenient 
shorthand notation 
\beq
a = \left(m_{ll}^{max}\right)^2, \quad
b = \left(m_{jll}^{max}\right)^2, \quad
c = \left(m_{jl(lo)}^{max}\right)^2, \quad
d = \left( m_{jl(hi)}^{max}\right)^2, \quad
e = \left( m_{jll(\theta>\frac{\pi}{2})}^{min}\right)^2 
\eeq
for the kinematical endpoints of the
mass {\em squared} distributions\footnote{Note that Ref.~\cite{Gjelsten:2004ki}
uses $a,b,c,d$ to label the same endpoints, but for the {\em linear} masses
instead of the masses squared.}.

One can see that the formulas (\ref{jlldef}-\ref{jlhidef}) are piecewise-defined:
they are given in terms of different expressions, 
depending on the parameter range for $R_{AB}$, $R_{BC}$ and $R_{CD}$.
This divides the $\{ R_{AB},R_{BC},R_{CD}\}$ parameter space 
into several distinct regions, illustrated in Fig.~\ref{fig:regions}.
\FIGURE[t]{
\epsfig{file=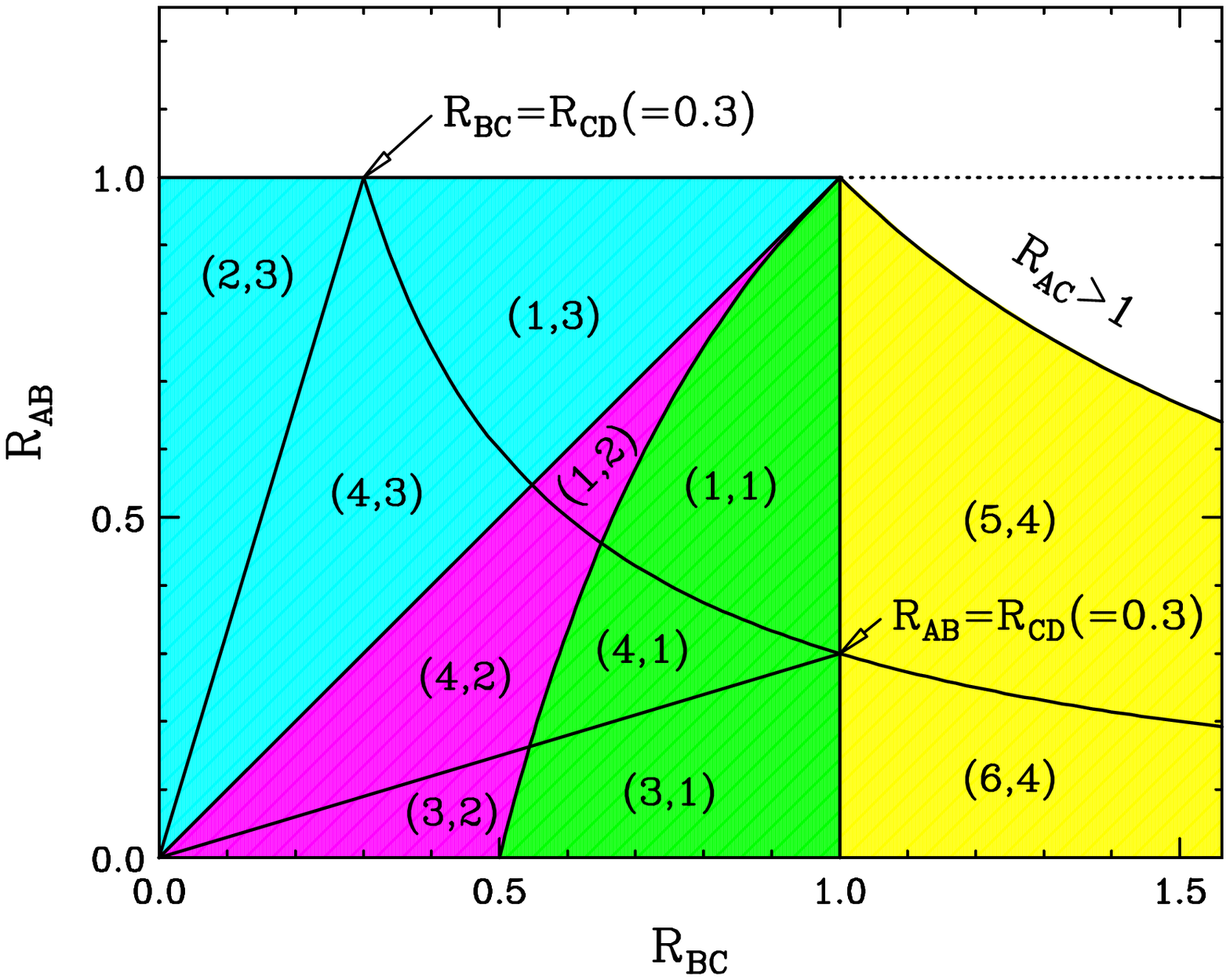,width=10cm}
\caption{A slice through the $\{ R_{AB},R_{BC},R_{CD}\}$
parameter space at a fixed $R_{CD}=0.3$, illustrating
the 11 parameter space regions $(N_{jll},N_{jl})$
resulting from the piecewise-definition of the $m_{jll}^{max}$ endpoint 
in eqs.~(\ref{jlldef}) and (\ref{jlloff}) and the $(m_{jl(lo)}^{max},m_{jl(hi)}^{max})$ 
endpoints in eqs.~(\ref{jllodef}), (\ref{jlhidef}), (\ref{jllooff}) and (\ref{jlhioff}).
The four $N_{jl}$ regions used later in our analysis are
color-coded as follows: $N_{jl} =1$ (green); 
$N_{jl} =2$ (magenta); $N_{jl} =3$ (cyan) and $N_{jl} = 4$ (yellow).
\label{fig:regions}
}
}
Following \cite{Gjelsten:2004ki}, we label those by a pair of integers $(N_{jll},N_{jl})$.
As already indicated in eqs.~(\ref{jlldef}-\ref{jlhidef}), the first integer $N_{jll}$ 
identifies the relevant case for $m_{jll}^{max}$, while the second integer $N_{jl}$ 
identifies the corresponding case for $(m_{jl(lo)}^{max},m_{jl(hi)}^{max})$.
In the on-shell case considered here, only 9 out of the 12 pairings
$(N_{jll},N_{jl})$ are physical, and they are all exhibited within the
unit square of Fig.~\ref{fig:regions}. The remaining two regions 
$(5,4)$ and $(6,4)$ seen in  Fig.~\ref{fig:regions} correspond to 
the off-shell case and will be introduced below in Sec.~\ref{sec:offshell}.
Notice at this point that the formula for the $ m_{jll(\theta>\frac{\pi}{2})}^{min}$
threshold is unique.

Using (\ref{lldef}), (\ref{jlldef}) and (\ref{jlhidef}),
it is easy to check that the relation (\ref{mjllcorrelation}), which
can be equivalently rewritten in the new notation as
\beq
b = a + d,
\label{bad}
\eeq
is identically satisfied in regions (3,1), (3,2) and (2,3) of Fig.~\ref{fig:regions}.
Therefore, in these regions one would necessarily have to rely on the additional 
information provided by the measurement of the $e$ endpoint (\ref{jllthetadef}).

\subsubsection{Off-shell scenario}
\label{sec:offshell}

We now list the relevant formulas \cite{Lester:2006cf}
for the off-shell scenario of Fig.~\ref{fig:chain}(b), 
in which $m_B>m_C$, i.e. $R_{BC}>1$:
\begin{eqnarray}
a&\equiv& \left(m_{ll}^{max}\right)^2 = m_D^2\, R_{CD}\,(1-\sqrt{R_{AC}})^2;
\label{lloff}
\\ [4mm]
b&\equiv& \left(m_{jll}^{max}\right)^2 = 
\left\{ 
\begin{array}{lll}
m_D^2 (1-R_{CD})(1-R_{AC}),  & ~{\rm for}\ R_{CD}<R_{AC}, & {\rm case}\ (5,-),~~~  \\[4mm] 
m_D^2 (1-\sqrt{R_{AD}})^2,   & ~{\rm otherwise},          & {\rm case}\ (6,-);~~~
\end{array}
\right. 
\label{jlloff}
\\ [4mm]
c &\equiv& \left(m_{jl(lo)}^{max}\right)^2=
\frac{1}{2}\, m_D^2 (1-R_{CD})(1-R_{AC}), \quad {\rm case}\ (-,4);
\label{jllooff}
\\ [4mm]
d &\equiv& \left( m_{jl(hi)}^{max}\right)^2 =
m_D^2 (1-R_{CD})(1-R_{AC}), \qquad {\rm case}\ (-,4);
\label{jlhioff}
\\ [4mm]
e &\equiv& \left( m_{jll(\theta>\frac{\pi}{2})}^{min}\right)^2 =
\frac{1}{4}m_D^2(1-\sqrt{R_{AC}}) 
\bigg\{ 2R_{CD}(1-\sqrt{R_{AC}}) \label{jllthetaoff}
\\[4mm]
&& \qquad\qquad\qquad +(1-R_{CD})\bigg(3+\sqrt{R_{AC}}-\sqrt{1+R_{AC}+6\sqrt{R_{AC}}}\bigg)\bigg\}.
\nonumber
\end{eqnarray}
Notice the absence of the $B$ index in these expressions, indicating
that they are indeed independent of the mass $m_B$ of the 
heavy (off-shell) particle $B$. Nevertheless,
the off-shell case can still be represented in the $(R_{BC},R_{AB})$
parameter plane of Fig.~\ref{fig:regions} as the right-most yellow-shaded region. 
Its left boundary is the line $R_{BC}=1$,
beyond which particle $B$ becomes on shell, while its upper boundary is the
line $R_{AB}R_{BC}=R_{AC}=1$, beyond which $A$ is heavier than $C$
and $C$ becomes the LSP, which contradicts our original assumption
(\ref{masshierarchy}). For consistency with the earlier notation 
$(N_{jll},N_{jl})$ for the on-shell parameter space regions, 
we shall simply use $N_{jl}=4$ to label the single off-shell case 
for $\{c,d\}=\{(m_{jl(lo)}^{max})^2,( m_{jl(hi)}^{max})^2\}$, 
and $N_{jll}=5,6$ to label the corresponding two 
off-shell expressions for $b=(m_{jll}^{max})^2$ given in eq.~(\ref{jlloff}).
This gives us a total of 11 allowed $(N_{jll},N_{jl})$
combinations, which are all exhibited in Fig.~\ref{fig:regions}.

\subsection{Inversion formulas}
\label{sec:inverse}

Having presented all the ``forward'' formulas for the 
five kinematic endpoints $a$, $b$, $c$, $d$ and $e$ 
in terms of the mass parameters $R_{ij}$ and $m_D$,
we are now in position to tackle the inverse problem:
deriving the inverse relations, which would give
the mass spectrum $m_A$, $m_B$, $m_C$ and $m_D$
in terms of the measured endpoints  $a$, $b$, $c$, $d$ and $e$.
Our goal will be to obtain the exact analytical inverse formulas 
for {\em each} of the relevant parameter space regions
of Fig.~\ref{fig:regions}. Until now, the inverse relations  
have been derived for only 6 of the 11 regions, namely (1,1), 
(1,2), (1,3), (4,1), (4,2) and (4,3), and have never included 
the $e$ measurement \cite{Gjelsten:2004ki}.

Before we begin, we need to make a decision about the following issue.
In general, the system appears to be over-constrained, since
we are trying to solve for four unknowns ($m_A$, $m_B$, $m_C$ and $m_D$)
in terms of five measurements ($a$, $b$, $c$, $d$ and $e$).
Therefore, for the purpose of inversion, we are allowed to 
drop one of the five measurements and use only the remaining four. 
Which measurement should we drop? This question actually turns out
to be quite important for the subsequent discussion.

The approach of Ref.~\cite{Gjelsten:2004ki} (which considered
only the on-shell case of Fig.~\ref{fig:chain}(a)) was to eliminate
$e$ and use only $a$, $b$, $c$, and $d$. The reasoning 
was that the ``forward'' expression for $e$ 
(\ref{jllthetadef}) appears to be too 
complicated to be tackled by analytic means.
However, the problem with this approach is that it cannot be applied 
in the three on-shell regions (3,1), (3,2) and (2,3), 
where the three measurements $a$, $b$ and $d$ are not 
independent, due to the relation (\ref{bad}).
Therefore, in order to obtain inverse relations valid over
the full parameter space of Fig.~\ref{fig:regions},
{\em we must make use of the $m_{jll(\theta>\frac{\pi}{2})}^{min}$ measurement}
(\ref{jllthetadef}).
For the same reason, we must also use the $m_{jl(lo)}^{max}$
measurement (\ref{jllodef}). Therefore, the choice of candidates
to be omitted is narrowed down to three: $a$, $b$ and $d$,
i.e. precisely the problematic ones entering the linear dependence
relation (\ref{bad}).

Leaving aside the experimental issues of precision, at this point
it should be clear that it is most convenient to drop
the $b$ measurement and {\em always} perform the inversion in terms of
$a$, $c$, $d$ and $e$. There are two important advantages of
our approach:
\begin{itemize}
\item Since we are never using the $b$ measurement, the
linear dependence (\ref{bad}) between $a$, $b$ and $d$
never becomes an issue, and the same four inputs 
$a$, $c$, $d$ and $e$ can be used in all 
parameter space regions $(N_{jl},N_{jll})$. 
\item More importantly, once we eliminate $b$ from the discussion,
we do not have to worry about the division of the parameter space 
into regions labelled by the integer $N_{jll}$. Instead, the full 
parameter space is now divided just into the four color-coded regions
of Fig.~\ref{fig:regions}, each of which is uniquely identified by 
the value of $N_{jl}$ and from now on will be labelled as ${\cal R}_{N_{jl}}$:
\begin{enumerate}
\item Region ${\cal R}_1$ ($N_{jl}=1$), defined by $\frac{1}{2-R_{AB}}<R_{BC}<1$
and shaded green in Fig.~\ref{fig:regions}.
\item Region ${\cal R}_2$ ($N_{jl}=2$), defined by $R_{AB}<R_{BC}<\frac{1}{2-R_{AB}}$
and shaded magenta in Fig.~\ref{fig:regions}.
\item Region ${\cal R}_3$ ($N_{jl}=3$), defined by $0<R_{BC}<R_{AB}$
and shaded cyan in Fig.~\ref{fig:regions}.
\item Region ${\cal R}_4$ ($N_{jl}=4$), defined by $1<R_{BC}$
and shaded yellow in Fig.~\ref{fig:regions}.
\end{enumerate}
\end{itemize}
In what follows, we sometimes refer to regions 
${\cal R}_1$ , ${\cal R}_2$  and ${\cal R}_3$ 
collectively as the ``on-shell'' region, and region ${\cal R}_4$  as 
the ``off-shell'' region, in reference to whether particle 
$B$ is on-shell or off-shell.
This distinction is in one-to-one correspondence with the 
distinction between the two-body scenario of Fig.~\ref{fig:chain}(a)
and the three-body scenario of Fig.~\ref{fig:chain}(b), respectively.
Note that the region identification only depends on the two mass parameters
$R_{AB}$ and $R_{BC}$. For comparison, the original endpoint 
method utilizing the $b$ measurement, required all eleven regions
of Fig.~\ref{fig:regions}, whose definitions depend also on $R_{CD}$, 
and one must check the solution for consistency in each 
region by trial and error \cite{Gjelsten:2004ki}.
Instead, we have now reduced the number of regions from 
eleven down to four. Furthermore, in Section~\ref{sec:hivslo}
we shall show that the shape of the kinematical boundaries of the 
$m_{jl(hi)}^2$ versus $m_{jl(lo)}^2$ distribution reveal the 
exact region ${\cal R}_i$  in which the mass spectrum occurs, 
thus eliminating the need for trial-and-error inversion altogether.
We consider this to be one of our most important results.

With those preliminaries, we are now ready to present our inversion formulas
which can be cast in the following form common to all regions:
\bea
m_A^2 &=& G_i \left(\alpha_i-1\right)\left(\beta_i-1\right)\left(\gamma_i-1\right), \label{mAon} \\ [2mm]
m_B^2 &=& G_i \left(\alpha_i-1\right)\left(\beta_i-1\right)       \gamma_i,         \label{mBon} \\ [2mm]
m_C^2 &=& G_i \left(\alpha_i-1\right)      \beta_i      \,        \gamma_i,         \label{mCon} \\ [2mm]
m_D^2 &=& G_i\,       \alpha_i      \,       \beta_i       \,       \gamma_i.         \label{mDon}
\eea
where the subscript $i=1,2,3,4$ is used to indicate the corresponding
(color-coded) region ${\cal R}_i$ of Fig.~\ref{fig:regions}.
The quantities $G_i$, $\alpha_i$, $\beta_i$, and $\gamma_i$ are 
functions of the measured endpoints $\{a,c,d,e\}$ and are region-dependent, 
just like the ``forward'' expressions for the endpoints in terms of the 
input masses (see Sec.~\ref{sec:forward}).
Before defining $G_i$, $\alpha_i$, $\beta_i$, and $\gamma_i$, we 
identify an ubiquitous combination of observables
\begin{equation}
g \equiv 2e - a
\end{equation}
and use it in place of $e$, so that our starting point is 
the equivalent set of four measurements $\{a,c,d,g\}$.
Then the quantities appearing on the right hand side of 
eqs.~(\ref{mAon}-\ref{mDon}) are defined by
\begin{eqnarray}
G_1&\equiv&\frac{g \left(2d-g\right) - 2c \left(d-g\right)}{g},
\qquad \alpha_1\equiv{}\frac{a+G_1}{G_1},
\quad \quad \beta_1\equiv{}\frac{d}{G_1},
\qquad \quad \gamma_1\equiv{}\frac{c}{G_1}; \label{G1def}  \\ [2mm]
G_2&\equiv&\frac{g\left(2d-g\right)\left(d-c\right)}{g\left(d-c\right)+2c\left(d-g\right)},
\qquad \quad \alpha_2\equiv{}\frac{a+G_2}{G_2},
\quad \beta_2\equiv{}\frac{d}{G_2},
\qquad\quad  \gamma_2\equiv{}\frac{c}{d-c};  \label{G2def}  \\ [2mm]
G_3&\equiv&\frac{\left(g \left(2d-g\right) - 2c\left(d-g\right)\right)d}{gd + 2c\left(d-g\right)},
\quad \alpha_3\equiv{}\frac{a+G_3}{G_3}, 
\quad \beta_3\equiv{}\frac{c\left(d+G_3\right)}{dG_3},
\quad \gamma_3\equiv{}\frac{d}{G_3};~~~  \label{G3def}  \\ [2mm]
G_4 &=& -d+g+\sqrt{(2d-g)g},
\qquad\quad \alpha_4=\frac{a+G_4}{G_4},
\qquad \beta_4 =\gamma_4 = \frac{d+G_4}{2 G_4}. \label{G4def}
\end{eqnarray}
A word of caution is in order regarding the off-shell scenario 
of Fig.~\ref{fig:chain}(b), i.e. eq.~(\ref{G4def}).
In that case, particle $B$ is far off-shell and its mass 
$m_B$ is not among the relevant parameters for the kinematic 
endpoints\footnote{Of course, (\ref{mBon}) should only be used in the on-shell case.},
so that we only need to determine three unknowns: $m_A$, $m_C$ and $m_D$. 
At the same time, we have one 
fewer independent inputs within our original set $\{a,c,d,g\}$, 
since eqs.~(\ref{jllooff}) and (\ref{jlhioff}) imply the additional relation
\beq
c = \frac{1}{2} \, d.
\label{cdoff}
\eeq
For the purpose of inversion, in the off-shell case we chose to omit $c$ and work 
only with $\{a,d,g\}$, which are the only three endpoints appearing 
in eq.~(\ref{G4def}). Finally, the appearance of the square root
in (\ref{G4def}) should not be a problem, since in the off-shell scenario
the ratio $\frac{d}{g}$ is bounded by
\begin{equation}
1<\frac{d}{g}<2+\sqrt2\ .
\end{equation}
The set of analytical inversion formulas (\ref{mAon}-\ref{mDon},\ref{G1def}-\ref{G4def})
is the first main result of this paper.

\section{Duplication analysis}
\label{sec:duplication}

Armed with the analytical results from the previous section, we are
now ready to address the problem of duplicate solutions and the 
potential discrete ambiguities in the determination of the mass spectrum.
Our procedure will be very simple and straightforward. We shall consider
the four (color-coded) parameter space regions ${\cal R}_i$  
in Fig.~\ref{fig:regions} one at a time,
and in each case we shall ask the question: Is it possible that identically 
the same values of the endpoints $\{a,c,d,e\}$ can be obtained from
another type of mass spectrum belonging to a {\em different} parameter 
space region ${\cal R}_j$, with $j\ne i$? 
And if the answer is ``yes'', we shall then ask
two follow-up questions: First, exactly in which parts of ${\cal R}_i$
and ${\cal R}_j$ does this duplication occur? Second, will the 
ambiguity get resolved by utilizing the additional endpoint measurement 
$b$ at our disposal?

Operationally we proceed as follows. First, it is important to realize that
the ``forward'' analytical formulas of Section~\ref{sec:forward} provide
a map ${\cal F}_i$ of the corresponding parameter space region ${\cal R}_i$ 
onto the space of values of the kinematic endpoints:
\beq
\{m_A,m_B,m_C,m_D \}_i \stackrel{{\cal F}_i}{\longmapsto}
\{a,c,d,e\}\, ,
\label{MFmap}
\eeq
or equivalently, using the reparametrization (\ref{Rspace})
\beq
\left\{m_D, R_{AB}, R_{BC}, R_{CD}\right\}_i \stackrel{{\cal F}_i}{\longmapsto}
\{a,c,d,e\}\, .
\label{RFmap}
\eeq
Similarly, the inverse formulas from Section~\ref{sec:inverse}
provide an inverse map ${\cal F}^{-1}_i$ from the space of kinematical
endpoints back onto the mass parameter space:
\beq
\{a,c,d,e\} \stackrel{{\cal F}^{-1}_j}{\longmapsto}
\left\{m_D, R_{AB}, R_{BC}, R_{CD}\right\}_j \ .
\label{RImap}
\eeq
The composite of the two maps (\ref{RFmap}) and (\ref{RImap})
for $i\ne j$, is a transformation
\beq
T_{ij} \equiv {\cal F}^{-1}_j \cdot {\cal F}_i 
\eeq
relating parameter space points belonging to two {\em different} regions,
${\cal R}_i$ and ${\cal R}_j$, yet resulting in 
{\em identical} kinematical endpoints $\{a,c,d,e\}$:
\beq
\left\{m_D, R_{AB}, R_{BC}, R_{CD}\right\}_i
\stackrel{T_{ij}}{\longmapsto}
\left\{m'_D, R'_{AB}, R'_{BC}, R'_{CD}\right\}_j \ .
\label{FImap}
\eeq
The transformation $T_{ij}$ described in (\ref{FImap}) will serve as the basis 
of our duplication analysis. The exact analytical formulas for this 
mapping can be trivially obtained from our analytical 
results above in Section~\ref{sec:analytical}, but
are rather lengthy and we shall not present them here explicitly. 
However, we note that in the three on-shell cases $i=1,2,3$
they have the generic form
\bea
R'_{AB} &=& f_{AB}(R_{AB},R_{BC}),           \label{RpAB} \\ [2mm]
R'_{BC} &=& f_{BC}(R_{AB},R_{BC}),           \label{RpBC} \\ [2mm]
R'_{CD} &=& f_{CD}(R_{AB},R_{BC},R_{CD}),    \label{RpCD} \\ [2mm]
m'_D    &=& m_D\, f_D(R_{AB},R_{BC},R_{CD}), \label{RpD}
\eea
where $f_{AB}$, $f_{BC}$, $f_{CD}$ and $f_{D}$ are the
functions defining the transformation $T_{ij}$.
One important feature of the $T_{ij}$
map (\ref{RpAB}-\ref{RpD}) is that it transforms the 
2-dimensional subspace of dimensionless parameters 
$\{R_{AB},R_{BC}\}$ into itself. 
Notice that $R_{AB}$ and $R_{BC}$ are precisely the parameters
definitng the four regions ${\cal R}_i$ in Fig.~\ref{fig:regions}.
Therefore, for the purposes of our duplication analysis it is 
sufficient to consider the simpler transformation of
\beq
\left\{R_{AB}, R_{BC}\right\}_i
\stackrel{T_{ij}}{\longmapsto}
\left\{R'_{AB}, R'_{BC} \right\}_j 
\label{RFImap}
\eeq
given by eqs.~(\ref{RpAB}) and (\ref{RpBC}) only,
instead of the more general mapping (\ref{FImap})
given by all four eqs.~(\ref{RpAB}-\ref{RpD}).

We are now ready to answer the main question posed at the beginning 
of this section: does a consistent mapping (\ref{RFImap}) exist
for {\em some} pair of regions ${\cal R}_i$ and ${\cal R}_j$?
Note that the transformation (\ref{RFImap}) is not necessarily 
always well defined: consistency requires that
the obtained values of $\left\{R'_{AB}, R'_{BC}\right\}_j$
belong to Region ${\cal R}_j$, which is not automatically
guaranteed and must be explicitly checked. To put this in more 
formal terms, we are only interested
in those cases where the intersection of the image of 
region ${\cal R}_i$ under the transformation $T_{ij}$ 
and the intended target region ${\cal R}_j$ is a non-empty set:
\beq
\left\{ T_{ij} \left( {\cal R}_i\right) \right\}
\cap \left\{ {\cal R}_j \right\} \ne \emptyset.
\label{check}
\eeq
In order to find all such occurrences, we
consider all possible transformations $T_{ij}$
with $i\ne j$ and enforce the consistency check (\ref{check}).

We begin with the on-shell case ($i,j=1,2,3$),
where there are 6 possible mappings $T_{ij}$. 
For the purposes of finding the duplicated portion of parameter space,
it is sufficient to consider only 3 of them, which for convenience of 
illustration we choose as $T_{13}$, $T_{23}$ and $T_{21}$.
The corresponding results are shown in 
Figs.~\ref{fig:123map} and \ref{fig:21map}.
\FIGURE[t]{
\epsfig{file=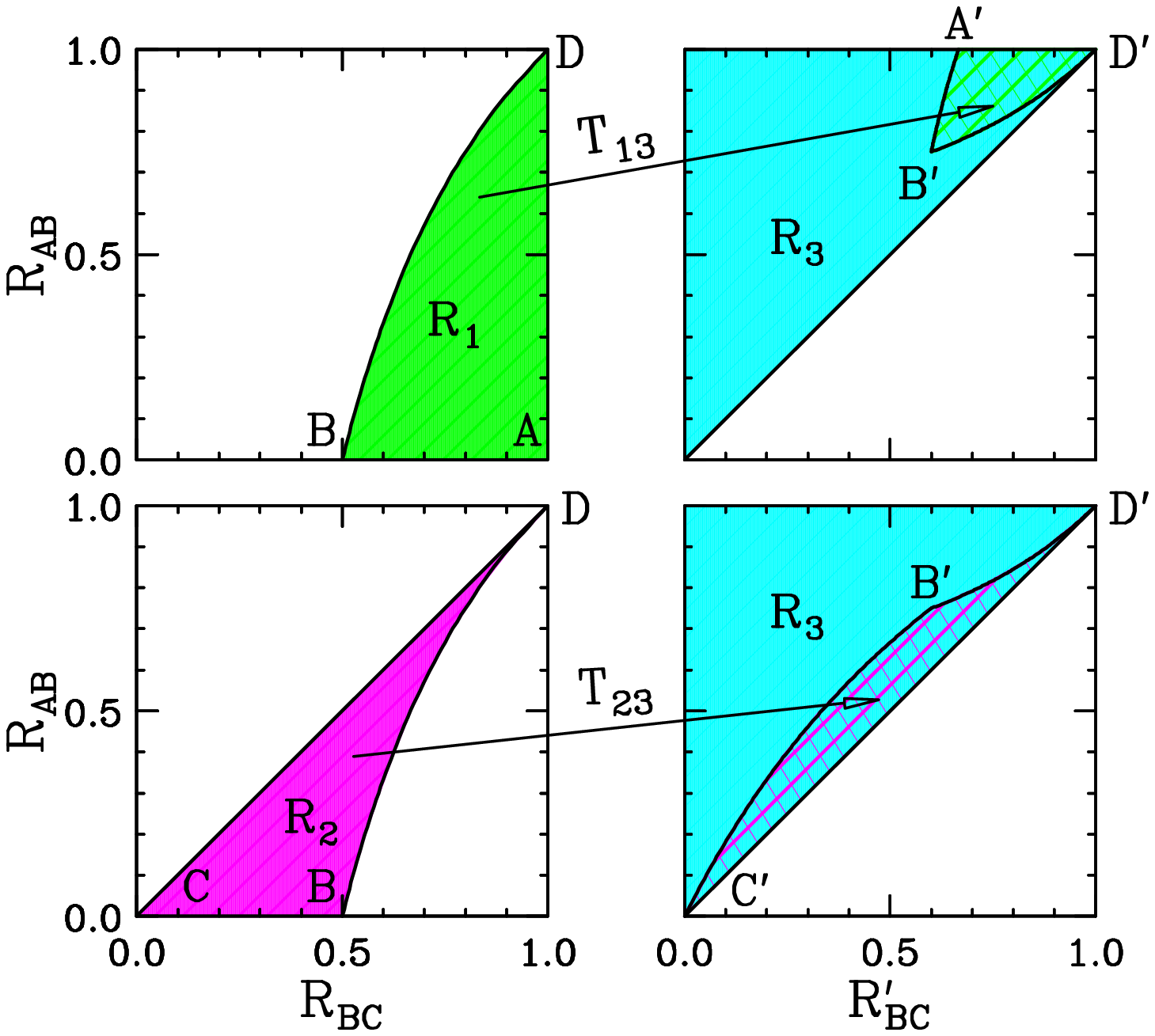,width=11.0cm}
\caption{The maps $T_{13}: {\cal R}_1 \longmapsto {\cal R}_3$ (top two panels)
and $T_{23}: {\cal R}_2 \longmapsto {\cal R}_3$ (bottom two panels),
which are implied by eq.~(\ref{RFImap}). In both cases the target 
region ${\cal R}_3$ is shaded in cyan. Under $T_{13}$, 
the green-shaded region $ABD$ in the top left panel
transforms into the green-hatched region $A'B'D'$ of the top right panel.
Under $T_{23}$, 
the magenta-shaded region $BCD$ in the bottom left panel
transforms into the magenta-hatched region $B'C'D'$ of the bottom right panel.
In both cases, the transformed (primed) region falls completely within the 
boundaries of the intended target (${\cal R}_3$).
\label{fig:123map}
}
}
Fig.~\ref{fig:123map} shows the effect of the transformation
$T_{13}: {\cal R}_1 \longmapsto {\cal R}_3$ (top two panels)
and $T_{23}: {\cal R}_2 \longmapsto {\cal R}_3$ (bottom two panels),
while Fig.~\ref{fig:21map} shows the map
$T_{21}: {\cal R}_2 \longmapsto {\cal R}_1$.
In both figures, the color-shaded areas in the left (right) panels 
exhibit the original regions ${\cal R}_i$ (the intended target regions $R_j$).
The cross-hatched areas in the right panels depict the 
actual image $T_{ij}({\cal R}_i)$ of the ${\cal R}_i$ region 
under the transformation $T_{ij}$. For example, in Fig.~\ref{fig:123map}~
$T_{13}$ maps the whole green-shaded region $ABD$ on the left
into the green-hatched region $A'B'D'$ on the right, while in
Fig.~\ref{fig:123map} (Fig.~\ref{fig:21map})
$T_{23}$ ($T_{21}$) maps the whole magenta-shaded region $BCD$ on the left
into the magenta-hatched region $B'C'D'$ on the right. In accordance with 
(\ref{check}), duplication occurs whenever the right panels 
in Figs.~\ref{fig:123map} and \ref{fig:21map} exhibit an overlap
between the cross-hatched area of the image and the 
solid color-shaded area of the intended target.
We see that duplication occurs in the case of $T_{13}$ and $T_{23}$,
but not for $T_{21}$, although in the latter case points 
which are on opposite sides, but close to the boundary line 
$BD$ will give rather similar values 
for the measured kinematic endpoints $\{a,c,d,e\}$.

\FIGURE[t]{
\epsfig{file=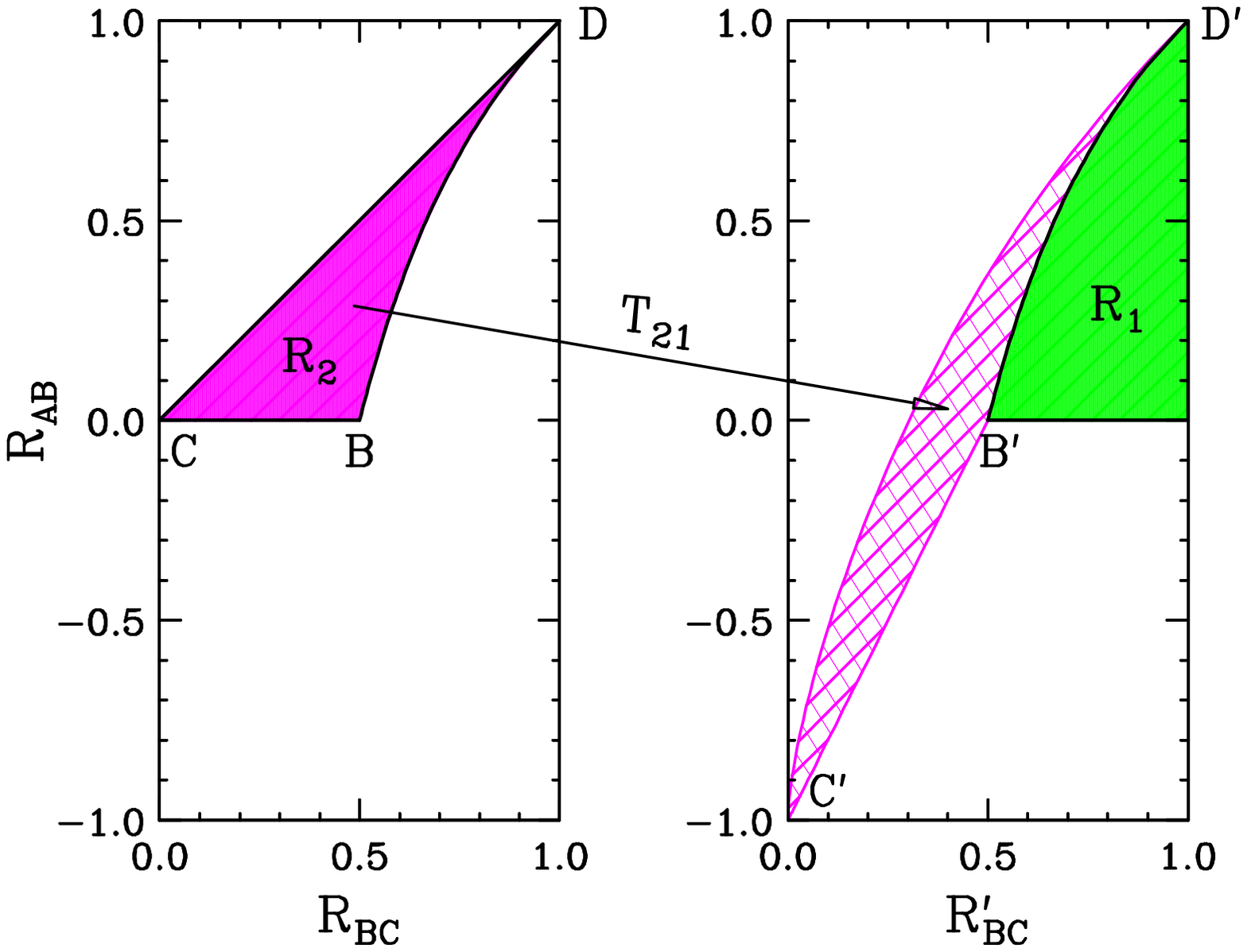,width=10.0cm}
\caption{The same as Fig.~\ref{fig:123map}, but for the map 
$T_{21}: {\cal R}_2 \longmapsto {\cal R}_1$, where 
the intended target is the green-shaded region ${\cal R}_1$.
Under $T_{21}$, the magenta-shaded region $BCD$ in the left panel
transforms into the magenta-hatched region $B'C'D'$ of the right panel.
The image $B'C'D'$ has no overlap with its intended target ${\cal R}_1$,
except along the $BD=B'D'$ boundary, which is left invariant under the 
$T_{12}$ transformation. 
\label{fig:21map}
}
}

At this point one may wonder whether the result of Fig.~\ref{fig:123map}
is sufficient to prove the existence of duplication. Indeed, 
Fig.~\ref{fig:123map} tells us nothing about the remaining two 
parameters $R_{CD}$ and $m_D$ and more specifically about their 
transformed values $R'_{CD}$ and $m'_D$ under the mappings
$T_{13}$ and $T_{23}$. Duplication will in fact not occur, if 
$R'_{CD}$ and $m'_D$ turn out to be unphysical, for example, if
$R'_{CD}<0$, $R'_{CD}>1$ or $(m'_D)^2<0$. Unfortunately, a
closer inspection of (\ref{RpCD}) reveals that
\bea
f_{CD}(R_{AB},R_{BC},R_{CD}=0) &=& 0,   \\ [2mm]
f_{CD}(R_{AB},R_{BC},R_{CD}=1) &=& 1, 
\eea
for any values of $R_{AB}$ and $R_{BC}$, 
so that $R_{CD}$ is always consistently mapped within its 
definition region. Similarly, we find no problem with eq.~(\ref{RpD}).
Therefore, the duplication examples shown in Fig.~\ref{fig:123map} 
truly represent a problem. 

We then perform a similar analysis involving the off-shell region ${\cal R}_4$
and find no occurrences of duplication, which is not surprising,
since the off-shell case is more restricted, due to (\ref{cdoff}). Therefore, 
Figs.~\ref{fig:123map} and \ref{fig:21map} already provide the final 
answer to the first question posed at the beginning of this Section:
which portions of the mass parameter space (\ref{Rspace})
exhibit exact duplication? We can summarize our result as follows:
\begin{quote}
For every point with $R_{AB}<R_{BC}<1$ (i.e. in region ${\cal R}_1$ 
or ${\cal R}_2$) and arbitrary values of $R_{CD}$ and $m_D$,
there exists another parameter space point with $R_{BC}<R_{AB}<1$ and 
certain (in general different) values of $R_{CD}$ and $m_D$, which 
would result in {\em identical} predictions for all four kinematic
endpoints $\{a,c,d,e\}$.
\end{quote}
The reverse statement is not true: not every point with $R_{BC}<R_{AB}<1$
(i.e. in region ${\cal R}_3$) is subject to duplication. Referring
to the right panels of Fig.~\ref{fig:123map}, only the cross-hatched
portions of the cyan-shaded region ${\cal R}_3$ are duplicated.

Having found duplication examples for the limited
set of measurements $\{a,c,d,e\}$, it is now time to ask
whether the additional measurement of the $b$ kinematic 
endpoint will help. We find that, as might have been expected, 
whenever the $b$ measurement is independent of the 
others, the duplication goes away. Unfortunately, as already 
mentioned in the discussion following eq.~(\ref{bad}), in the three
subregions (3,1), (3,2) and (2,3), $b$ is not an independent measurement, and 
thus the duplication will persist even for the full set of 5 measurements 
$\{a,b,c,d,e\}$! In terms of the subregions of Fig.~\ref{fig:regions}, 
the two cases of duplication found in Fig.~\ref{fig:123map}
can then be summarized as
\bea
(3,1) &\stackrel{T_{13}}{\longrightarrow}& (2,3),  \label{3123dup} \\  [2mm]
(3,2) &\stackrel{T_{23}}{\longrightarrow}& (2,3). \label{3223dup}
\eea
As long as the original parameter space point and its image belong to these 
particular subregions, the resulting two sets of endpoints (\ref{5meas})
will be identical.

We caution the reader not 
to get the impression from (\ref{3123dup}) and (\ref{3223dup}) that 
{\em every} parameter space point in regions (3,1), (3,2) and (2,3)
is duplicated with something. Recall that 
the boundaries of the dangerous subregions
(3,1), (3,2) and (2,3) depend on $R_{CD}$, 
thus the range of $R_{CD}$ values resulting 
in duplication will now be restricted.
Therefore, in the presence of the additional $b$ measurement,
our previous statement about duplication is now modified as follows:
\begin{quote}
For every point with $R_{AB}<R_{BC}<1$ 
and {\em any} $m_D$, there exists a range of $R_{CD}$ 
for which exactly the same values of the five kinematic endpoint 
measurements $\{a,b,c,d,e\}$ can also be obtained from a different
parameter space point with $R_{BC}<R_{AB}<1$ and some 
other (generally different) values of $R_{CD}$ and $m_D$.
\end{quote}

\FIGURE[t]{
\epsfig{file=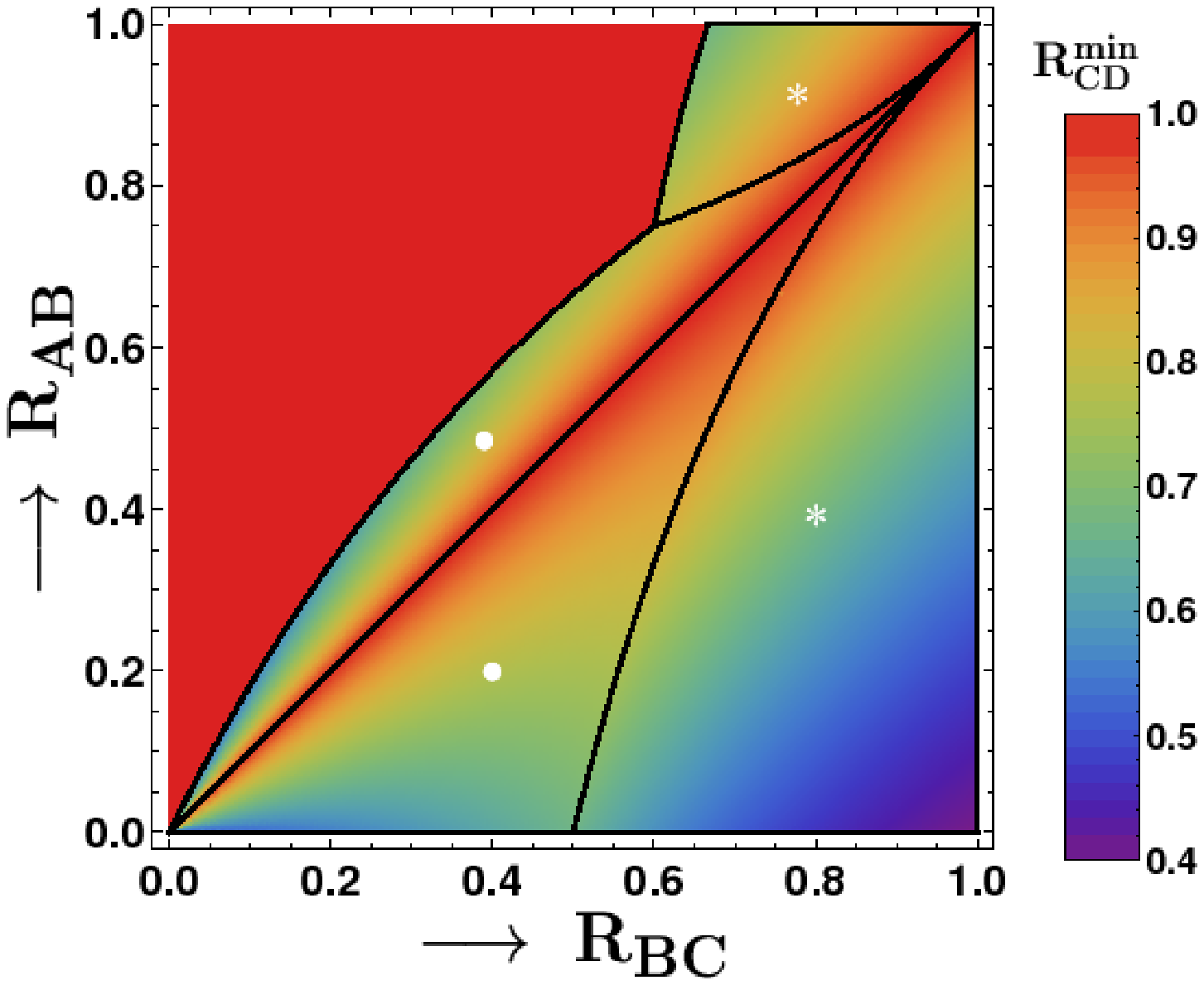,width=10cm}
\caption{The minimum value $R_{CD}^{min}(R_{BC},R_{AB})$ 
required for duplication, as a function of $R_{BC}$ and $R_{AB}$.
The white asterisks (circles) mark the duplicate pair of points 
$P_{31}$ and $P_{23}$ ($P_{32}$ and $P'_{23}$) in 
Table~\ref{table:dup2}.
\label{fig:xmin}
}
}

We shall now describe the duplicated parameter space implied by
(\ref{3123dup}) and (\ref{3223dup}) a bit more quantitatively.
As we just mentioned, for any given point in the $(R_{BC},R_{AB})$ 
plane, there may exist a range of values for $R_{CD}$ which 
would cause duplication. Let us denote the minimum and 
maximum values of that range by $R_{CD}^{min}$ and $R_{CD}^{max}$, 
correspondingly. Clearly, both $R_{CD}^{min}$ and $R_{CD}^{max}$
are in general functions of $R_{BC}$ and $R_{AB}$. Then, the 
``duplicated'' parameter space can be simply described as the 
set of all points $\{R_{AB},R_{BC},R_{CD}\}$,
which satisfy the two inequalities
\beq
R_{CD}^{min}(R_{BC},R_{AB}) < R_{CD} < R_{CD}^{max}(R_{BC},R_{AB}).
\label{dupspace}
\eeq
If, on the other hand, the values of $R_{BC}$ and $R_{AB}$ 
are such that duplication does not occur for {\em any}
value of $R_{CD}$, we can simply take $R_{CD}^{min}=R_{CD}^{max}$,
resulting in a set of zero measure for (\ref{dupspace}).

Now, in order to delineate the duplicated parameter space,
we only need to supply its boundaries $R_{CD}^{min}(R_{BC},R_{AB})$ 
and $R_{CD}^{max}(R_{BC},R_{AB})$. Our analysis reveals that within 
the duplication region we always find
\beq
R_{CD}^{max}(R_{BC},R_{AB}) = 1,
\label{RCDmaxeq1}
\eeq
while the function $R_{CD}^{min}(R_{BC},R_{AB})$ is plotted in 
Fig.~\ref{fig:xmin}.
Duplication does not occur in the uniformly red region in 
the upper left corner, so there we choose to plot 
$R_{CD}^{min}=R_{CD}^{max}=1$, in accordance with our convention.
Within the rainbow-colored region in Fig.~\ref{fig:xmin}, 
duplication will exist for any value of $m_D$, as long as 
$R_{CD}$ is larger than the $R_{CD}^{min}$ value shown in the figure, 
i.e. for
\beq
R_{CD}^{min}(R_{BC},R_{AB}) < R_{CD} < 1 ,
\eeq
where we have made use of (\ref{RCDmaxeq1}).
Fig.~\ref{fig:xmin} reveals that the duplication region is typically 
characterized by a rather high\footnote{The minimum 
value of $R_{CD}$ that we find over the whole
parameter region in Fig.~\ref{fig:xmin} is $R_{CD}=0.4$
and is found at $R_{BC}= 1$, $R_{AB}= 0$.} 
value of $R_{CD}=m_C^2/m_D^2$.
This implies that in order to have duplication, 
particle $D$ cannot be too much heavier than particle $C$.
%, i.e. we must have $m_D\gsim m_C$ as opposed to $m_D >> m_C$.
This situation does not often arise in typical SUSY models, 
where $D$ is a squark $\tilde q$ and $C$ is the second-lightest 
(wino-like) neutralino $\tilde\chi^0_2$. In models with a high
SUSY breaking scale like SUGRA, the Renormalization Group
Equation (RGE) running tends to split the squark and electroweak 
gaugino masses, so that the hierarchy $m_D\sim m_C$ is rather 
unlikely. On the other hand, minimal UED models predict a rather 
degenerate spectrum, since the mass splittings arise mostly 
at the loop level, so that $m_D\sim m_C$ is rather natural 
in this case \cite{Cheng:2002iz}.

\TABULAR[!ht]{|l|c||r|r||r|r|}{
\hline
\multicolumn{2}{|c||}{}
         &  \multicolumn{2}{c||}{${\cal R}_1 \leftrightarrow {\cal R}_3$}
         &  \multicolumn{2}{c|} {${\cal R}_2 \leftrightarrow {\cal R}_3$}    \\ [0.5mm]
\cline{3-6}
\multicolumn{2}{|c||}{}
         &  (3,1) &  (2,3)   &  (3,2)  &  (2,3)     \\  [0.5mm] 
\cline{3-6}
\multicolumn{2}{|c||}{Variable}
         &$P_{31}$& $P_{23}$ & $P_{32}$ & $P'_{23}$     \\  [0.5mm] \hline \hline
\multicolumn{2}{|c||}{$m_A$ (GeV)}
         & 236.643& 915.618  & 126.491 & 241.618    \\ [0.5mm] \hline 
\multicolumn{2}{|c||}{$m_B$ (GeV)}
         & 374.166& 954.747  & 282.843 & 346.073    \\ [0.5mm] \hline 
\multicolumn{2}{|c||}{$m_C$ (GeV)}
         & 418.33 & 1083.10  & 447.214 & 554.133    \\ [0.5mm] \hline 
\multicolumn{2}{|c||}{$m_D$ (GeV)}
         & 500.00 & 1172.57  & 500.00  & 610.443    \\ [0.5mm] \hline 
\multicolumn{2}{|c||}{$R_{AB}$}
         &  0.400 &  0.920   &   0.200 &  0.487     \\ [0.5mm] \hline 
\multicolumn{2}{|c||}{$R_{BC}$}
         &  0.800 &  0.777   &   0.400 &  0.390     \\ [0.5mm] \hline 
\multicolumn{2}{|c||}{$R_{CD}$}
         &  0.700 &  0.853   &   0.800 &  0.824     \\ [0.5mm] \hline 
\multicolumn{2}{|c||}{$R_{CD}^{min}$}
         & 0.686  &  0.845   &   0.774 &  0.800     \\ [0.5mm] \hline 
\hline
$m_{ll}^{max}$ (GeV)  & $\sqrt{a}$
         & \multicolumn{2}{c||}{145 } 
                             & \multicolumn{2}{c|}{ 310 } \\ [0.5mm] \hline
$m_{jll}^{max}$ (GeV) & $\sqrt{b}$
         & \multicolumn{2}{c||}{257 } 
                             & \multicolumn{2}{c|}{ 369 }  \\ [0.5mm] \hline
$m_{jl(lo)}^{max}$ (GeV)  & $\sqrt{c}$
         & \multicolumn{2}{c||}{122 } 
                             & \multicolumn{2}{c|}{ 149 } \\ [0.5mm] \hline
$m_{jl(hi)}^{max}$ (GeV)  & $\sqrt{d}$
         & \multicolumn{2}{c||}{212} 
                             & \multicolumn{2}{c|}{ 200 } \\ [0.5mm] \hline
$m_{jll(\theta>\frac{\pi}{2})}^{min}$ (GeV)  & $\sqrt{e}$
         & \multicolumn{2}{c||}{132 } 
                             & \multicolumn{2}{c|}{ 248}  \\ [0.5mm] \hline
\hline
$m_{jl_f}^{max}$ (GeV)  & $\sqrt{f}$
         &212     & 127      & 200     & 183                   \\ [0.5mm] \hline
$m_{jl_f}^{(p)}$ (GeV)  & $\sqrt{p}$
         &190     & 112      & 126     & 115                   \\ [0.5mm] \hline
$m_{jl_n}^{max}$ (GeV)  & $\sqrt{n}$
         &122     & 212      & 173     & 200                   \\ [0.5mm] \hline
$m_{jl(eq)}^{max}$ (GeV)& $\sqrt{q}$
         &  NA   %167.705 no equal point in region 1 
                  & 122      & 149     & 149                   \\ [0.5mm] \hline
\hline
$m_{jll(+)}(0)$ (GeV)  & $\sqrt{s}$
         & 226    & 240      & 214     & 230            \\ [0.5mm] \hline
$m_{jll(+)}(a_{t})$ (GeV)  & $\sqrt{t}$
         & 263    & 257      & 374     & 369                   \\ [0.5mm] \hline
$m_{jll(+)}(a_{\rm on})$ (GeV)  & $\sqrt{u}$
         & 257    & 257      & 369     & 369                   \\ [0.5mm] \hline
$m_{jll(-)}(a_{\rm on})$ (GeV)  & $\sqrt{v}$
         & 190    & 193      & 355     & 360                   \\ [0.5mm] \hline
$m_{jll(+)}(a_{\rm off})$ (GeV)  & $\sqrt{w}$
         & 256    & 243      & 372     & 367                   \\ [0.5mm] \hline
\hline
}{\label{table:dup2} Two examples of {\em exact} duplication
as implied by (\ref{3123dup}) and (\ref{3223dup}). The pairs 
of study points $P_{31}$ and $P_{23}$, as well as 
$P_{32}$ and $P'_{23}$, exhibit identical values for all five
kinematic endpoints $m_{ll}^{max}$, $m_{jl(lo)}^{max}$, $m_{jl(hi)}^{max}$,
$m_{jll}^{max}$ and $m_{jll(\theta>\frac{\pi}{2})}^{min}$.
Point $P_{31}$ belongs to Region ${\cal R}_1$,
point $P_{32}$ belongs to Region ${\cal R}_2$, while
points $P_{23}$ and $P'_{23}$ belong to Region ${\cal R}_3$.
In Fig.~\ref{fig:xmin}, the duplicate pair of points 
$P_{31}$ and $P_{23}$ ($P_{32}$ and $P'_{23}$)
is marked with white asterisks (white circles).
The second-to-last and last blocks in the table contain 
the endpoint measurements which are available from
the two-dimensional distributions $(m_{jl(lo)}^2,m_{jl(hi)}^2)$
and $(m_{ll}^2,m_{jll}^2)$, discussed below in Sections~\ref{sec:hivslo}
and \ref{sec:jllvsll}, correspondingly.
}
As an illustration of the whole duplication discussion so far, 
let us now choose two specific examples of duplicate mass spectra, 
one for the case of (\ref{3123dup}) and another for the case of 
(\ref{3223dup}). The corresponding input masses and mass ratios,
as well as the resulting kinematic endpoints, are shown in
Table~\ref{table:dup2}. The first five  kinematic endpoints shown in the 
Table were already discussed, while the rest are new and will be 
introduced below in Sections~\ref{sec:hivslo} and \ref{sec:jllvsll}.
As an application of our previous results, let us outline our
procedure of selecting each pair of study points in Table~\ref{table:dup2}.
Let us start with the case of (\ref{3123dup}). Since we know 
from Fig.~\ref{fig:xmin} that the whole region (3,1) is duplicated, 
it is convenient to first choose the point from that region.
We select nice round numbers like $R_{AB}=0.4$ and $R_{BC}=0.8$.
This choice is indicated in Fig.~\ref{fig:xmin} with the 
white asterisk inside region ${\cal R}_1$. Then Fig.~\ref{fig:xmin}
shows that $R_{CD}^{min}=0.686$, therefore we choose a somewhat larger value:
$R_{CD}=0.7$. This choice of $R_{AB}$, $R_{BC}$ and $R_{CD}$
already guarantees duplication for any value of $m_D$, and we 
choose $m_D=500$ GeV (another nice round number).
The resulting masses $m_A$, $m_B$ and $m_C$ can be readily 
computed in terms of $m_D$ and the mass ratios. We call the
resulting spectrum ``study point $P_{31}$'', which is listed
in the second column of Table~\ref{table:dup2}. Given $P_{31}$,
one can use the transformation $T_{13}$ to obtain the matching spectrum in
region (2,3), which is listed in the third column of Table~\ref{table:dup2}
under the name of ``study point $P_{23}$''. 
In  the case of (\ref{3223dup}), we follow a similar
procedure, except we start with a point in region ${\cal R}_2$
(indicated with a white circle in Fig.~\ref{fig:xmin})
and then use the $T_{23}$ transformation to obtain the 
corresponding point in region (2,3). The two resulting mass spectra 
(called $P_{32}$ and $P'_{23}$) are given in the fourth and fifth 
column of Table~\ref{table:dup2}, respectively.

In the on-shell case the parameters $\{R_{AB},R_{BC},R_{CD}\}$ 
belong to a unit cube, due to the restrictions (\ref{Rcond}).
The volume of the unit cube is 1. Fig.~\ref{fig:xmin} then allows us 
to calculate the volume fraction of this unit cube which corresponds 
to a duplicated parameter space region. The result that we find is 0.158.
Then one might be tempted to say that if new physics like 
supersymmetry or UED with a cascade 
decay of the type shown in Fig.~\ref{fig:chain}(a) is discovered 
at the LHC, there would be roughly a $15.8\%$ probability that 
endpoint measurements alone would result in a duplicate spectrum,
even under ideal experimental conditions. However, it is rather
difficult to justify such probabilistic statements, since 
they are not invariant under reparametrizations, and furthermore,
they depend on the assumed (usually uniform) prior for the probability 
distribution of new physics models in mass parameter space. 
Any given model of SUSY breaking, for example, 
would select a preferred parameter space within the unit cube, 
and may lower or increase this naively calculated probability.
The generic mass degeneracy in UED, on the other hand, 
would prefer the region $R_{AB}\sim R_{BC}\sim R_{CD}\sim 1$,
and the duplication is much more likely. 
The important result from our point of view is that there {\em exists}
a non-vanishing duplication region, and this fact alone is sufficient to 
motivate us to look for alternative methods for mass determination,
which we shall undertake in the following two sections.

\section{Kinematic boundary lines for the $m_{jl(lo)}^2$ versus $m_{jl(hi)}^2$ distribution}
\label{sec:hivslo}

%%%%%%%%%%%%%%%%%%%%%%%%%%%%%%%%%%%%%%%%%%%%%%%
% introducing the single lepton distributions %
%%%%%%%%%%%%%%%%%%%%%%%%%%%%%%%%%%%%%%%%%%%%%%%

In this section we shall analyze the shape of the 
{\em two-dimensional} invariant mass distribution 
\beq
\frac{d^2 \Gamma}{dm_{jl(lo)}^2\, dm_{jl(hi)}^2}\ ,
\label{d2Gammalohi}
\eeq
which we imagine plotted as either a scatter plot or a two-dimensional 
histogram with $m_{jl(lo)}^2$ on one axis and 
$m_{jl(hi)}^2$ on the other\footnote{In practice, as we shall 
demonstrate below, it may be more 
convenient to plot the {\em linear} masses $m_{jl(lo)}$ and 
$m_{jl(hi)}$ on the two axes, but use a quadratic power scale instead
of the conventional linear or logarithmic scales.}.
The purpose of our analysis will be twofold. On the one hand, we
shall be interested whether we can use the shape of this 
two-dimensional distribution to resolve the mass spectrum
duplication problem encountered in the previous section. 
But more importantly, we shall investigate what {\em additional}
kinematic endpoint measurements besides those already considered 
in (\ref{5meas}), may become available in this case.

Recall that the variables $m_{jl(lo)}$ and $m_{jl(hi)}$
were introduced in eqs.~(\ref{mjllodef}) and (\ref{mjlhidef})
as a way to deal with the ambiguity in the experimental 
identification of the ``near'' and ``far'' leptons 
$l_n^{\pm}$ and $l_f^{\mp}$ in Fig.~\ref{fig:chain}.
It is therefore not very surprising that the shape of
the $m_{jl(lo)}$ versus $m_{jl(hi)}$ distribution
(\ref{d2Gammalohi}) that we are interested in, is very 
closely related to the corresponding $m_{jl_n}^2$ versus 
$m_{jl_f}^2$ distribution
\beq
\frac{d^2 \Gamma}{dm_{jl_n}^2\, dm_{jl_n}^2}\ .
\label{d2Gammanf}
\eeq
In principle, both distributions (\ref{d2Gammalohi})
and (\ref{d2Gammanf}) depend not only on the mass spectrum, 
but also on the spins and on the chiralities of the coupling constants 
of the particles $A$, $B$, $C$ and $D$ involved 
in the cascade \cite{Athanasiou:2006ef,Wang:2006hk,Burns:2008cp}.
However, the location and the shape of the boundary lines 
in the scatter plots (\ref{d2Gammalohi})
and (\ref{d2Gammanf}) are determined
purely by kinematics, and do not depend on the spin and type of couplings.
To the extent that we are only interested in these boundary lines, 
it is therefore sufficient to ignore spin effects and consider only 
pure phase space decays, in which case the analytical results 
for the distributions (\ref{d2Gammanf}) and (\ref{d2Gammalohi}) 
are in principle already available \cite{Miller:2005zp}. 
From now on, the term ``shape'' will therefore refer to the 
location and shape of the boundary lines, and will otherwise 
have nothing to do with the probability density of the 
two-dimensional distributions such as (\ref{d2Gammalohi}) or (\ref{d2Gammanf}).

%%%%%%%%%%%%%%%%%%%%%%%%%%%%%%%%%%%%%%%%%%%%%%%%%%%%%
% description of the shape of the far vs. near plot %
%%%%%%%%%%%%%%%%%%%%%%%%%%%%%%%%%%%%%%%%%%%%%%%%%%%%%

In the on-shell case, the shape of the $m_{jl_n}^2$ versus 
$m_{jl_f}^2$ invariant mass distribution (\ref{d2Gammanf})
is extremely simple, and is illustrated in Fig.~\ref{fig:nearfar}.
The scatter plot in the $(m_{jl_n}^2,m_{jl_f}^2)$ plane
fills the right-angle trapezoid $ONPF$, whose 
corner points are defined as follows.
\FIGURE[t]{
\epsfig{file=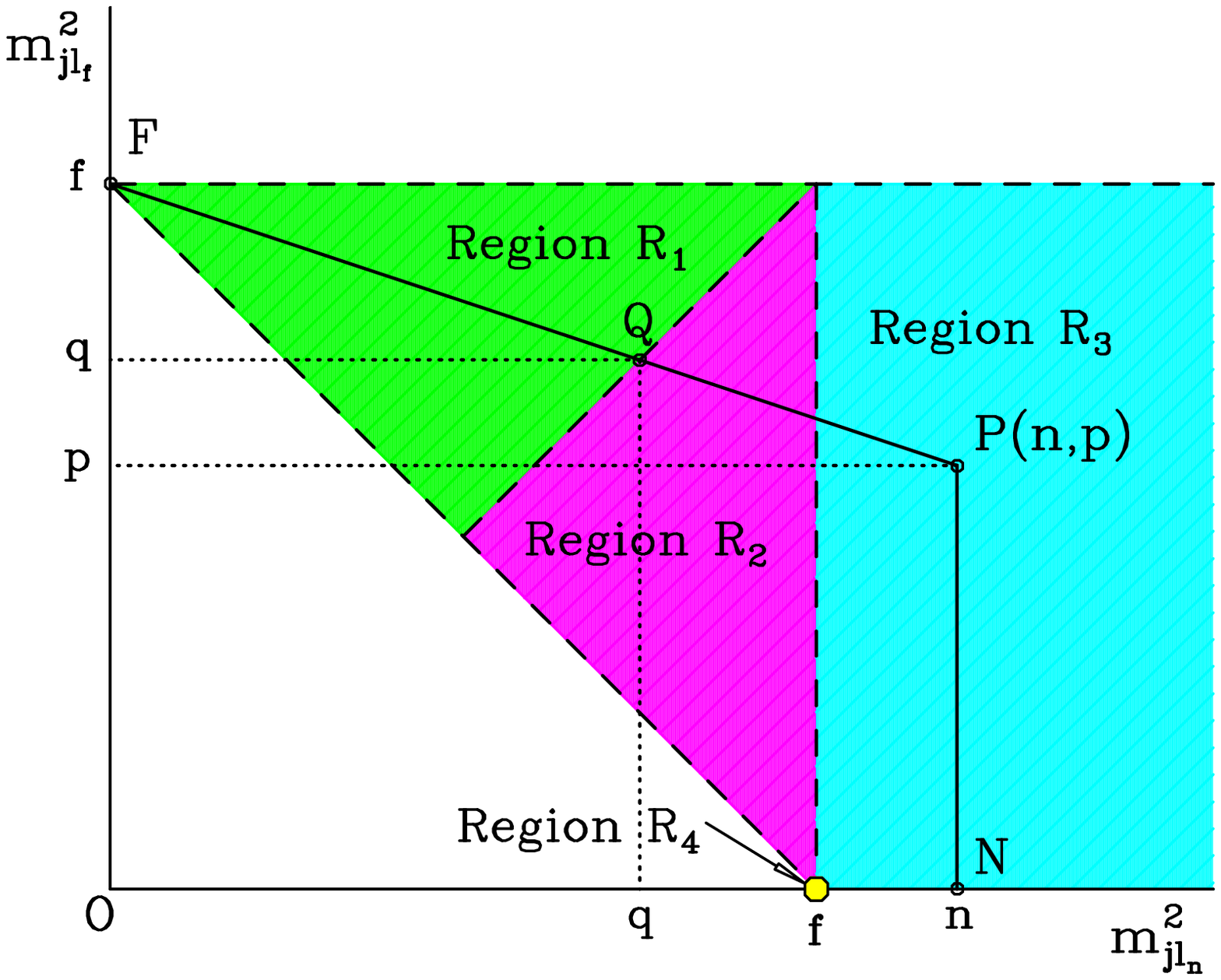,width=10cm}
\caption{The generic shape $ONPF$ of the bivariate distribution
(\ref{d2Gammanf}) in the $(m_{jl_n}^2,m_{jl_f}^2)$ plane.
\label{fig:nearfar}
}
}
Point $O$ is simply the origin of the coordinate system.
Point $N$ (for ``near'') lies on the $m_{jl_n}^2$ axis, and its
coordinate is nothing but the maximum possible value 
of the jet-near lepton invariant mass 
\beq
n\equiv \left(m_{jl_n}^{max}\right)^2  = m_D^2\, (1-R_{CD})\, (1-R_{BC})\, , \label{ndef}
\eeq
which was already introduced in eq.~(\ref{mjlnmax}).
Similarly, point $F$ (for ``far'') lies on the $m_{jl_f}^2$ axis, 
and its coordinate is nothing but the maximum possible value 
of the jet-far lepton invariant mass 
\beq
f\equiv \left(m_{jl_f}^{max}\right)^2 = m_D^2\, (1-R_{CD})\, (1-R_{AB})\, , \label{fdef}
\eeq
which was already defined in eq.~(\ref{mjlfmax}).
Finally, the point $P$ is the most important of the four corners,
since it defines the actual shape of the trapezoid, once points 
$N$ and $F$ are fixed. The coordinates of point $P$ in the  
$(m_{jl_n}^2,m_{jl_f}^2)$ plane are $(n,p)$, where 
$n$ was already defined in (\ref{ndef}), while  
$p$ is a {\em new} quantity:
\beq
p \equiv \left(m_{jl_f}^{(p)}\right)^2 \equiv f R_{BC} = m_D^2\, (1-R_{CD})\, R_{BC}\,(1-R_{AB})\, . \label{f0def}
\eeq
Since $N$ and $P$ share the same $m_{jl_n}^2$ coordinate $n$, point $P$ always 
lies directly above point $N$. At the same time, the definition of $p$ 
implies that
\beq
p < f\ , \label{f0<f}
\eeq
so that point $P$ always lies lower than point $F$, as illustrated in 
Fig.~\ref{fig:nearfar}. Finally, in Regions ${\cal R}_2$ and ${\cal R}_3$,
there is one more special point, $Q$,
which can be seen in Fig.~\ref{fig:nearfar}: it is the point where the $FP$ 
side of the trapezoid intersects the $45^\circ$ line $m_{jl_n}^2=m_{jl_f}^2$.
The two coordinates of point $Q$ are equal by definition, and are given by
\beq
q\equiv \left(m_{jl(eq)}^{max}\right)^2 = m_D^2\, (1-R_{CD})\, \frac{1-R_{AB}}{2-R_{AB}} \, , \label{qdef}
\eeq
which is nothing but the quantity previously defined in eq.~(\ref{mjleqmax}).
The four quantities $n$, $p$, $f$ and $q$ just introduced are not all independent, 
but obey the relation
\beq
\frac{f}{q} = 1 + \frac{f-p}{n}\ .
\label{fenf0}
\eeq
With those conventions, the trapezoid $ONPF$ can be equivalently defined 
through the parametric equation of the boundary line segment $FP$.
A convenient choice for the line parameter is the running value of $m_{jl_n}^2$. 
Then the parametric equation of the line $FP$ is given by
\beq
FP:\quad m^2_{jl_f}(m^2_{jl_n}) = f - \frac{f-p}{n}\, m^2_{jl_n}\ .
\label{FPline}
\eeq
In terms of this parametrization, the three special $m^2_{jl_f}$ values
introduced in Fig.~\ref{fig:nearfar} are given as follows:
\bea
f   &=& m^2_{jl_f}(0)\, ,   \\ [2mm]
p   &=& m^2_{jl_f}(n)\, ,  \\ [2mm]
q   &=& m^2_{jl_f}(q) \ .
\eea
The last equation is exactly the relation (\ref{fenf0}).

The color-coded regions in Fig.~\ref{fig:nearfar} show the allowed 
locations of point $P$, and are in one-to-one correspondence with 
the color-coded parameter space regions of Fig.~\ref{fig:regions}
(in both cases we use the same color coding). This correspondence 
is most easily seen as follows. First, note that the two white areas
in Fig.~\ref{fig:nearfar} are not accessible to point $P$. The region
with $m_{jl_f}^2>f$ is forbidden due to the relation (\ref{f0<f}). 
Similarly, the white triangular area near the origin, defined by
\beq
m_{jl_f}^2 < f - m_{jl_n}^2
\eeq
is also not allowed, which can be seen by using the inequality 
\beq
p \ge f - n\, ,
\eeq
following from the defining relations (\ref{ndef}-\ref{f0def}).
Therefore, point $P$ must belong to one of the three colored regions in 
Fig.~\ref{fig:nearfar}. As can be seen from the figure,
these three regions are distinguished based on the
value of $n$ relative to $f$ and $p$: recall that eq.~(\ref{f0<f})
already determines the hierarchy $p<f$, so that for $n$ there are 
only three possible options: $n$ can be smaller than $p$,
$n$ can be larger than $f$, or $n$ can fall in between $p$ and $f$.
Let us consider each case in turn.
\begin{enumerate}
\item The case $n<p<f$. Point $P$ then lies somewhere within the green-shaded 
area in Fig.~\ref{fig:nearfar}. Using (\ref{ndef}-\ref{f0def}), it is easy to 
see that the conditions $n<p<f$ imply
\beq
n<p<f \quad \Longrightarrow \quad \frac{1}{2-R_{AB}} < R_{BC} <1\ ,
\label{defreg1}
\eeq
which was precisely the defining relation for region ${\cal R}_1$
in Fig.~\ref{fig:regions}. Therefore, in Fig.~\ref{fig:nearfar} we 
have labelled and color-coded the area with $n<p<f$ to match the notation 
for region ${\cal R}_1$ used in Fig.~\ref{fig:regions}.
\item The case $p<n<f$. In this case point $P$ would belong to the 
magenta-shaded triangular area in Fig.~\ref{fig:nearfar}.
The conditions $p<n<f$ now imply
\beq
p<n<f \quad \Longrightarrow \quad R_{AB}<R_{BC}<\frac{1}{2-R_{AB}}\ ,
\label{defreg2}
\eeq
which is the definition of region ${\cal R}_2$ in Fig.~\ref{fig:regions}.
Once again, we label and color-code this region to match
the notation used in Fig.~\ref{fig:regions}.
\item The case $p<f<n$. Now the point $P$ should fall somewhere 
within the cyan-shaded semi-infinite rectangular strip in Fig.~\ref{fig:nearfar}.
Using (\ref{ndef}-\ref{f0def}), the constraints $p<f<n$ 
now translate into
\beq
p<f<n \quad \Longrightarrow \quad 0< R_{BC} < R_{AB}\ ,
\label{defreg3}
\eeq
which is the definition of region ${\cal R}_3$ in Fig.~\ref{fig:regions},
again justifying the notation and color-coding used in Fig.~\ref{fig:nearfar}.
\end{enumerate}
Unlike the three on-shell cases just discussed, the off-shell scenario
of Fig.~\ref{fig:chain}(b) should be handled with care, since the
``near'' and ``far'' lepton distinctions become meaningless in 
that case. Nevertheless, the off-shell scenario can still be represented 
in Fig.~\ref{fig:nearfar}, and in fact this representation is unique:
there is a single allowed location for point $P$ at $n=f$ and $p=0$. 
In Fig.~\ref{fig:nearfar} this unique location is indicated with 
a yellow-shaded circle, which corresponds to the whole yellow-shaded region 
${\cal R}_4$ in Fig.~\ref{fig:regions}. In other words, in the off-shell case 
we can randomly assign ``near'' and ``far'' labels to the two leptons in 
each event, and then the shape $ONF$ of the resulting $(m_{jl_n}^2, m_{jl_f}^2)$ 
scatter plot will be an isosceles right triangle.

From the preceding discussion it should be clear that the two-dimensional
theoretical distribution (\ref{d2Gammanf}) contains a great deal of 
useful information: its shape uniquely identifies the on-shell parameter space 
region ${\cal R}_i$, and yields the four measurements $\{n,f,p,q\}$
given in eqs.~(\ref{ndef}-\ref{f0def}) and (\ref{qdef}) instead of the 
usual two ($m_{jl(lo)}^{max}$ and $m_{jl(hi)}^{max}$).
Ideally, one would like to preserve and subsequently extract this
additional information from the experimentally observable two-dimensional 
distribution (\ref{d2Gammalohi}) as well.
We shall now show that this is in fact possible, using the simple
intuitive understanding of the shape exhibited in Fig.~\ref{fig:nearfar}.

The key is to realize that the reordering (\ref{mjllodef}) and (\ref{mjlhidef})
of the $(m_{jl_n}^2,m_{jl_f}^2)$ pair into a $(m_{jl(lo)}^2,m_{jl(hi)}^2)$
pair in geometrical terms simply corresponds to ``folding'' the trapezoid 
$ONPF$ in Fig.~\ref{fig:nearfar} along the $45^\circ$ line $m_{jl_n}^2=m_{jl_f}^2$.
\FIGURE[t]{
\epsfig{file=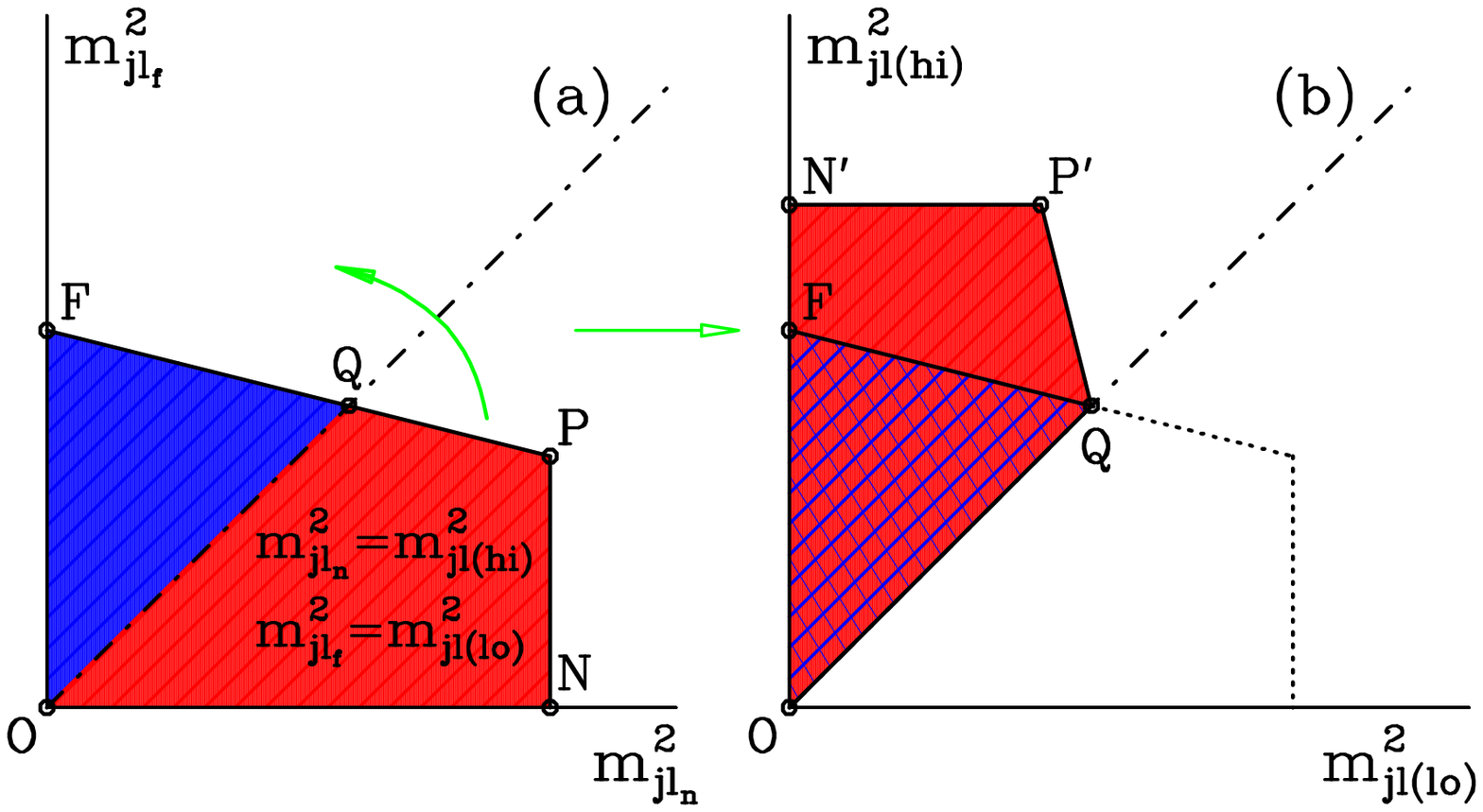,width=13cm}
\caption{Obtaining the shape of the $m_{jl(lo)}^2$ versus $m_{jl(hi)}^2$ 
bivariate distribution by folding the $m_{jl_n}^2$ versus $m_{jl_f}^2$  
distribution across the line $m_{jl_n}^2=m_{jl_f}^2$. This 
particular example applies to region ${\cal R}_3$. For the 
other three regions, refer to Figs.~\ref{fig:lohi}(a), \ref{fig:lohi}(b)
and \ref{fig:lohi}(d).
\label{fig:flip}
}
}
This procedure is shown pictorially in Fig.~\ref{fig:flip},
where for illustration we use an example from region ${\cal R}_3$,
i.e. $p<f<n$.
Panel (a) shows the trapezoidal shape of the original 
$m_{jl_n}^2$ versus $m_{jl_f}^2$ invariant mass distribution
from Fig.~\ref{fig:nearfar}. Now suppose that we want to convert 
this $(m_{jl_n}^2,m_{jl_f}^2)$ scatter plot
into a $(m_{jl(lo)}^2,m_{jl(hi)}^2)$ scatter plot, simply by 
reinterpreting the $m_{jl_n}^2$ axis as $m_{jl(lo)}^2$ and 
the $m_{jl_f}^2$ axis as $m_{jl(hi)}^2$. From that point of view,
the trapezoid $ONPF$ in Fig.~\ref{fig:flip} divides into two adjacent regions: 
$OQF$ (blue-shaded) and $ONPQ$ (red-shaded).
Within the blue-shaded area $OQF$ we have $m_{jl_n}^2<m_{jl_f}^2$,
so that the coordinate pair $(m_{jl_n}^2,m_{jl_f}^2)$ can be 
directly identified with $(m_{jl(lo)}^2,m_{jl(hi)}^2)$.
Thus the blue-shaded area $OQF$ in panel (a) remains unchanged
and appears identically in panel (b), where it is marked with a blue 
cross-hatch. In contrast, within the red-shaded area $ONPQ$ 
of panel (a), the coordinates $m_{jl_n}^2$ and $m_{jl_f}^2$ are 
in the wrong order, and need to be reversed when going to
the $(m_{jl(lo)}^2,m_{jl(hi)}^2)$ scatter plot of panel (b).
In layman terms, this reversal corresponds to ``folding'' the 
trapezoid $ONPF$ along the $45^\circ$ line $OQ$, 
as shown in Fig.~\ref{fig:flip}. The resulting image $ON'P'Q$
in Fig.~\ref{fig:flip}(b) is then overlayed on the original 
region $OQF$. We see that any $(m_{jl(lo)}^2,m_{jl(hi)}^2)$ 
scatter plot will therefore exhibit two characteristic types of
population density. For example, in the 
{\em blue-hatched} red area of Fig.~\ref{fig:flip}(b) 
we expect the density of points to roughly double, since
the folded distribution $ON'P'Q$ is overlaid on top of the
existing distribution $OQF$ underneath. In Fig.~\ref{fig:lohi}
below, we shall mark such ``double-density'' regions 
with a blue cross-hatch in addition to the solid red shading.
In contrast, region $FQP'N'$ in Fig.~\ref{fig:flip}(b)
is a ``single density'' region, since the folded distribution 
happened to fall onto empty space, where originally there were 
no points to begin with. A single density region can also be 
obtained when portions of the original $(m_{jl_n}^2,m_{jl_f}^2)$ 
scatter plot are not overlaid in the process of folding.
In either case, we shall denote a single-density region by a solid
(red) color-shading, but no cross-hatch.

We are now ready to apply the intuition gained from
Figs.~\ref{fig:nearfar} and \ref{fig:flip} and identify the
characteristic shapes of the $(m_{jl(lo)}^2,m_{jl(hi)}^2)$ 
distribution for each parameter space region ${\cal R}_i$.
Our results are displayed in Fig.~\ref{fig:lohi}, where we show the 
four characteristic shapes for each case: the on-shell cases of
(a) Region ${\cal R}_1$,
(b) Region ${\cal R}_2$, 
(c) Region ${\cal R}_3$, and the off-shell case of
(d) Region ${\cal R}_4$.
\FIGURE[t]{
\epsfig{file=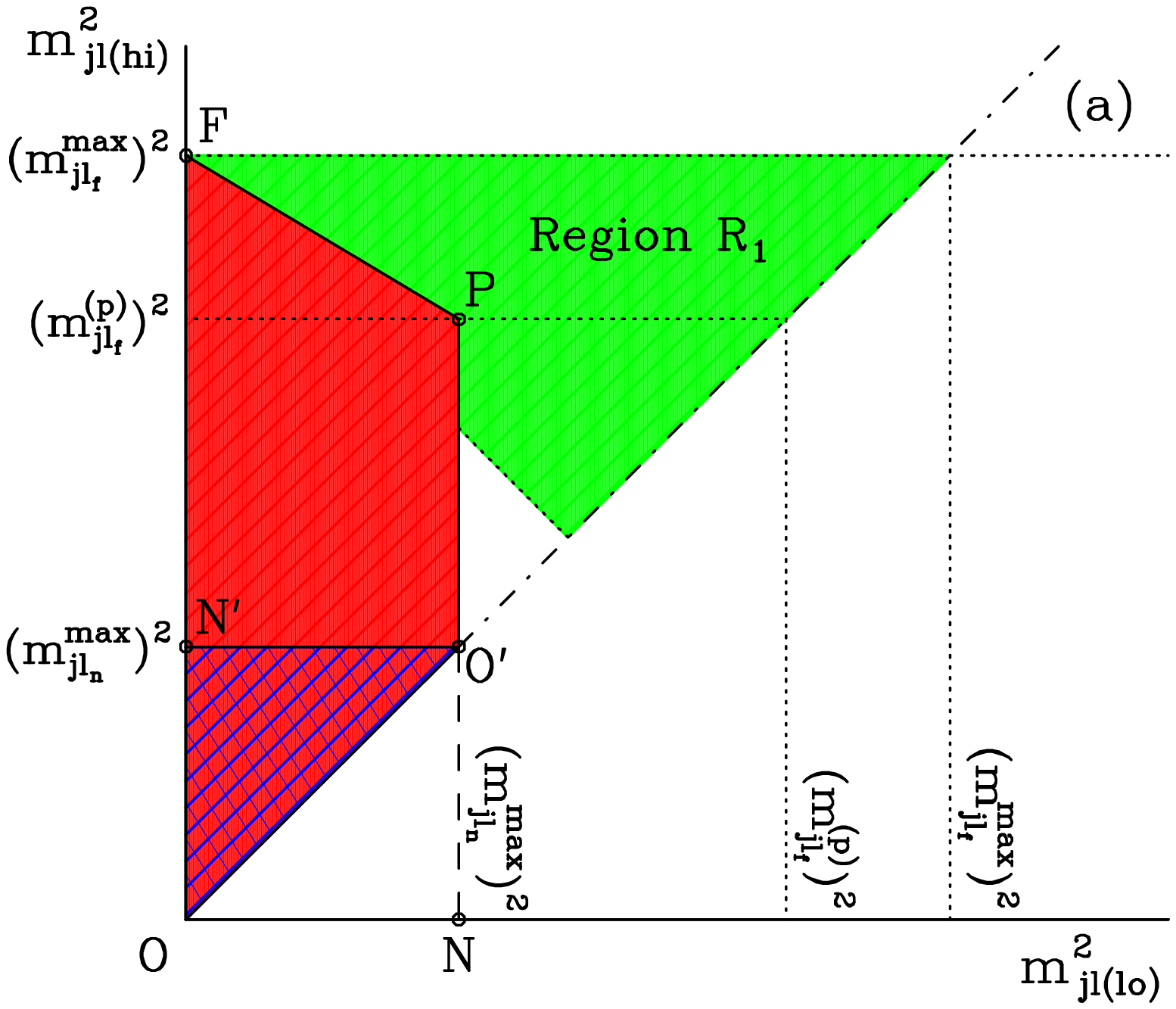, width=7.5cm}~~
\epsfig{file=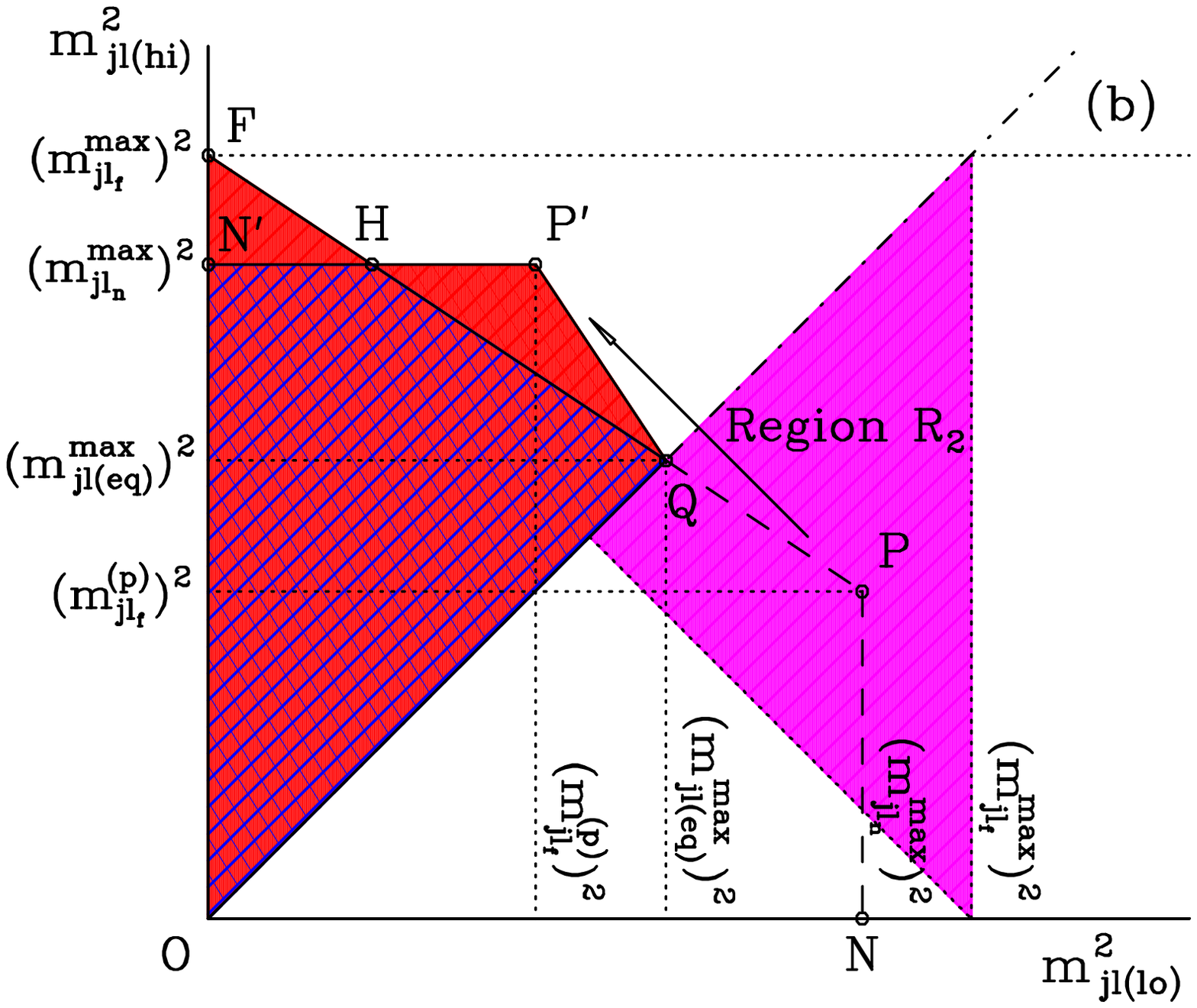, width=7.5cm}\\ [0.5cm]
\epsfig{file=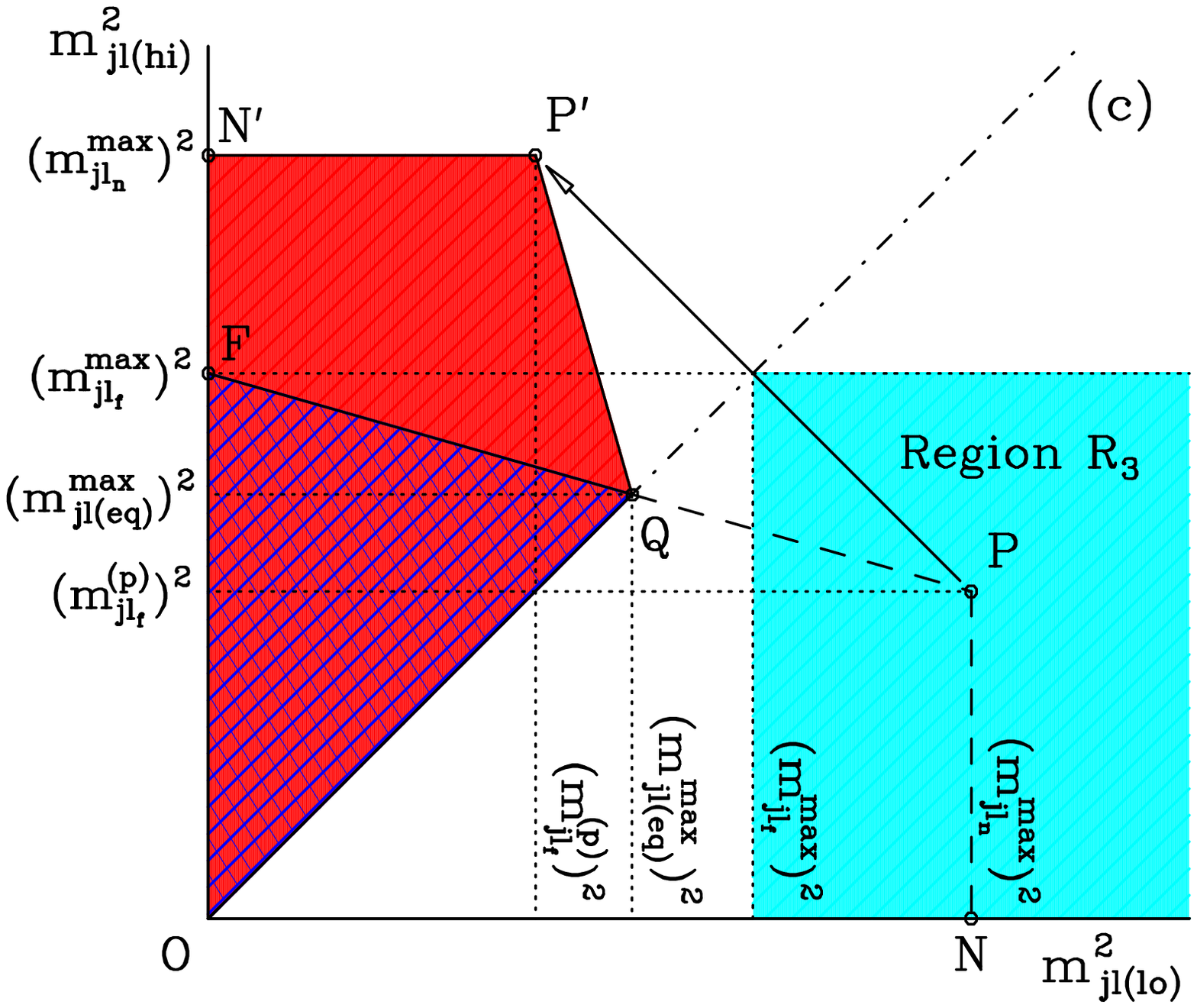, width=7.5cm}~~
\epsfig{file=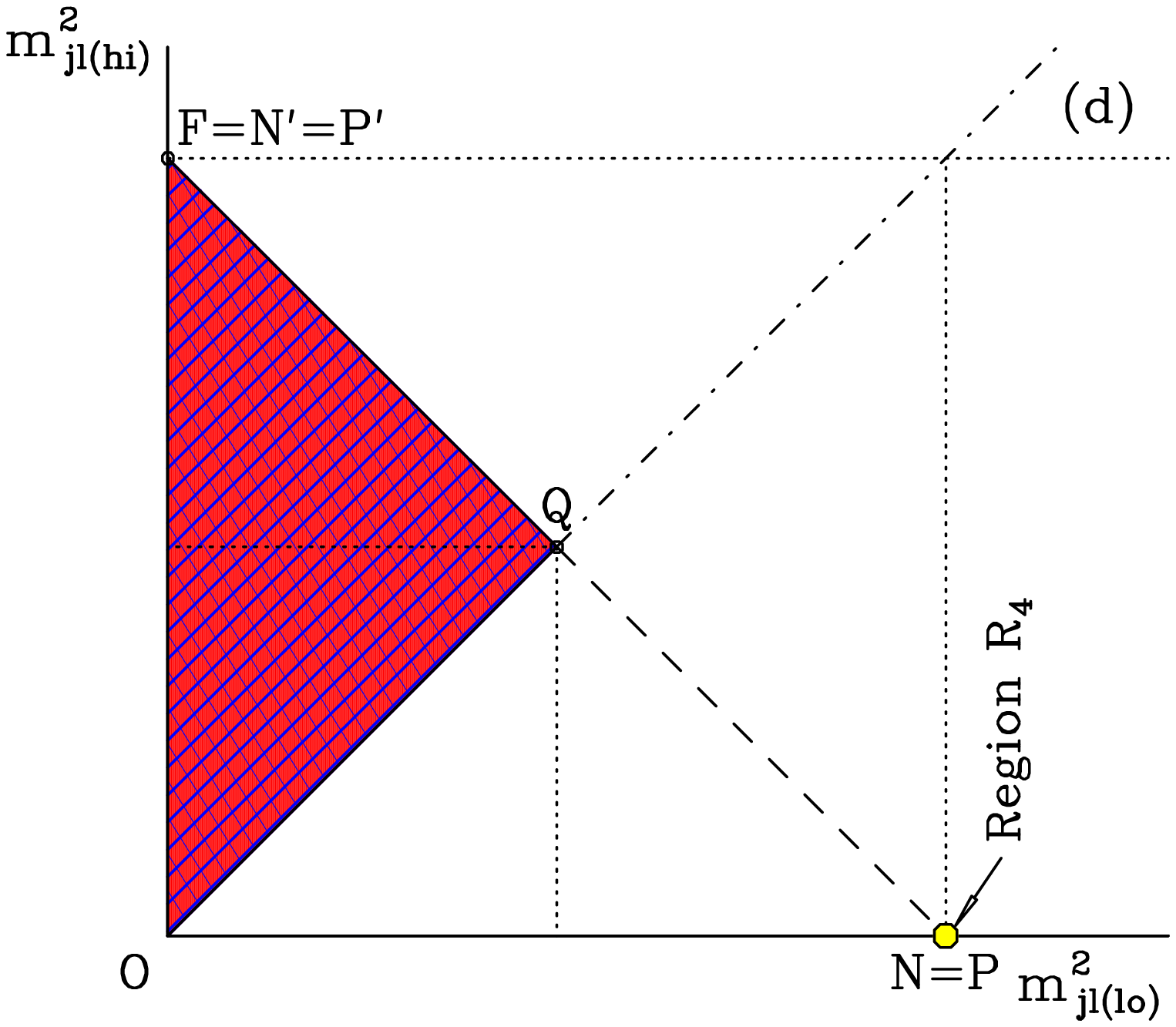, width=7.5cm}
\caption{The generic shape of the bivariate distribution $m_{jl(lo)}^2$
versus $m_{jl(hi)}^2$ for each of the four parameter space regions:
(a) Region ${\cal R}_1$,
(b) Region ${\cal R}_2$, 
(c) Region ${\cal R}_3$ and 
(d) the off-shell case of Region ${\cal R}_4$.
Each panel shows the typical shape (red-shaded) of the resulting 
$(m_{jl(lo)}^2,m_{jl(hi)}^2)$ distribution, after the ``folding'' 
in Fig.~\ref{fig:flip}. Blue-hatched (unhatched)
areas correspond to double-density (single-density) regions.
Each panel also shows the original location of the point $P$ in the
$(m_{jl_n}^2,m_{jl_f}^2)$ plot, as well as the allowed 
positions of point $P$, following the color conventions
of Figs.~\ref{fig:regions} and \ref{fig:nearfar}.
\label{fig:lohi}
}
}
Each panel shows the typical shape (red-shaded) of the resulting 
$(m_{jl(lo)}^2,m_{jl(hi)}^2)$ distribution, after the ``folding'' 
in Fig.~\ref{fig:flip}. Blue-hatched (unhatched)
areas correspond to double-density (single-density) regions.
In addition, we show the original location of the point 
$P$ in the $(m_{jl_n}^2,m_{jl_f}^2)$ plot. The allowed 
positions of point $P$ in each case are color-shaded, following the color 
conventions of Figs.~\ref{fig:regions} and \ref{fig:nearfar}.

The nice feature of all the plots in
Fig.~\ref{fig:lohi} is that they are composed entirely 
of straight lines. This is a consequence of the fact that the original
trapezoid in Fig.~\ref{fig:nearfar} is made up of straight lines,
and then the ``folding'' of Fig.~\ref{fig:flip}
does not curve the boundaries.
Notice also the presence of internal kinematic boundaries, 
marking abrupt changes in the density of the distribution, 
e.g.~$O'N'$ in Fig.~\ref{fig:lohi}(a),
$QHN'$ in Fig.~\ref{fig:lohi}(b) and $QF$ in Fig.~\ref{fig:lohi}(c).

It is clear from Fig.~\ref{fig:lohi} that the shape of the
$(m_{jl(lo)}^2,m_{jl(hi)}^2)$ scatter plot allows us to uniquely
determine the parameter space region at hand. For example, 
the typical shape for Region ${\cal R}_1$, exhibited in
Fig.~\ref{fig:lohi}(a), consists of a right-angle triangular 
region $OO'N'$ of double density and a right-angle trapezoidal 
region $N'O'PF$ of single density.
In this case, point $P$ is directly observable, 
and its coordinates immediately yield the quantities 
$n$ and $p$ defined in (\ref{ndef}) and (\ref{f0def}).
In addition, one can also measure the location $f$ 
of point $F$ along the $m_{jl(hi)}^2$ axis, given by eq.~(\ref{fdef}). 
This gives a total of three independent measurements: $n$, $p$ and $f$,
which should be ordered as $n<p<f$, in accordance with (\ref{defreg1}). 
Now we can clearly see the benefit of considering the two-dimensional 
$(m_{jl(lo)}^2,m_{jl(hi)}^2)$ distribution as opposed to the
two individual one-dimensional distributions $m_{jl(lo)}^2$
and $m_{jl(hi)}^2$. Those one-dimensional distributions 
are obtained by projecting the $(m_{jl(lo)}^2,m_{jl(hi)}^2)$ 
scatter plot shown in Fig.~\ref{fig:lohi}(a) onto the two axes.
It is easy to see from Fig.~\ref{fig:lohi}(a) that in this case
the endpoint of the one-dimensional $m_{jl(lo)}^2$ distribution 
will be given by $c=n$, while the endpoint of the one-dimensional 
$m_{jl(hi)}^2$ distribution will be given by $d=f$, and neither of 
those will reveal the quantity $p$. In contrast, $p$ can be easily 
identified on the scatter plot, and provides an additional 
{\em independent} measurement. 
 
The case of Region ${\cal R}_3$, which is shown in Fig.~\ref{fig:lohi}(c),
is rather similar: the double-density region is still a triangle
($OQF$), while the single-density region is a quadrilateral
($FQP'N'$). This time instead of point $P$ we can clearly see its image
$P'$, whose coordinates nevertheless still reveal the values of $n$ and $p$.
Point $F$ is now hidden within the scatter plot, but may still be identifiable, 
since it corresponds to an abrupt change in density of points.
Finally, now we have an additional measurement $q$ of point $Q$, which is where
the original line segment $FP$ was folded in the reordering process of 
eqs.~(\ref{mjllodef},\ref{mjlhidef}). As a result, in region ${\cal R}_3$
we have a total of 4 measurements of kinematic endpoints, $n$, $p$, $f$ and $q$, 
ordered as follows: $p<q<f<n$ (see also eq.~(\ref{defreg3})).
Later, when we project onto the two axes, the endpoint of the 
one-dimensional $m_{jl(lo)}^2$ distribution will be given by $c=q$, 
while the endpoint of the one-dimensional $m_{jl(hi)}^2$ distribution 
will be given by $d=n$. This now leaves out {\em two} additional potential 
measurements, $p$ and $f$, which can be accessed on the two-dimensional 
scatter plot. 

In the third on-shell case of  Region ${\cal R}_2$, shown in Fig.~\ref{fig:lohi}(b),
the shape is more complex: the double density region is now a 
quadrilateral $OQHN'$, while there are two disjoint single density 
triangular regions $QP'H$ and $N'HF$. The point $H$ appears on the
intersection of the original $FP$ boundary of the $(m_{jl_n}^2,m_{jl_f}^2)$
scatter plot in Fig.~\ref{fig:nearfar} and the (horizontal) image $P'N'$ 
of the (vertical) $PN$ boundary in Fig.~\ref{fig:nearfar}.
Once again, the coordinates of point $P'$ reveal $p$ and $n$, while points $F$
and $Q$ reveal $f$ and $q$, correspondingly. In region ${\cal R}_2$,
therefore, there are 4 potential measurements, $p$, $n$, $f$ and $q$,
ordered as follows: $p<q<n<f$. When the scatter plot of Fig.~\ref{fig:lohi}(b)
is projected onto the axes, the endpoint of the 
one-dimensional $m_{jl(lo)}^2$ distribution will be given by $c=q$, 
while the endpoint of the one-dimensional $m_{jl(hi)}^2$ distribution 
will be given by $d=f$. Once again, this leaves out {\em two} additional potential 
measurements, $p$ and $n$, which can be extracted from the two-dimensional 
scatter plot. 

Let us now turn to the off-shell scenario of Region ${\cal R}_4$, 
which is represented in Fig.~\ref{fig:lohi}(d).
Because of the symmetry between the ``near'' and ``far'' leptons in the off-shell 
case, the folded region $ONQ$ has an identical triangular shape as the 
underlying region $OFQ$, so that after the fold the two match perfectly
and we obtain a single triangular region of double density, and no 
single-density areas. As can be seen from Fig.~\ref{fig:lohi}(d),
the off-shell scenario offers only one nontrivial endpoint measurement, 
which can be taken as $f$. The latter appears as the endpoint $d$ in the 
one-dimensional $m_{jl(hi)}^2$ distribution, while the endpoint $c$ of the
other one-dimensional distribution, $m_{jl(lo)}^2$, is then simply given as
$c=f/2$, in agreement with eq.~(\ref{cdoff}). 

We have just seen that in the off-shell scenario of Fig.~\ref{fig:chain}(b)
the two-dimensional scatter plot $(m_{jl(lo)}^2,m_{jl(hi)}^2)$
does not yield any additional kinematic endpoint measurements. 
However, it can still be helpful in discriminating
a potential regional ambiguity which may arrise as follows. 
Notice that the triangular double-density shape of the 
scatter plot in the off-shell case of Fig.~\ref{fig:lohi}(d) can in principle 
also be obtained in the on-shell cases of Fig.~\ref{fig:lohi}(b) 
and Fig.~\ref{fig:lohi}(c), provided that the image 
$P'$ of point $P$ ends up very close to point $F$. In terms of the 
$(m_{jl_n}^2,m_{jl_f}^2)$ scatter plot of Fig.~\ref{fig:nearfar}, this
situation corresponds to the on-shell cases of Regions ${\cal R}_2$
or ${\cal R}_3$, with point $P$ lying very close to the yellow-shaded dot 
representing Region ${\cal R}_4$. In spite of having the same shape
of their boundary lines, the two scatter plots will be quite different,
as they will exhibit a different point density. In particular, for all three
on-shell cases, the pure phase space two-dimensional differential distribution 
(\ref{d2Gammanf}) is given by the following (unit-normalized) formula
\beq
\frac{d^2 \Gamma}{dm_{jl_n}^2\, dm_{jl_f}^2}
= \frac{1}{n\left( m^2_{jl_f}(m^2_{jl_n}) \right)}
= \frac{1}{fn-(f-p)\, m_{jl_n}^2}\ , \qquad ({\rm for} \ R_{BC}<1).
\label{ondensity}
\eeq
Notice that within the kinematically allowed region, 
the density is independent of $m_{jl_f}^2$. 
In the limit $p\to 0$, the expression (\ref{ondensity}) becomes 
singular when $m_{jl_n}^2\to n$. This singularity is regularized by the 
width of particle $B$ and the branching fraction for the $C\to B$ decay.
In contrast, the corresponding density in the off-shell case 
is quite different, and in particular 
does not exhibit such singular features. 

We are now ready to revisit the duplication problem discussed in 
Section~\ref{sec:duplication}. We have just seen that the two-dimensional 
distribution of $m_{jl(lo)}^2$ versus $m_{jl(hi)}^2$ can help resolve the
duplication in two very different ways. First, the shape 
of the kinematic boundary lines in the $(m_{jl(lo)}^2,m_{jl(hi)}^2)$
scatter plot uniquely identifies the region, as shown in Fig.~\ref{fig:lohi}.
Since the duplicate solutions that we found always appear in two different regions, 
this is in principle sufficient to eliminate the wrong solution. 
Secondly, the scatter plots offer the possibility of additional measurements,
and at the very least a measurement of the quantity $p$. 
As can be seen from Table~\ref{table:dup2},
the value of $p$ is already different for each pair of duplicate spectra, and,
provided that it can be measured with sufficient precision, 
can also be used to remove the ambiguity. 

\FIGURE[t]{
\epsfig{file=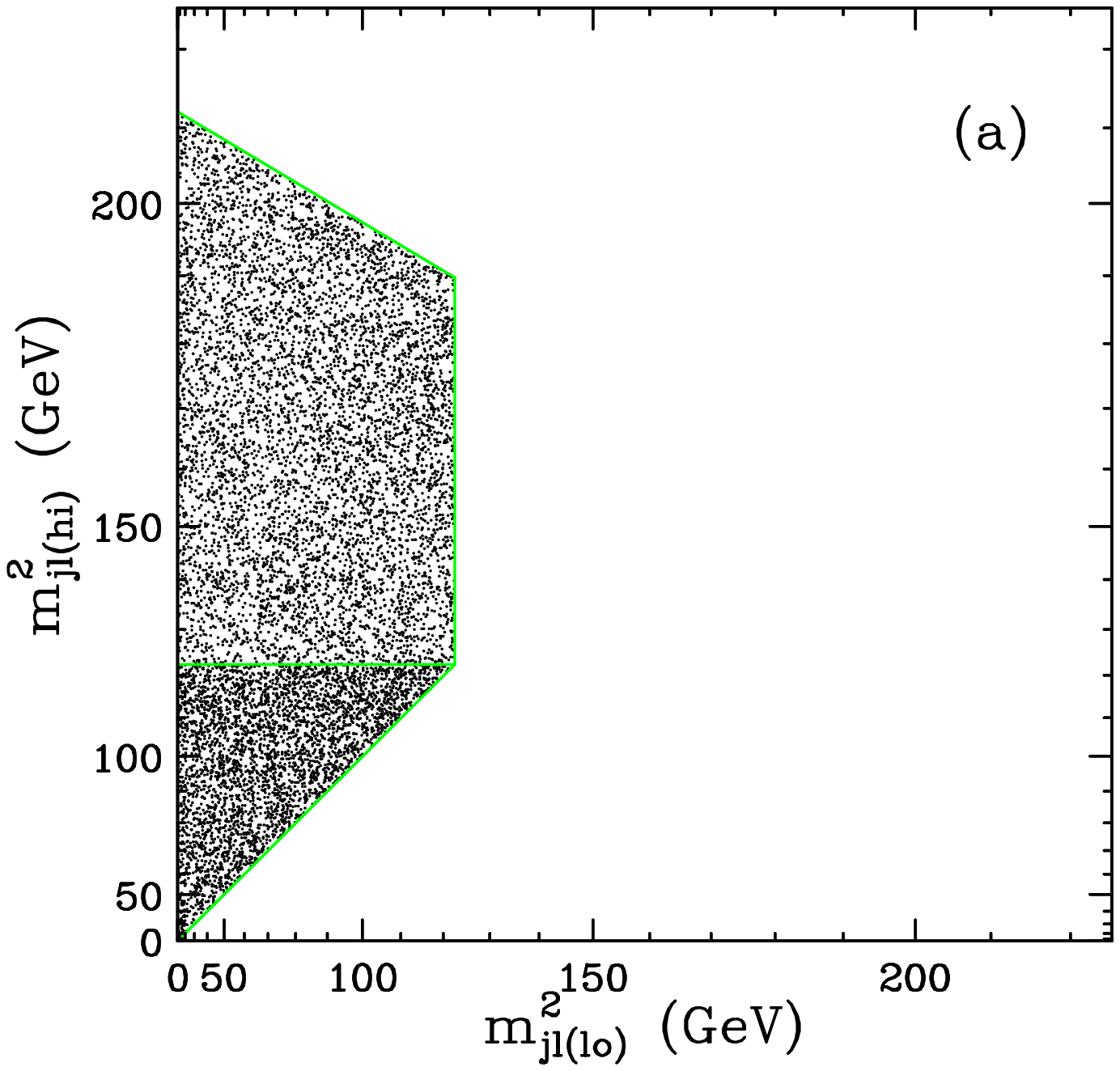, width=7.5cm}~~
\epsfig{file=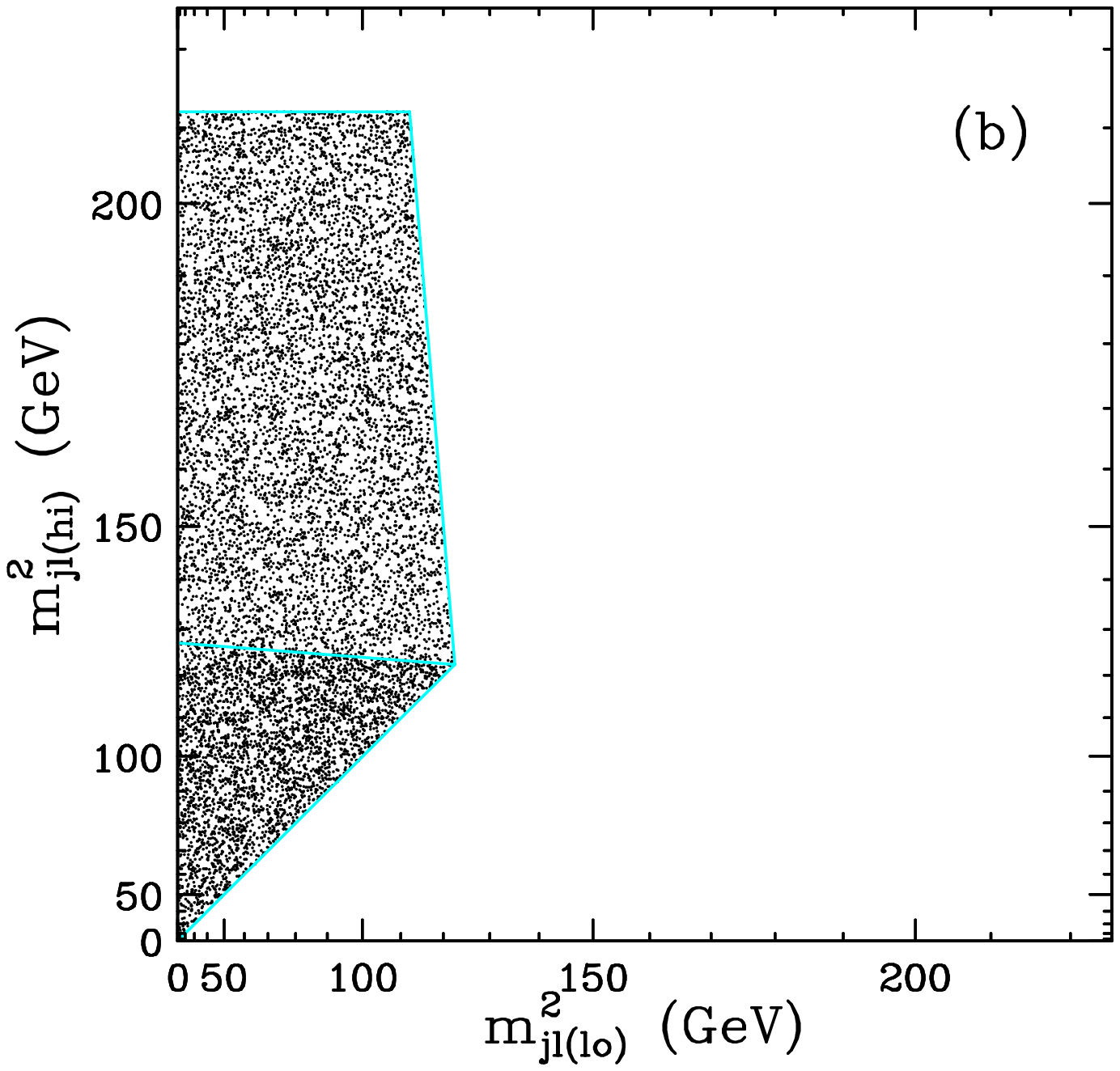, width=7.5cm}\\ [0.5cm]
\epsfig{file=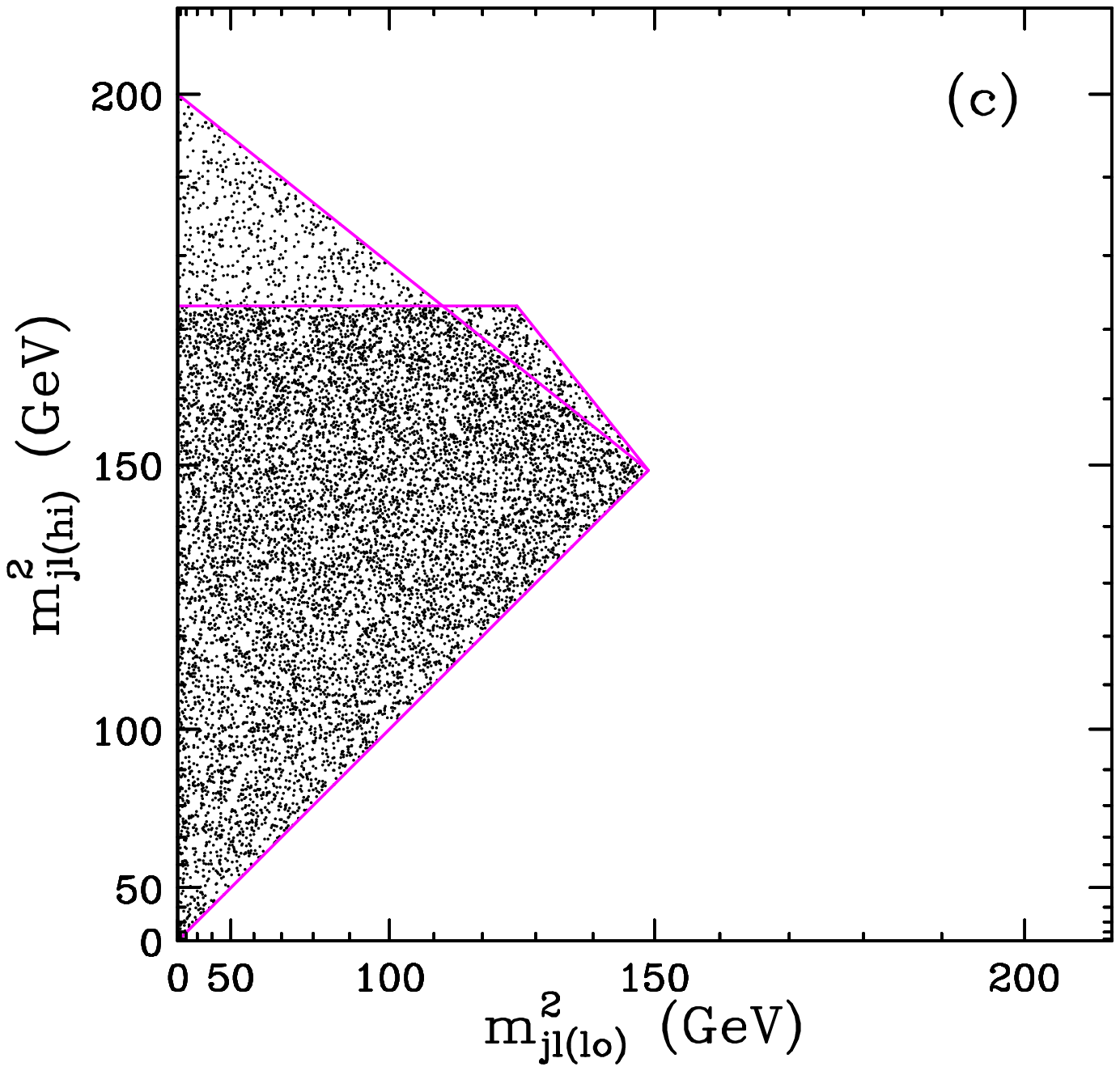, width=7.5cm}~~
\epsfig{file=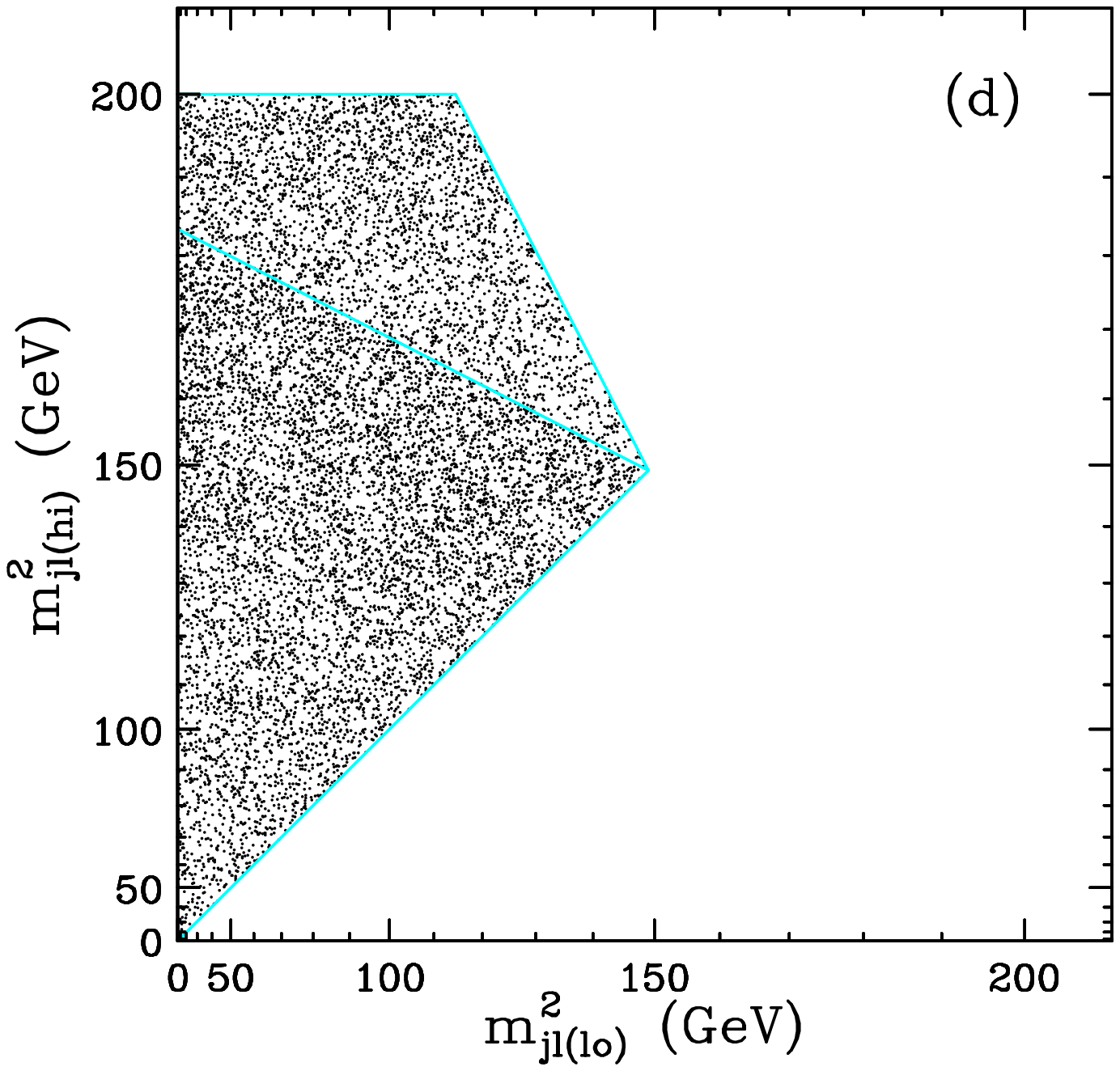, width=7.5cm}
\caption{Scatter plots of $m_{jl(lo)}^2$ versus $m_{jl(hi)}^2$
for the four study points from Table~\ref{table:dup2} exhibiting duplication:
(a) point $P_{31}$, (b) point $P_{23}$, 
(c) point $P_{32}$ and (d) point $P'_{23}$.
Notice the quadratic scale used on both axes.
The kinematic boundary lines are outlined 
with the corresponding color for each region,
following the color coding conventions of 
Figs.~\ref{fig:regions} and \ref{fig:nearfar}.
Each plot has 10,000 data points.
We assume that all particles $A$, $B$, $C$ and $D$ 
are exactly on-shell.
\label{fig:HL}
}
}

Our conclusions are confirmed by Fig.~\ref{fig:HL}, which shows the 
$(m_{jl(lo)}^2,m_{jl(hi)}^2)$ scatter plots for the four study points 
from Table~\ref{table:dup2} exhibiting duplication:
(a) point $P_{31}$, (b) point $P_{23}$, 
(c) point $P_{32}$ and (d) point $P'_{23}$.
The figure indeed shows that each pair of duplicate points has identical 
values for the endpoints of the separate {\em one-dimensional} invariant 
mass distributions $m_{jl(lo)}^2$ and $m_{jl(hi)}^2$.
However, the shapes of the scatter plots are very different, 
and so are the values of the corresponding $p$ endpoints.
We therefore conclude that the duplication encountered in Section
\ref{sec:duplication} ceases to be a problem, once we generalize
the analysis to two-dimensional (bivariate) distributions as
discussed here.

In conclusion of this section, we point out that when the two-dimensional 
scatter plots like those in Fig.~\ref{fig:HL} are projected onto the
axes to obtain the corresponding one-dimensional distributions 
of either $m_{jl(lo)}^2$ or $m_{jl(hi)}^2$,
the latter often exhibit some peculiar features near their endpoints,
which were classified as either ``feet'' or ``drops'' in ref.~\cite{Miller:2005zp}.
The origin of these features is now easy to understand in terms of
the original two-dimensional scatter plot. For example, consider the 
scatter plots in Figs.~\ref{fig:HL}(b) and \ref{fig:HL}(d). 
When projected onto the $m_{jl(hi)}^2$ axis, both of them will exhibit 
a classic ``drop'' at the $m_{jl(hi)}^2$ endpoint, which is simply due to
the flat upper boundary $P'N'$ in Fig.~\ref{fig:lohi}(c).
Similarly, the projection of the scatter plot in Fig.~\ref{fig:HL}(c) 
onto the $m_{jl(hi)}^2$ axis will exhibit a classic 
``foot'' extending from $n$ to $f$. The ``foot'' can be easily understood 
in terms of the generic shape of Fig.~\ref{fig:lohi}(b), where it 
arises from the projection of the single density area $N'HF$.

\section{Kinematic boundary lines for the $m_{ll}^2$ vs. $m_{jll}^2$ distribution}
\label{sec:jllvsll}

Following the logic of the previous section, we shall now proceed to analyze the 
two-dimensional distribution
\beq
\frac{d^2 \Gamma}{dm_{jll}^2\, dm_{ll}^2}\ ,
\label{d2Gammajll}
\eeq
whose generic shape $OVUS$ is shown in Fig.~\ref{fig:jll}.
The kinematic boundary lines of the scatter plot (\ref{d2Gammajll})
generally consist of four segments. The upper ($SU$) and lower ($OV$) 
curved boundaries are parts of a hyperbolic curve $OWS$, 
while the left ($OS$) and right ($UV$) boundaries are straight lines.
Therefore, in order to describe the shape of the $(m_{ll}^2,m_{jll}^2)$
scatter plot, it is sufficient to provide the parametric equations for 
the upper and lower curved boundaries $SW$ and $OW$, plus the location
of the vertical line $UV$. In analogy with (\ref{FPline}), we choose 
the variable on the horizontal axis, in this case $m_{ll}^2$, as the
line parameter describing the hyperbola $OWS$. Then the upper boundary line 
$SUW$ is given by the parametric equation \cite{Lester:2006cf}
\bea
m_{jll(+)}^2(m_{ll}^2) &=& \frac{1+R_{CD}}{2} \frac{m_{ll}^2}{R_{CD}} 
                     + \frac{1}{2}\, m_D^2 (1-R_{CD})(1-R_{AC}) \label{upper} \\ [2mm]
&+& \frac{1-R_{CD}}{2} \left\{ \left[ \left(\frac{m_{ll}^2}{R_{CD}}\right)-m_D^2(1+R_{AC})\right]^2 
                               -4m_D^4R_{AC} \right\}^{\frac{1}{2}}, \nonumber
\eea
while the lower boundary line $OVW$ is given by \cite{Lester:2006cf}
\bea
m_{jll(-)}^2(m_{ll}^2) &=& \frac{1+R_{CD}}{2} \frac{m_{ll}^2}{R_{CD}} 
                     + \frac{1}{2}\, m_D^2 (1-R_{CD})(1-R_{AC}) \label{lower} \\ [2mm]
&-& \frac{1-R_{CD}}{2} \left\{ \left[ \left(\frac{m_{ll}^2}{R_{CD}}\right)-m_D^2(1+R_{AC})\right]^2 
                               -4m_D^4R_{AC} \right\}^{\frac{1}{2}}. \nonumber
\eea
The vertical straight line segment $UV$ is in general located at $m_{ll}^2=a$,
where $a$ is the value of the dilepton invariant mass endpoint $(m_{ll}^{max})^2$
already introduced in Section~\ref{sec:forward}. 
\FIGURE[t]{
\epsfig{file=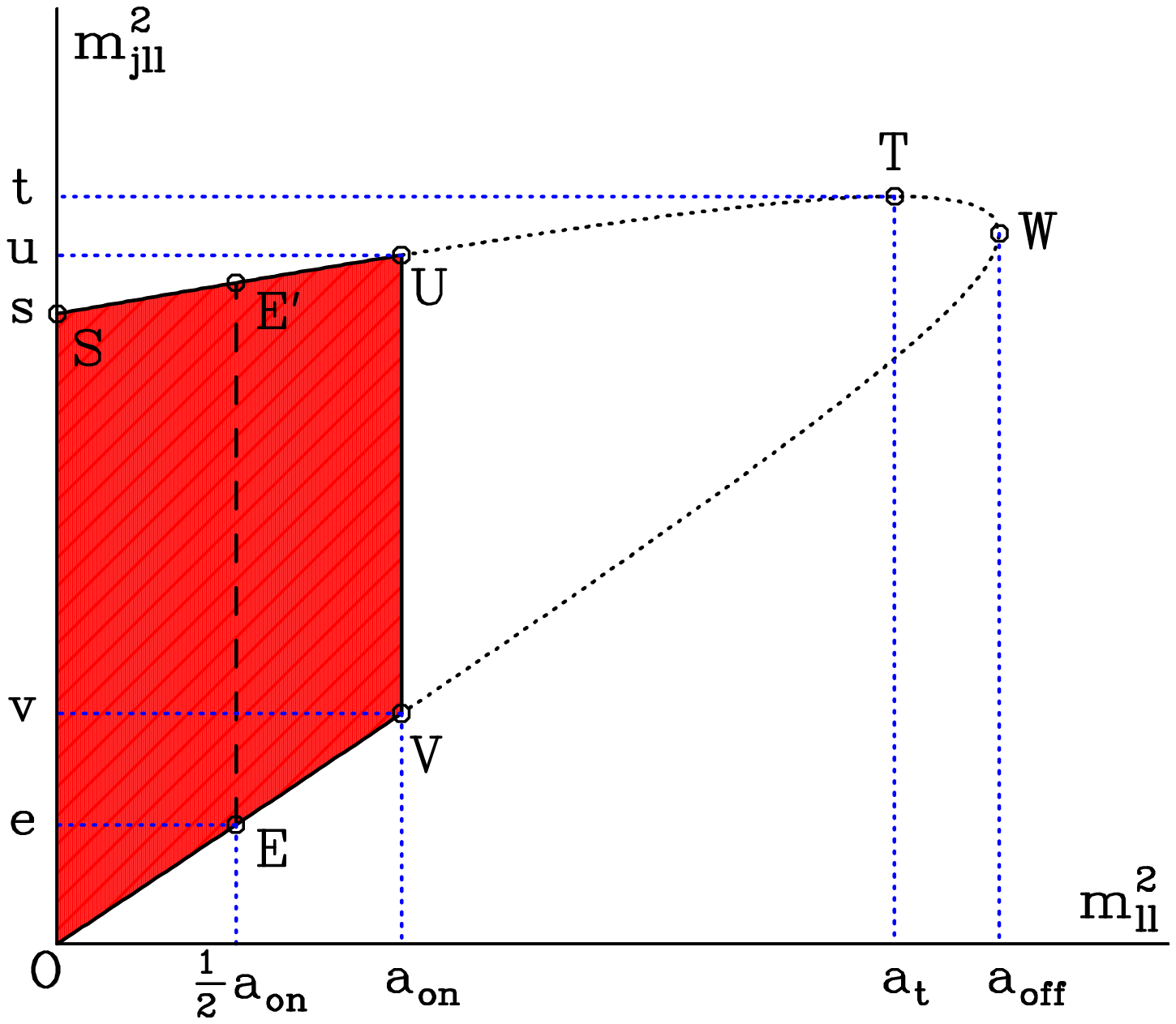,width=10cm}
\caption{The generic shape $OVUS$ of the bivariate distribution
(\ref{d2Gammajll}) in the $(m_{ll}^2,m_{jll}^2)$ plane.
\label{fig:jll}
}
}
As we saw in Section~\ref{sec:forward}, the expression for $a=(m_{ll}^{max})^2$
depends on whether we are dealing with the on-shell scenario of Fig.~\ref{fig:chain}(a)
or the off-shell scenario of Fig.~\ref{fig:chain}(b). Therefore we shall now introduce 
separate notation for the endpoint $a$ in each of these two cases.
In the on-shell scenario of Fig.~\ref{fig:chain}(a) we shall use $a_{\rm on}$ to designate 
our previous eq.~(\ref{lldef})
\begin{equation}
a_{\rm on} \equiv m_D^2\, R_{CD}\, (1-R_{BC})\, (1-R_{AB}),
\label{aon}
\end{equation}
while in the off-shell scenario of Fig.~\ref{fig:chain}(b) 
we shall use $a_{\rm off}$ for the previous result (\ref{lloff})
\begin{equation}
a_{\rm off} \equiv m_D^2\, R_{CD}\,(1-\sqrt{R_{AC}})^2.
\label{aoff}
\end{equation}
From these two equations, it is not difficult to see that
\beq
a_{\rm on} \le a_{\rm off},
\label{aonoff}
\eeq
as indicated in Fig.~\ref{fig:jll}. As an aside, we mention that 
with the help of eqs.~(\ref{aon}) and (\ref{aoff})
it is easy to show that the equal sign in (\ref{aonoff})
is achieved when $R_{AB}=R_{BC}$, i.e. when the on-shell spectrum happens 
to lie exactly on the border between regions ${\cal R}_2$ and ${\cal R}_3$
in Fig.~\ref{fig:regions}. In physical terms this means that the 
mass $m_B$ of particle $B$ is equal to the geometric mean of $m_A$ and $m_C$:
\beq
m_B = \sqrt{m_A m_C}\quad \Rightarrow \quad a_{\rm on} = a_{\rm off}.
\label{onspecial}
\eeq
This represents another potential source of confusion
in extracting the mass spectrum -- the measurement of the 
dilepton invariant mass endpoint $a$ alone tells us nothing about 
whether the intermediate particle $B$ is on-shell or off-shell, 
leading to two possible solutions \cite{Birkedal:2005cm}. 
Fortunately, with the inclusion of the additional measurements
$c$, $d$ and $e$, we did not encounter this type of duplication
in the course of our analysis in Sec.~\ref{sec:duplication}.

Returning now to our discussion of the $(m_{ll}^2,m_{jll}^2)$ scatter 
plot in Fig.~\ref{fig:jll}, 
eq.~(\ref{aonoff}) implies that in the on-shell case of Fig.~\ref{fig:chain}(a),
the data points do not fill up the whole region $OWS$, but only extend up to
the vertical boundary $UV$. The region to the right of the $UV$ line
is kinematically inaccessible. On the other hand, in the off-shell case of 
Fig.~\ref{fig:chain}(b), the whole region $OWS$ is filled up.
The only exception to this rule is the very special on-shell case of
(\ref{onspecial}), when the $UV$ line moves to the very tip $W$ of the hyperbola, 
thus allowing the whole region $OWS$, as if this were an off-shell scenario.

In analogy with our discussion in Section~\ref{sec:hivslo} of the 
kinematic boundaries in the two-dimensional $(m_{jl(lo)}^2,m_{jl(hi)}^2)$ 
distribution, we now identify several special points along the hyperbola $OWS$
in Fig.~\ref{fig:jll}. Point $O$ is simply the origin $(0,0)$ of the 
$(m_{ll}^2,m_{jll}^2)$ coordinate system.
Point $W$ is the tip of the hyperbola, where the upper branch $m_{jll(+)}^2(m_{ll}^2)$
meets the lower branch $m_{jll(-)}^2(m_{ll}^2)$. By definition, the $m_{ll}^2$ coordinate 
of point $W$ is $a_{\rm off}$, while its $m_{jll}^2$ coordinate is
\beq
w \equiv m_{jll(+)}^2(a_{\rm off}) \equiv m_{jll(-)}^2(a_{\rm off}) = 
m_D^2\, \left(1-R_{CD}\sqrt{R_{AC}}\right)\left(1-\sqrt{R_{AC}}\right).
\label{wdef}
\eeq
Point $S$ is where the upper kinematic boundary line $m_{jll(+)}^2(m_{ll}^2)$
intersects the $m_{jll}^2$ coordinate axis. The $m_{jll}^2$ coordinate 
of point $S$ is therefore
\beq
s \equiv m_{jll(+)}^2(0) = m_D^2 \left(1-R_{CD}\right)\left(1-R_{AC}\right).
\label{sdef}
\eeq
Points $U$ and $V$ label the intersections of the vertical boundary $UV$ with the
upper and lower hyperbolic branches (\ref{upper}) and (\ref{lower}), 
respectively. They share the same $m_{ll}^2$ coordinate $a_{\rm on}$, 
while their $m_{jll}^2$ coordinates are correspondingly given by
\bea
u &\equiv& m_{jll(+)}^2(a_{\rm on})  \label{udef}  \\[2mm]
&=& \frac{1}{2}\, m_D^2 \biggl[(1+R_{CD})(1-R_{BC})(1-R_{AB})
+(1-R_{CD})(1-R_{AC}+|R_{BC}-R_{AB}|)  \biggr], \nonumber \\[2mm]
v &\equiv& m_{jll(-)}^2(a_{\rm on})  \label{vdef}  \\[2mm]
&=& \frac{1}{2}\, m_D^2 \biggl[(1+R_{CD})(1-R_{BC})(1-R_{AB})
+(1-R_{CD})(1-R_{AC}-|R_{BC}-R_{AB}|)  \biggr]. \nonumber 
\eea
Finally, there is one more special point on the upper branch $SUW$: 
it is the point $T$ where $m_{jll(+)}^2(m_{ll}^2)$ has a local maximum. 
The $m_{ll}^2$ coordinate $a_t$ of point $T$ can be found from the 
minimization condition
\beq
\left(\frac{d m_{jll(+)}^2}{d m_{ll}^2} \right)_{m_{ll}^2=a_t} = 0
\eeq
and is given by
\beq
a_t \equiv m_D^2 \left(R_{CD}- \sqrt{R_{AD}}\,\right)\left( 1- \sqrt{R_{AD}}\,\right).
\label{atdef}
\eeq
Then, the $m_{jll}^2$ coordinate $t$ of point $T$ is easily found by
substituting (\ref{atdef}) into (\ref{upper}):
\beq
t \equiv m_{jll(+)}^2(a_t) 
=m_D^2 \left(1-\sqrt{R_{AD}}\, \right)^2.
\label{tdef}
\eeq
We should point out that point $T$ as we have defined it here,
does not exist in all parameter space regions. To see this, let us 
calculate the slope of the upper branch $m_{jll(+)}^2(m_{ll}^2)$
at point $S$:
\beq
\left(\frac{d m_{jll(+)}^2}{d m_{ll}^2} \right)_{m_{ll}^2=0} = \frac{R_{CD}-R_{AC}}{R_{CD}(1-R_{AC})}\ .
\eeq
Since the denominator is always positive, the 
sign of the derivative is determined by the 
relative size of $R_{CD}$ and $R_{AC}$. 
When $R_{CD}<R_{AC}$, the slope is negative, and $T$ does not exist.
In that case, the maximum value of $m_{jll}^2$ over the whole 
scatter plot $OVUS$ is obtained exactly at $S$, and is 
given by $s$ in eq.~(\ref{sdef}). Comparing to the first lines
in eqs.~(\ref{jlldef}) and (\ref{jlloff}),
we see that this happens precisely for the cases of $N_{jll}=1$ and $N_{jll}=5$.
In contrast, for the other four cases $N_{jll}=2,3,4,6$,
the slope at point $S$ is positive and point $T$ is well defined.
However, this does not mean that point $T$ would then necessarily 
belong to the scatter plot $OVUS$. In the off-shell case of $N_{jll}=6$, 
point $T$ clearly belongs to the scatter plot, and 
the maximum value of $m_{jll}^2$ is given by $t$ in eq.~(\ref{tdef}),
in agreement with the second line in eq.~(\ref{jlloff}).
However, in the remaining three on-shell cases $N_{jll}=2,3,4$
one has to be more careful. Since the scatter plot
is then limited by the $UV$ vertical boundary, point $T$ will be included  
only if it lies to the left of the $UV$ line, i.e. we must have
\beq
a_t < a_{\rm on}.
\eeq
Using (\ref{atdef}) and (\ref{aon}), this condition 
can be equivalently rewritten as 
\beq
(R_{BC}-R_{AB}R_{CD})(R_{AB}-R_{BD}) > 0.
\label{Njll4condition}
\eeq
Alternatively, the point $T$ will fall outside the scatter plot, whenever
\beq
a_t > a_{\rm on},
\eeq
or equivalently,
\beq
(R_{BC}-R_{AB}R_{CD})(R_{AB}-R_{BD}) < 0.
\label{Njll23condition}
\eeq
We see that whether point $T$ is included or not, depends on the sign of
the expression
\beq
(R_{BC}-R_{AB}R_{CD})(R_{AB}-R_{BD}).
\label{factors}
\eeq
Notice that the two factors entering this expression
cannot be simultaneously negative: if that were the case, we would have
\beq
\left.
\begin{array}{lcl}
R_{BC}-R_{AB}R_{CD} < 0 &\Rightarrow & m_B^2 < \frac{m_A}{m_D}\, m_C^2 \\ [3mm]
R_{AB}-R_{BD}<0         &\Rightarrow & m_A\, m_D < m_B^2
\end{array}
\right\} \Rightarrow m_D < m_C,
\eeq
which contradicts our basic assumption (\ref{masshierarchy}).
Therefore, whenever one of the two factors in (\ref{factors}) 
is negative, the other is guaranteed to be positive. Of course, 
it is also possible that both factors in (\ref{factors}) are positive to begin with.
Altogether, this leads to three different possibilities, which are 
related to the $N_{jll}=2,3,4$ cases of eq.~(\ref{jlldef}).
\begin{itemize}
\item $N_{jll}=2$. In this case, the first factor in eq.~(\ref{factors})
is negative, leading to the following logical chain
\bea
&& N_{jll}=2:~ R_{BC}-R_{AB}R_{CD} < 0 ~\Rightarrow~ R_{AB}-R_{BD} > 0   \nonumber   \\
&&\qquad\qquad \Rightarrow~ (R_{BC}-R_{AB}R_{CD})(R_{AB}-R_{BD}) < 0 ~\Rightarrow~ a_t>a_{\rm on}, 
\label{Njll2logic}
\eea
placing point $T$ outside the scatter plot. Then, the maximum value of 
$m^2_{jll}$ is obtained at point $U$ and is given by eq.~(\ref{udef}).
Since in this case $R_{BC}<R_{AB}R_{CD}<R_{AB}$, the absolute value
sign in (\ref{udef}) can be resolved as $|R_{BC}-R_{AB}|=R_{AB}-R_{BC}$
and then eq.~(\ref{udef}) simplifies to
\beq
u = m_D^2 (1-R_{BC})(1-R_{AB}R_{CD}),
\eeq
confirming the result on the second line of eq.~(\ref{jlldef}).
\item $N_{jll}=3$. In this case, it is the second factor in eq.~(\ref{factors})
which is negative:
\bea
&& N_{jll}=3:~ R_{AB}-R_{BD}<0 ~ \Rightarrow~  R_{BC}-R_{AB}R_{CD} > 0  \nonumber \\
&& \qquad\qquad \Rightarrow~ (R_{BC}-R_{AB}R_{CD})(R_{AB}-R_{BD}) < 0 ~\Rightarrow~  a_t>a_{\rm on}.
\eea
Once again, point $T$ is outside the scatter plot, and
the maximum value of $m_{jll}^2$ is obtained at point $U$
and is given by (\ref{udef}).
This time, however, $R_{AB}<R_{BD}=R_{BC}R_{CD}<R_{BC}$, and
correspondingly, $|R_{BC}-R_{AB}|=R_{BC}-R_{AB}$. 
Then, eq.~(\ref{udef}) simplifies to
\beq
u = m_D^2 (1-R_{AB})(1-R_{BD}),
\eeq
agreeing with the third line of eq.~(\ref{jlldef}).
\item $N_{jll}=4$. This is the case when both factors in
eq.~(\ref{factors}) are positive, leading to
\beq
N_{jll}=4:~
\left.
\begin{array}{l}
R_{BC}-R_{AB}R_{CD} > 0 \\ [3mm]
R_{AB}-R_{BD}>0         
\end{array}
\right\} \Rightarrow~
(R_{BC}-R_{AB}R_{CD})(R_{AB}-R_{BD}) > 0 ~\Rightarrow~  a_t<a_{\rm on}.
\eeq
Point $T$ now belongs to the scatter plot, and its coordinate
$t$ defined in (\ref{tdef}) gives the maximum value of the 
$m^2_{jll}$ distribution, in agreement with the fourth line of 
(\ref{jlldef}).
\end{itemize}

\FIGURE[t]{
\epsfig{file=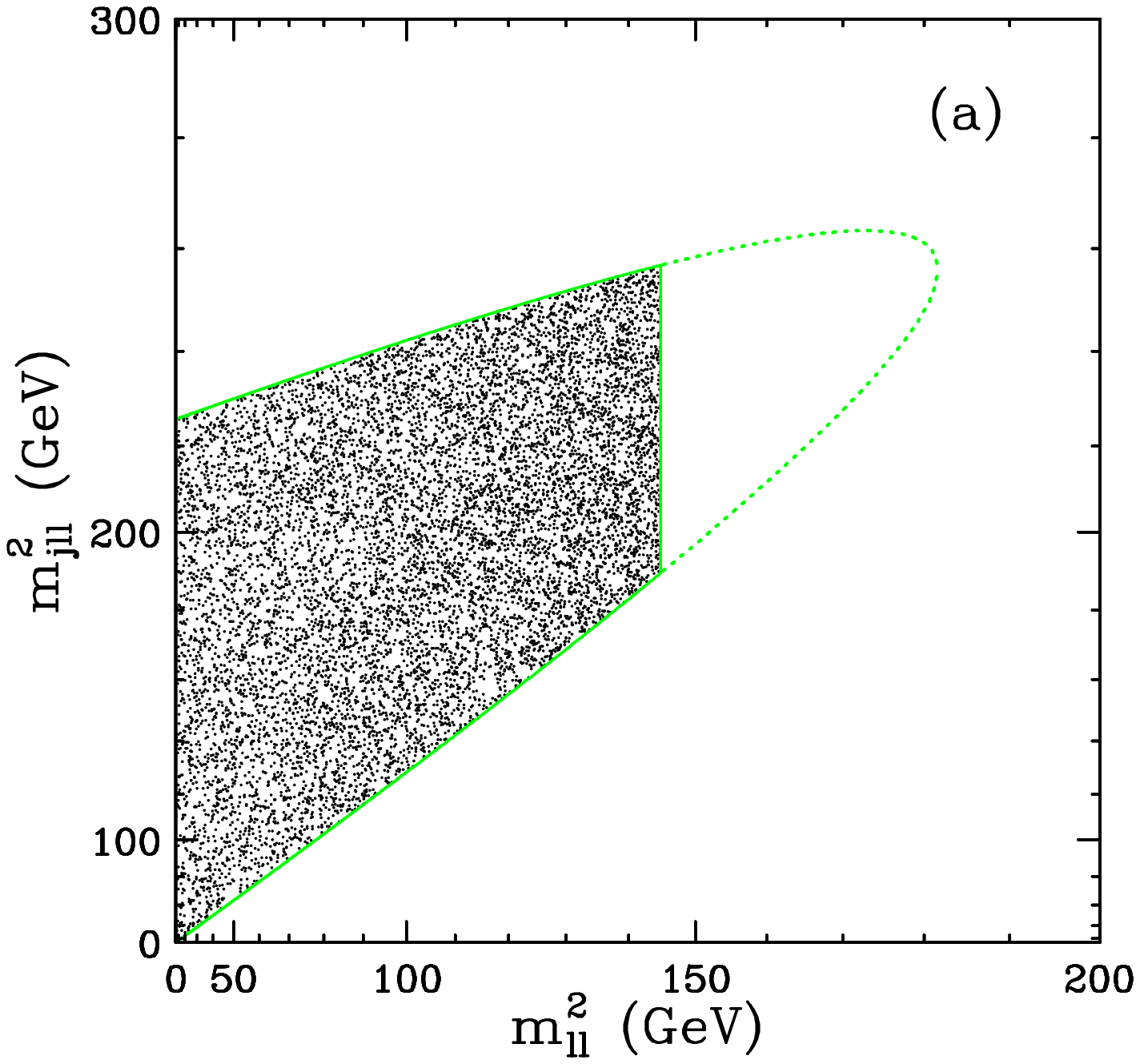, width=7.5cm}~~
\epsfig{file=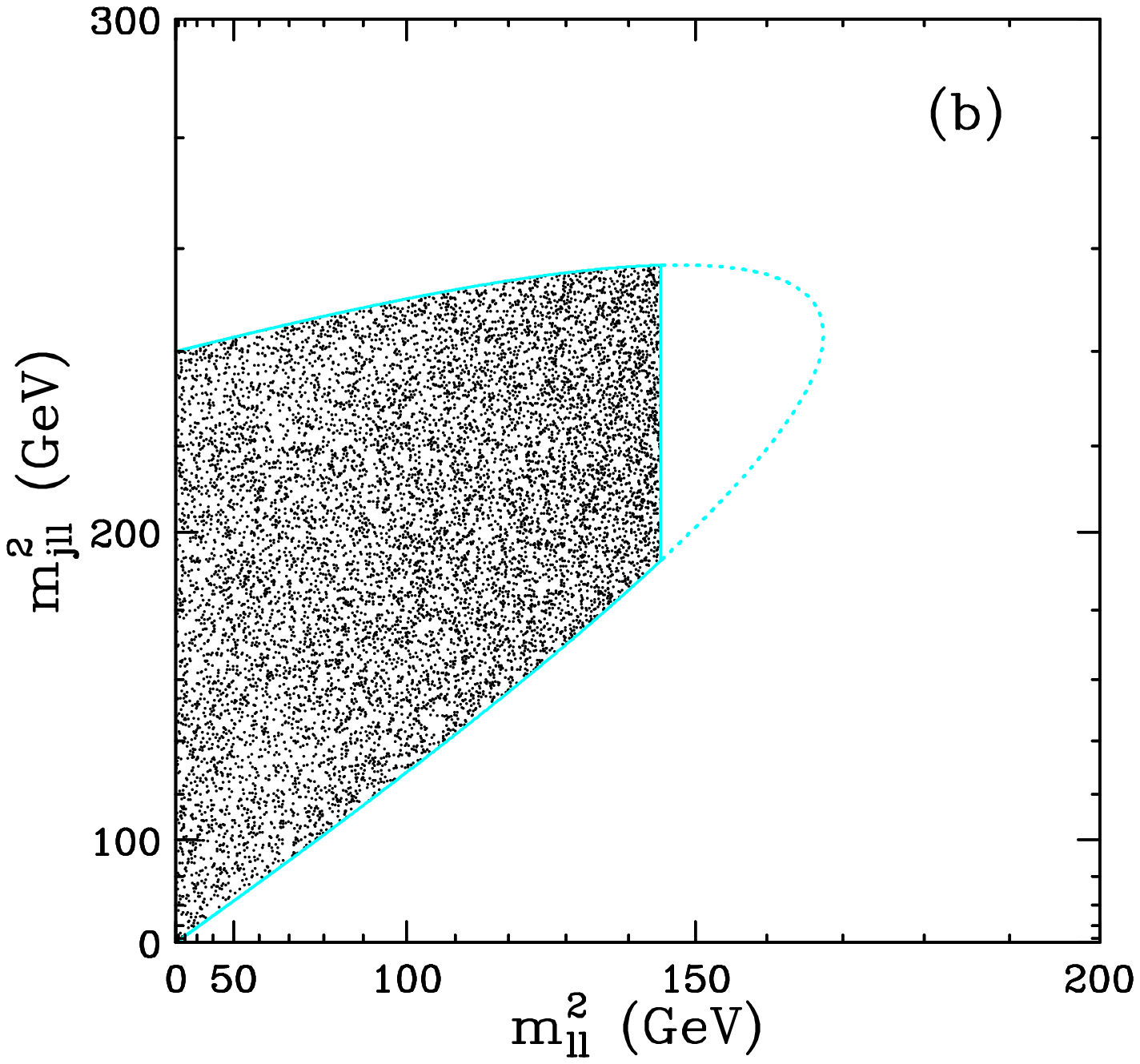, width=7.5cm}\\ [0.5cm]
\epsfig{file=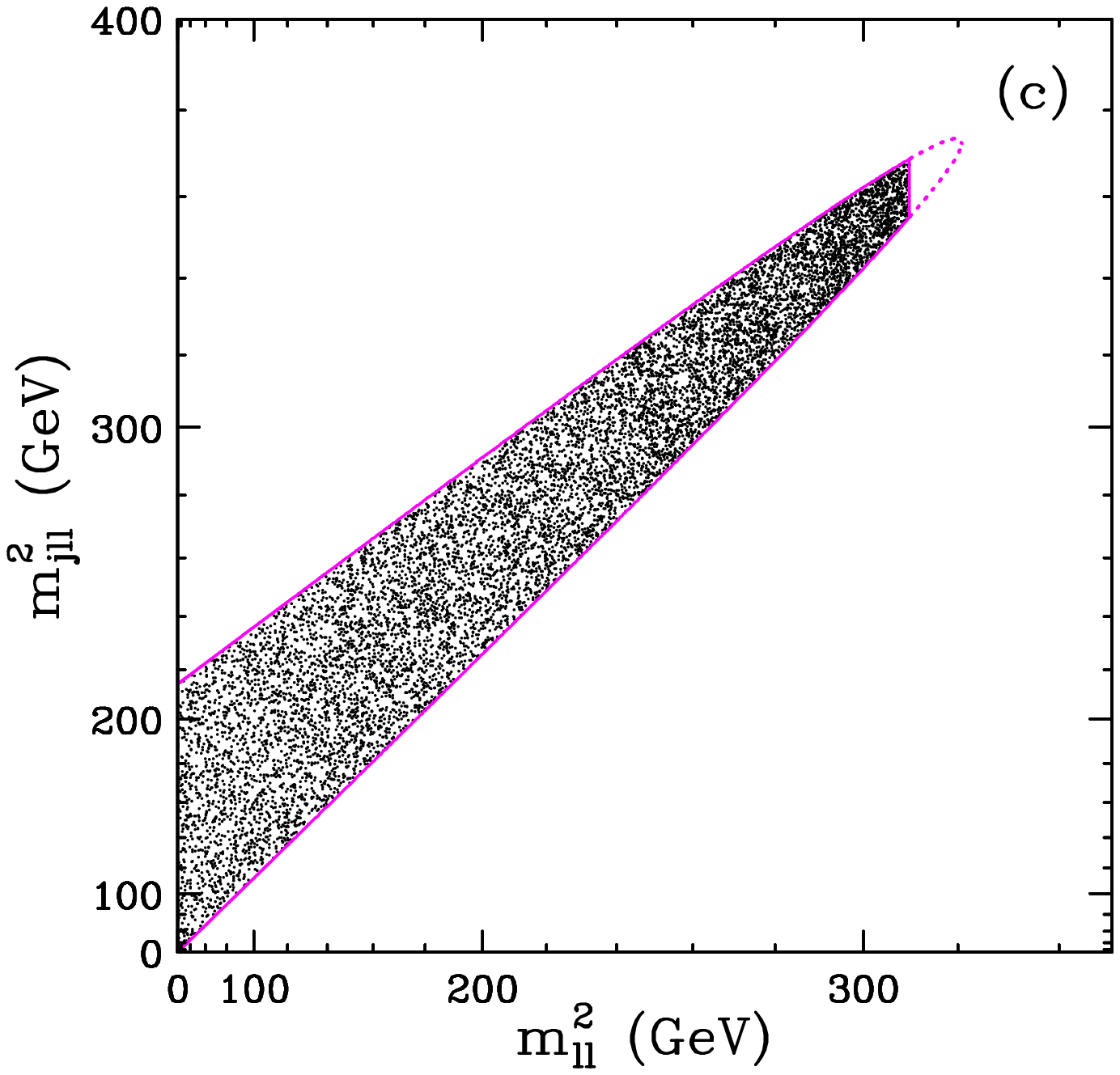, width=7.5cm}~~
\epsfig{file=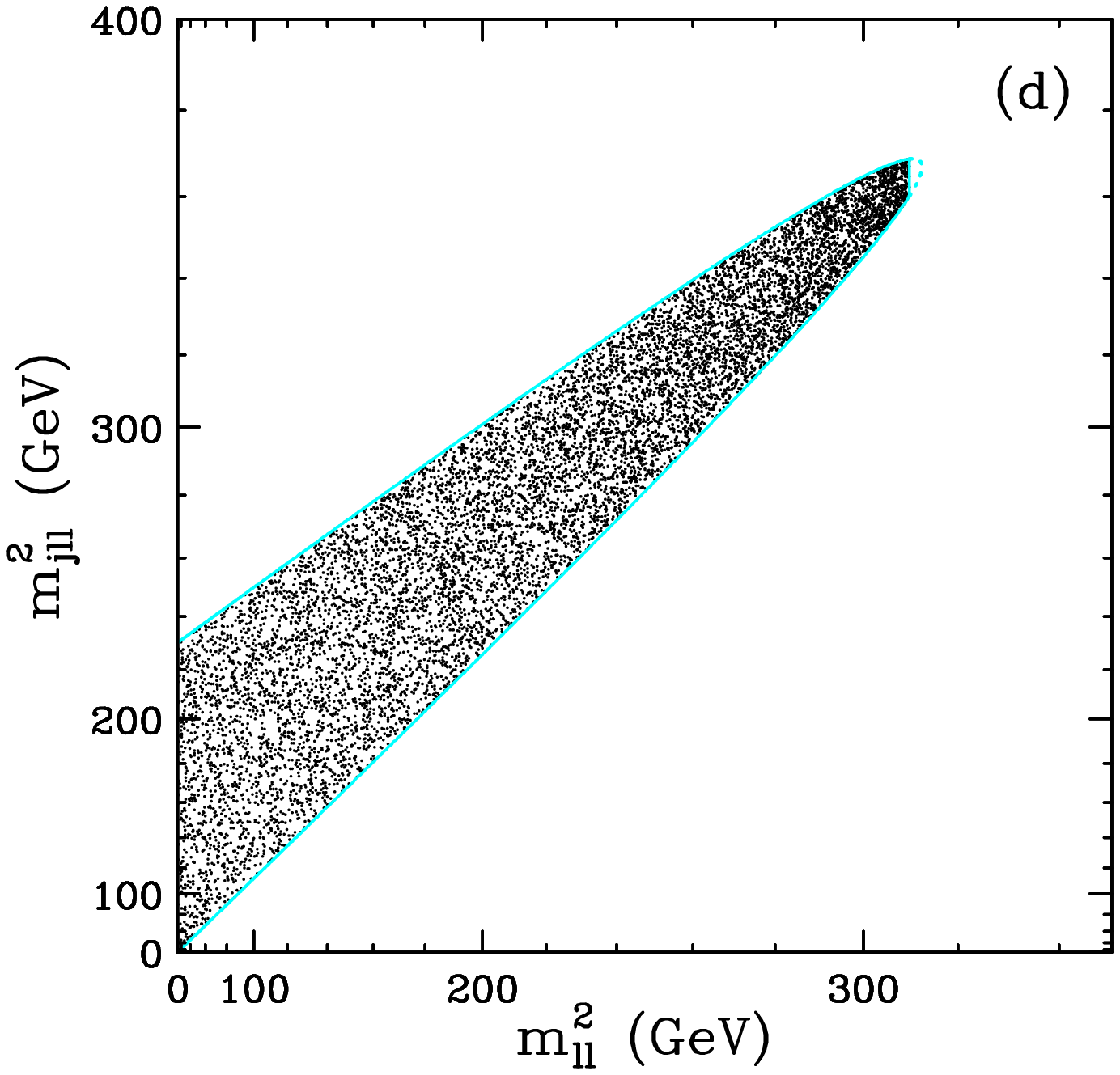, width=7.5cm}
\caption{The same as Fig.~\ref{fig:HL}, but for
$m_{ll}^2$ versus $m_{jll}^2$.
\label{fig:QLL}
}
}

Note that Fig.~\ref{fig:jll} now allows us to understand geometrically
the physical meaning of the {\em lower} threshold 
$e=(m_{jll(\theta>\frac{\pi}{2})}^{min})^2$ 
introduced in Section~\ref{sec:analytical}. 
If we restrict ourselves only to points with $m_{ll}^2>\frac{1}{2}a_{\rm on}$,
i.e. to the right of the dashed line $EE'$, the one-dimensional
$m_{jll}^2$ distribution will exhibit a {\em lower} endpoint, whose
value $e$ is given by the $m_{jll}^2$ coordinate of point $E$ in 
Fig.~\ref{fig:jll}. In the on-shell case, $e$ is given by
\bea
e &\equiv& m_{jll(-)}^2(a_{\rm on}/2)     
=  \frac{1}{4}m_D^2 \Biggl\{ (1+R_{CD})(1-R_{AB})(1-R_{BC}) \label{edef} \\
  &+& 2 (1-R_{CD})(1-R_{AC})
   -(1-R_{CD})\sqrt{(1+R_{AB})^2 (1+R_{BC})^2-16 R_{AC}}\Biggr\} , \nonumber
\eea
while in the off-shell case $e$ is given by
\bea
e \equiv m_{jll(-)}^2(a_{\rm off}/2) 
&=& \frac{1}{4}m_D^2(1-\sqrt{R_{AC}}) 
\bigg\{ 2R_{CD}(1-\sqrt{R_{AC}}) \label{edefoff}
\\[4mm]
&+& (1-R_{CD})\bigg(3+\sqrt{R_{AC}}-\sqrt{1+R_{AC}+6\sqrt{R_{AC}}}\bigg)\bigg\}.
\nonumber
\eea
It is not difficult to see that eqs.~(\ref{edef}) and (\ref{edefoff})
are identical to (\ref{jllthetadef}) and (\ref{jllthetaoff}), correspondingly.

The newly introduced quantities $s$, $t$, $u$ and $v$ 
can be directly observed experimentally\footnote{In principle, 
one can also measure indirectly the locations of points $W$ and 
$T$, even when they fall outside the observable scatter 
plot. Since the analytical expressions (\ref{upper}) and (\ref{lower})
for the boundary lines are already known, one can fit them to the 
observable portions on the scatter plot, and then extrapolate the 
obtained analytical fit into the kinematically inaccessible region, 
thus obtaining the ``would-be'' positions of $T$ and $W$.} 
on the scatter plot of Fig.~\ref{fig:jll}.
Table~\ref{table:dup2} lists their square root values
for our four duplicate study points $P_{31}$, $P_{23}$, 
$P_{32}$ and $P'_{23}$. As expected, the value of $u$ 
is matched identically for each pair. However, the two other 
directly observable quantities $s$ and $v$ differ, and in principle 
can be used to resolve the duplication. This is illustrated in Fig.~\ref{fig:QLL},
where we plot the two-dimensional distribution (\ref{d2Gammajll})
for each duplicated example: (a) $P_{31}$, (b) $P_{23}$, 
(c) $P_{32}$ and (d) $P'_{23}$.
Unlike Fig.~\ref{fig:HL}, here the differences between 
the scatter plots for each duplicated pair
are only quantitative, and may be difficult to
observe in practice. The fact that the plots look 
similar is not very surprising, given our earlier discussion.
Notice that duplication occurs only in regions with
$N_{jll}=2$ or $N_{jll}=3$. In both cases, the 
shape of the $(m_{ll}^2,m_{jll}^2)$ scatter plot 
is rather similar: the slope at point $S$ is positive, 
and the upper boundary $SU$ is cut off before it reaches 
the local maximum at point $T$. Furthermore, the duplication 
analysis ensures that the rightmost vertical boundary $UV$ 
occurs in the same location $a_{\rm on}$.

\section{Summary and outlook}

We now summarize the main results of the paper and 
discuss possible directions for future investigations.
Our main results are as follows:
\begin{itemize}
\item In Section~\ref{sec:inverse} we provided analytical 
inversion formulas which allow the immediate calculation of 
the mass spectrum $m_A$, $m_B$, $m_C$ and $m_D$
in terms of a set of four measured invariant mass endpoints $\{a,c,d,e\}$.
Our formulas are valid in {\em all} parameter space regions,
since we do not use the endpoint $b=m_{jll}^{max}$, which 
is problematic in regions (3,1), (3,2) and (2,3), see eq.~(\ref{bad}).
\item Once the endpoint $m_{jll}^{max}$ is eliminated 
from the discussion, we only need to consider 4 different 
cases, ${\cal R}_i$, $i=1,2,3,4$, as
illustrated with the color-coded regions in Fig.~\ref{fig:regions}.
In contrast, previous studies which made use of the $m_{jll}^{max}$
endpoint \cite{Gjelsten:2004ki,Gjelsten:2005aw,Miller:2005zp} 
were forced to consider all 11 different 
possibilities $(N_{jll},N_{jl})$ shown in Fig.~\ref{fig:regions}.
\item We investigated analytically the possibility of finding 
multiple solutions for the mass spectrum, even when a perfect experiment
can measure the values for {\em all five} invariant mass endpoints
$\{a,b,c,d,e\}$ with zero error bars. Although we still had to 
consider separately each of the four different cases ${\cal R}_i$, 
we found that in most of the parameter space the spectrum is uniquely
determined. Unfortunately, there is also a certain portion of parameter 
space, illustrated in Fig.~\ref{fig:xmin}, where one finds an exact 
duplication, i.e. two very different mass spectra yield identical values 
for all five measurements $\{a,b,c,d,e\}$. The situation is only
going to get worse, once we take into account the inevitable 
experimental errors on the endpoint measurements, which can only
proliferate the number of candidate solutions. Our results 
show that the conventional method of invariant mass endpoints 
may not be sufficient and one needs to look for new fresh ideas.
\item The main goal of this paper is to advertise a new approach
to the study of the usual invariant mass distributions. In particular,
we point out that the multivariate invariant mass distributions
contain a lot more useful information than the individual 
one-dimensional histograms, which are usually considered. 
As two illustrative examples, we discussed the two-dimensional 
$\{m^2_{j\ell (lo)},m^2_{j\ell (hi)}\}$ distribution in
Section~\ref{sec:hivslo} and the two-dimensional 
$\{m^2_{\ell\ell},m^2_{j\ell\ell}\}$ distribution 
in Section~\ref{sec:jllvsll}. The former is always bounded by
straight lines (see Fig.~\ref{fig:lohi}), while the latter is bounded by the 
hyperbola given by (\ref{upper}) and (\ref{lower}), and
(in the on-shell case only) by the straight line $UV$ in Fig.~\ref{fig:jll}.
\item The two-dimensional distributions exhibit two useful features.
First, their shapes, i.e.~the locations and orientations of their boundary 
lines, are characteristic of the corresponding parameter space region 
${\cal R}_i$, as shown in Figs.~\ref{fig:lohi} and \ref{fig:jll}.
This observation can be used to identify the relevant parameter 
space region, and resolve potential ambiguities in the extraction 
of the mass spectrum. Second, the boundary lines exhibit a number 
of special points, whose coordinates can in principle be measured,
providing additional experimental information about the mass spectrum.
For example, in the $\{m^2_{j\ell (lo)},m^2_{j\ell (hi)}\}$ scatter 
plots of Fig.~\ref{fig:lohi} one may identify points $F$, $P$, $N$ and $Q$,
and correspondingly measure their $m^2_{jl(hi)}$ coordinates, which 
(in the on-shell case) are given by 
\bea
f &=& m_D^2\, (1-R_{CD})\, (1-R_{AB})\, , \label{f} \\ [2mm]
p &=& m_D^2\, (1-R_{CD})\, R_{BC}\,(1-R_{AB})\, , \label{p} \\ [2mm]
n &=& m_D^2\, (1-R_{CD})\, (1-R_{BC})\, , \label{n} \\ [2mm]
q &=& m_D^2\, (1-R_{CD})\, \frac{1-R_{AB}}{2-R_{AB}} \, . \label{q}
\eea
Similarly, on the $\{m^2_{\ell\ell},m^2_{j\ell\ell}\}$ scatter plot in 
Fig.~\ref{fig:jll} one may identify the points $S$, $U$, $V$, $E$, and (sometimes)
$T$ and $W$. Their $m^2_{jll}$ coordinates are given by
\bea
s   &=&  m_D^2 \left(1-R_{CD}\right)\left(1-R_{AC}\right), \label{s} \\ [2mm]
u   &=&  \frac{1}{2}\, m_D^2 \biggl[(1+R_{CD})(1-R_{BC})(1-R_{AB})  \label{u} \\ [2mm]
    && \qquad\quad +\, (1-R_{CD})(1-R_{AC}+|R_{BC}-R_{AB}|)  \biggr], \nonumber \\ [2mm]
v   &=&  \frac{1}{2}\, m_D^2 \biggl[(1+R_{CD})(1-R_{BC})(1-R_{AB})  \label{v}\\ [2mm]
    && \qquad\quad +\, (1-R_{CD})(1-R_{AC}-|R_{BC}-R_{AB}|)  \biggr], \nonumber \\ [2mm]
e   &=&  \frac{1}{4}m_D^2 \biggl[ (1+R_{CD})(1-R_{AB})(1-R_{BC})+ 2 (1-R_{CD})(1-R_{AC}) \label{e} \\
    && \qquad\quad -(1-R_{CD})\sqrt{(1+R_{AB})^2 (1+R_{BC})^2-16 R_{AC}}\biggr] , \nonumber \\ [2mm]
t   &=&  m_D^2 \left(1-\sqrt{R_{AD}}\, \right)^2, \label{t} \\ [2mm]
w   &=&  m_D^2\, \left(1-R_{CD}\sqrt{R_{AC}}\right)\left(1-\sqrt{R_{AC}}\right). \label{w}
\eea
The advantage of the new approach is apparent from eqs.~(\ref{f}-\ref{w}).
Including the dilepton invariant mass endpoint $a$, the set of 
potential invariant mass endpoint measurements has now expanded
to 11:
\beq
\{ a,\, f,\, p,\, n,\, q,\, s,\, u,\, v,\, e,\, t,\, w \} 
\label{11meas}
\eeq
instead of the original five:
\beq
\{ a,\, b,\, c,\, d,\, e \} .
\label{new5meas}
\eeq
Of course, the endpoints in (\ref{11meas}) are not independent
from each other, since they are all given in terms of only 4
input parameters (\ref{Rspace}). Nevertheless, it is certainly 
preferable to have as many measurements as possible. The 
redundancy of information is a virtue, since it helps to
improve the precision of the mass determination.
\item The inversion formulas may simplify considerably, if we replace 
$e$, whose analytical expression (\ref{e}) is rather complicated, 
with some of the other measurements in (\ref{f}-\ref{w}).
One such example is shown in Appendix~\ref{app:nearfar}, where
we start from the set $\{a,f,p,n\}$, and obtain 
a very simple result (\ref{app:ma}-\ref{app:md}) for the inversion.
\item An important advantage of the two-dimensional approach is that
one can readily resolve the ambiguity between the endpoints 
of the $m_{jl_f}^2$ and the $m_{jl_n}^2$ distributions.
Indeed, notice that the endpoints (\ref{f}) and (\ref{n}),
are region-independent, and can be directly observed from the 
boundary lines. This removes the need to consider the different 
parameter space regions ${\cal R}_i$ one by one. The possibility 
of distinguishing the $jl_n$ and $jl_f$ invariant mass endpoints 
from two-dimensional scatter plots was also suggested in
Refs.~\cite{Luc} and \cite{Costanzo:2009mq}, where
the $\{m_{ll}^2,m_{jl}^2\}$ distribution was used instead.
\item Another advantage of the two-dimensional representation
of the data is that one can then perform a fit to the 
boundary lines of the scatter plot instead of 
a fit to the endpoints in the one-dimensional distributions.
This improves the precision of the mass determination, 
as demonstrated in \cite{Costanzo:2009mq} for the SPS1a 
SUSY benchmark example.
\end{itemize}

In conclusion, we outline several directions for future investigations.
\begin{itemize}
\item[$\star$] Perhaps the most pressing question is whether and how well
the method proposed here will survive the experimental
complications of a full-blown analysis including detector
simulation, backgrounds from Standard Model as well as SUSY combinatorics,  
the finite widths of the particles $B$, $C$ and $D$,
the varying population density of the scatter plots, etc.
This is currently under study in the CMS SUSY working group
and results will be presented in a separate publication.
\item[$\star$] In this paper we limited ourselves to the analysis of the 
boundary {\em lines} of two-dimensional distributions. However, the method 
can be easily generalized by including one more dimension and studying
the boundary {\em surface} of the three-dimensional distribution
(\ref{d3Gdm}). A similar generalization was already shown to be
beneficial in the case of spin measurements \cite{Athanasiou:2006hv}. 
\item[$\star$] One could also consider other choices of 
two-dimensional distributions, for example
$\{m^2_{ll},m^2_{jl}\}$, $\{m^2_{jl},m^2_{jll}\}$
\cite{Luc}, or $\{m^2_{ll},m^2_{jl(lo)}\}$,
$\{m^2_{ll},m^2_{jl(hi)}\}$ \cite{Costanzo:2009mq}.
Those distributions also allow to discriminate between
the ``near'' and ``far'' lepton endpoints, and 
will contribute even more data points to the set (\ref{11meas}).
\item[$\star$] One could also generalize the method to a 
longer decay chain, e.g. one which starts with a gluino 
\cite{Gjelsten:2005aw}.
\end{itemize}

\acknowledgments
We thank A.~Barr, H.-C.~Cheng, P.~Konar, K.~Kong, C.~Lester, 
F.~Moortgat, M.~Nojiri, L.~Pape and M.~Peskin for useful discussions.
This work is supported in part by a US Department of Energy 
grant DE-FG02-97ER41029. 

\appendix
%%%%%%%%%%%%%%%%%%%%%%%%%%%%%%%%%%%%%%%%%%%%%%%%%%%%%%%%%%%%%%%
\section{Appendix: \ Simple inversion formulas 
in regions ${\cal R}_1$, ${\cal R}_2$  and ${\cal R}_3$}
\label{app:nearfar}
%\addcontentsline{toc}{section}{Appendix:  \ }
\allowdisplaybreaks
\renewcommand{\theequation}{A.\arabic{equation}}
\setcounter{equation}{0}

In Sec.~\ref{sec:hivslo} have saw that the shape analysis of a 
$(m_{jl(lo)}^2,m_{jl(hi)}^2)$ scatter plot alone
reveals the values of 
\bea
f  &\equiv& (m_{jl_f}^{max})^2 = m_D^2 (1-R_{CD})(1-R_{AB}), \\ [2mm]
p&\equiv& \left(m_{jl_f}^{(p)}\right)^2   = m_D^2 R_{BC}(1-R_{CD})(1-R_{AB}) = fR_{BC}, \\ [2mm]
n  &\equiv& (m_{jl_n}^{max})^2 = m_D^2 (1-R_{CD})(1-R_{BC}), 
\eea
in each of the three on-shell regions ${\cal R}_1$, ${\cal R}_2$  and ${\cal R}_3$.
In addition, in regions ${\cal R}_2$  and ${\cal R}_3$ one also has a fourth measurement
\beq
q  \equiv \left(m_{jl(eq)}^{max}\right)^2 = m_D^2\, (1-R_{CD})\, \frac{1-R_{AB}}{2-R_{AB}} \, .
\eeq
Given these four measurements, it is worth asking whether the spectrum of four masses
$m_A$, $m_B$, $m_C$ and $m_D$ can be uniquely determined based on the 
$(m_{jl(lo)}^2,m_{jl(hi)}^2)$ scatter plot alone. Unfortunately, this is not possible, 
since the four measurements $f$, $p$, $n$ and $q$ are not all independent, due to the
constraint (\ref{fenf0}). Therefore, one more independent measurement is needed.

Fortunately, the dilepton mass edge measurement is both robust and 
on-shell-region-independent. Thus adding
\beq
a\equiv (m_{ll}^{max})^2 = m_D^2 R_{CD} (1-R_{BC})(1-R_{AB}),
\eeq
we obtain a set of 4 measurements 
\beq
\{a, f, p, n\}\equiv \{(m_{jl_f}^{max})^2,\left(m_{jl_f}^{(p)}\right)^2,(m_{jl_n}^{max})^2,(m_{ll}^{max})^2\},
\eeq
which can be easily inverted to obtain the spectrum:
\bea
R_{AB} &=& 1 - \frac{f-p}{n},  \\ [2mm]
R_{BC} &=& \frac{p}{f},  \\ [2mm]
R_{CD} &=& \left(1+\frac{f-p}{a}\right)^{-1},  \\ [2mm]
m_D^2  &=& \frac{a\,f\,n}{(f-p)^2} \left(1 + \frac{f-p}{a}\right).
\eea
In terms of the actual masses we get
\bea
m_A^2 &=& \frac{a\,n\,p}{(f-p)^2} \left(1 - \frac{f-p}{n}\right), \label{app:ma}\\ [2mm]
m_B^2 &=& \frac{a\,n\,p}{(f-p)^2},    \label{app:mb}  \\ [2mm]
m_C^2 &=& \frac{a\,n\,f  }{(f-p)^2},  \label{app:mc}  \\ [2mm]
m_D^2 &=& \frac{a\,n\,f  }{(f-p)^2} \left(1 + \frac{f-p}{a}\right). \label{app:md}
\eea
Notice the simplicity of these formulas in comparison to 
(\ref{mAon}-\ref{mDon}) and (\ref{G1def}-\ref{G3def}). The simplicity is mostly due to 
the fact that we are not using the measurement (\ref{jllthetadef})
whose analytical expression is rather complicated.

\end{document}